\documentclass[twocolumn,english,groupedaddress, superscriptaddress,aps,pre,10pt,floatfix]{revtex4-2}

\usepackage[T1]{fontenc}
\usepackage[utf8]{inputenc}
\usepackage{amsmath,amsthm,mathtools,amssymb}
\usepackage{graphicx}
\usepackage{dsfont}
\usepackage{url}

\usepackage[breaklinks]{hyperref}
\usepackage{todonotes}
\usepackage{orcidlink}
\usepackage{tikz}
\usepackage{xcolor}
\definecolor{mplblue}{HTML}{1f77b4}

\usepackage{bm}
\newcommand{\matr}[1]{\bm{\mathrm{#1}}}
\renewcommand{\vec}[1]{\bm{\mathrm{#1}}}

\newcommand{\EE}{\mathcal{E}}
\newcommand{\VV}{\mathcal{V}}
\newcommand{\braket}[1]{\left\langle {#1} \right\rangle}

\newtheorem{thm}{Theorem}

\begin{document}

\title{Symmetry breaking in minimum dissipation networks}

\author{Aarathi Parameswaran \orcidlink{0009-0005-8255-2152}}
\affiliation{Institute of Climate and Energy Systems: Energy Systems Engineering (ICE-1), Forschungszentrum J\"ulich, 52428 J\"ulich, Germany}
\affiliation{Bonn-Cologne Graduate School of Physics and Astronomy, University of Bonn, Bonn, Germany}
\affiliation{Center for Complex Networks and Systems Research, Luddy School of Informatics, Computing, and Engineering, Indiana University, Bloomington, Indiana 47408, USA}

\author{Pezhman Ebrahimzadeh \orcidlink{0000-0002-9919-3900}}
\affiliation{Neuromorphic Compute Nodes  (PGI-14), Forschungszentrum J\"ulich, 52428 J\"ulich, Germany}

\author{Andrea Benigni \orcidlink{0000-0002-2475-7003}}
\affiliation{Institute of Climate and Energy Systems: Energy Systems Engineering (ICE-1), Forschungszentrum J\"ulich, 52428 J\"ulich, Germany}
\affiliation{RWTH Aachen University, Aachen, Germany}

\author{Dirk Witthaut \orcidlink{0000-0002-3623-5341}}
\email{d.witthaut@fz-juelich.de}
\affiliation{Institute of Climate and Energy Systems: Energy Systems Engineering (ICE-1), Forschungszentrum J\"ulich, 52428 J\"ulich, Germany}
\affiliation{Institute for Theoretical Physics, University of Cologne, 50937 K\"oln, Germany}

\author{Iva Ba\v{c}i\'{c}  \orcidlink{0000-0003-2987-5065}}
\email{i.bacic@fz-juelich.de}
\affiliation{Institute of Climate and Energy Systems: Energy Systems Engineering (ICE-1), Forschungszentrum J\"ulich, 52428 J\"ulich, Germany}
\affiliation{Center of Excellence for Study of Complex Systems, Institute of Physics Belgrade, University of Belgrade, Belgrade, Serbia}

\begin{abstract}
Natural and engineered supply networks exhibit universal structural patterns, such as the formation of loops, yet the principles governing optimal structures remain unclear.
These patterns can be interpreted as solutions of optimization problems in which dissipation is minimized under a constraint on the available resources.
Optimal networks break the symmetry of such a problem in two distinct ways: weakly, where the topology stays symmetric but the edge capacities do not, and strongly, altering the topology itself. Both scenarios are organized by the symmetry group of the optimization problem, which determines the pattern of capacities in the weakly broken state and may constrain the residual symmetry of the strongly broken one. 
Symmetric structures are favored at intermediate fluctuation levels, i.e., noise both induces and suppresses symmetry breaking. 
Here, we show that symmetry breaking, influenced by network meshing, resource constraints and fluctuations, is a generic organizing principle of optimal transport networks.
\end{abstract}
  
\maketitle

\section*{Introduction}

Supply networks are fundamental to natural and engineered systems, including leaf venation and vascular networks~\cite{sack2013leaf,pittman2011regulation,kirkegaard2020optimal}, power grids~\cite{witthaut2022collective}, transportation systems~\cite{ganin2017resilience}, and hydraulic networks~\cite{houghtalen2016fundamentals}. Despite their diversity, these systems exhibit recurring structural patterns, notably radial and meshed architectures. Understanding the principles that govern these structures remains a central challenge in network science~\cite{havlin_challenges_2012}.

A common approach is to interpret network structures as solutions of optimization problems~\cite{katifori2010damage,ronellenfitsch2016global,cho2022recent}. In particular, minimizing dissipation under a global resource constraint can reproduce characteristic features such as leaf venation patterns~\cite{katifori2010damage}, transitions between radial and meshed networks~\cite{corson2010fluctuations,kaiser2020discontinuous, Boyd2001ISPD}, and community structure~\cite{kaiser2022dual}. Yet the role of symmetry and fluctuations in shaping optimal network architecture remains poorly understood.

Symmetry breaking is a central concept in statistical physics~\cite{schwabl2006statistical} and has recently emerged as an organizing principle in transport systems. Studies indicate that symmetry breaking can occur in optimized transport networks~\cite{patwardhan2024symmetry}. Symmetry and non-uniqueness of optima have been analyzed in structural topology optimization~\cite{Rozvany2011,Guo2012,Watada2011}. Broken symmetries can generate directional flow in fluidic vessels~\cite{winn2026unidirectional}. Such system-specific models provide detailed insight into individual mechanisms, but the general principles governing symmetry breaking across different network structures remain less clear. Group-theoretic methods provide a natural framework for this analysis. Symmetries of a network correspond to transformations that leave its structure unchanged, and these symmetries divide perturbations into independent blocks with distinct transformation properties~\cite{pecora2014, sanchezgarcia2020, sorrentino2016complete}. In convex optimization problems on graphs, symmetries guarantee symmetric optima and reduce the number of independent variables~\cite{gatermann2004symmetry,boyd2009fastest}. In dynamical systems, the block that first becomes unstable constrains the symmetry of the emerging state, an idea formalized in equivariant bifurcation theory~\cite{golubitsky2002}. Many real-world networks, including power grids, have large automorphism groups~\cite{macarthur2008, xiao2008}. Fluctuations provide a second key ingredient, as noise is known to alter network structure~\cite{folz2023noise} and can induce non-monotonic responses in nonlinear systems~\cite{gammaitoni1998stochastic,lindner2004effects}.

Here, we show that symmetry breaking provides a general organizing principle for optimal transport networks. Depending on the resource constraint, optimal networks can remain symmetric, develop unequal capacities while preserving their topology, or lose connections altogether. The network symmetries determine which symmetry-breaking patterns are possible, while the degree of meshing controls the freedom available to redistribute flows. Fluctuations provide an additional control parameter and can have a non-monotonic effect: increasing their strength can first restore symmetry and subsequently break it again. We demonstrate these mechanisms in several model networks and show how the results translate beyond idealized systems through an application to renewable-energy networks. Together, our results reveal how symmetry, meshing, resource constraints and fluctuations jointly shape optimal network architecture.

\section*{Results}

\subsection*{Optimal flow networks with fluctuating injections}
\label{sec:fundamentals}

A flow network is modeled as a weighted graph $G=(\VV,\EE)$, where edges $e \in \EE$ may represent pipes in a hydraulic network, veins in a biological supply network, or transmission lines in a power grid. Each node $n\in\VV$ has an injection $S_n$, with $S_n>0$ for sources (generators) and $S_n<0$ for sinks (consumers). The injections may fluctuate in time but are assumed to remain balanced, $\sum_n S_n=0$.

In many applications, including DC (direct current) electric circuits, hydraulic and vascular networks~\cite{houghtalen2016fundamentals,katifori2010damage}, and high-voltage power grids in the DC approximation~\cite{witthaut2022collective}, the edge flows are linear in the potential difference,
\begin{align}
    F_{n\rightarrow m}
    =
    k_{nm}(\theta_n-\theta_m).
    \label{eq:linflow}
\end{align}
Together with Kirchhoff's current law (KCL),
\begin{align}
    \sum_{m\in\VV}F_{n\rightarrow m}=S_n,
    \label{eq:kcl1}
\end{align}
these equations determine the network flows.

We collect flows $\vec F$, injections $\vec S$, potentials $\vec\theta$ and edge capacities $\vec k$ in the respective vectors, and define $\matr K=\mathrm{diag}(\vec k)$. With the node-edge incidence matrix $\matr E$~\cite{newman2018networks}, the network equations become
\begin{align}
    \vec F = \matr K\matr E^\top\vec\theta, 
    \qquad
    \vec S = \matr E\vec F,
    \label{eq:network-equations}
\end{align}
Introducing the weighted Laplacian $\matr L=\matr E\matr K\matr E^\top$, the balanced solution is
\begin{align}
    \vec\theta=\matr L^+\vec S,
    \qquad
    \vec F=\matr K\matr E^\top\matr L^+\vec S ,
    \label{eq:F-L-X}
\end{align}
where $\matr L^+$ denotes the Moore--Penrose pseudoinverse.

We consider a model of optimal supply networks introduced in~\cite{bohn2007structure,durand2007structure,corson2010fluctuations,katifori2010damage}. The model is based on the observation that the linear flows minimize the dissipation
\begin{align}
    D = \sum_{\substack{e \in \EE \\ k_e>0}} \frac{F_e^2}{k_e}\label{eq:D-definition}
\end{align}
under the constraint $\vec S = \matr E \vec F$. For a connected graph, KCL imposes $|\VV|-1$ independent constraints on the $|\EE|$ edge flows, leaving $\varsigma=|\EE|-|\VV|+1$ degrees of freedom, which can be interpreted as loop flows~\cite{kaiser2022dual}. The optimal network model assumes that the edge capacities minimize the average dissipation
\begin{align}
    \bar D
    =
    \sum_{\substack{e\in\EE\\k_e>0}}
    \frac{\langle F_e^2\rangle}{k_e}
    =
    \mathrm{Tr}(\matr L^+\matr\Sigma),
    \qquad
    \matr\Sigma=\langle\vec S\vec S^\top\rangle ,
    \label{eq:D-from-second-moments}
\end{align}
where $\langle\cdot\rangle$ denotes the expectation over injection fluctuations and $\matr \Sigma$ denotes their second moment matrix; see Methods for details. Thus, the injection statistics enter the optimization only through their second moments. We focus on quadratic dissipation, cf. Eq.~\eqref{eq:D-definition} because it arises naturally from linear flow laws. Alternative cost functions are discussed in Ref.~\cite{whittle2007networks}. The extension to mixed boundary conditions with prescribed node potentials is discussed in Supplementary Note~1.

The available resources constrain the capacities to
\begin{align}
    \mathcal K_\gamma
    =
    \left\{
        \vec k\geq0:
        \sum_{e\in\EE}k_e^\gamma\leq\kappa^\gamma
    \right\},
    \label{eq:resource-constraint}
\end{align}
where $\kappa^\gamma$ is the total budget. Since $\bar D$ decreases monotonically with increasing capacities, every optimum exhausts the budget, $\sum_e k_e^\gamma=\kappa^\gamma$. The value of the exponent $\gamma>0$ depends on the physical system. For Poiseuille flow through a cylindrical pipe, $k_e\sim r_e^4$ while the material of a thin-walled pipe of fixed wall thickness scales as $r_e$, giving $\gamma=1/4$. For a DC electric cable, both conductance and material scale with cross-sectional area, giving $\gamma=1$. Further examples are discussed in Supplementary Note~2.

The character of the optimization problem changes at $\gamma=1$. The dissipation $\bar D$ is convex in the edge capacities, and for $\gamma\geq1$ the feasible set $\mathcal K_\gamma$ is convex. For $\gamma>1$, the optimum is unique and therefore retains all symmetries of the problem, as shown below. For $0<\gamma<1$, the feasible set is non-convex, allowing multiple local minima and symmetry-broken capacity configurations. For a single, fixed injection pattern, eliminating the capacities turns the problem into the minimization of a cost function of the flows~\cite{bohn2007structure}. For $\gamma<1$, this cost is concave in the flow magnitudes, and a minimum is attained at an extreme point of the feasible set~\cite{Bazaraa2013}, such that the optimal network is loop-free~\cite{banavar2000topology,bohn2007structure,durand2007structure}. Fluctuating injections change this picture, sometimes stabilizing loops~\cite{katifori2010damage,corson2010fluctuations}.

\begin{figure}[tb]
    \centering
   \includegraphics[width = \linewidth]{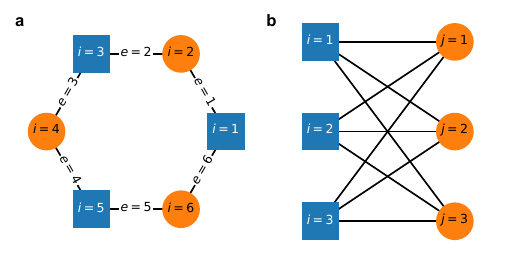}
    \caption{\textbf{Elementary networks with $3$ sources (squares) and $3$ sinks (circles).} 
\textbf{a}~Ring network with $2N = 6$ alternating sources and sinks. The symmetry group of the optimization problem is $\Gamma \cong D_3$.
\textbf{b}~Complete bipartite network $K_{3,3}$, with every source connected to every sink. The symmetry group is $\Gamma \cong S_3 \times S_3$, which permutes sources and sinks independently. 
In both cases, $\Gamma$ maps every edge onto every other edge, making all edges equivalent.}
    \label{fig:ring-networks_schem}
\end{figure}

\subsection*{General theory of symmetry and stability of optimal networks}

A symmetry of the optimization problem is a relabeling of the nodes that leaves the optimization problem invariant, i.e., unchanged. In the ring networks in Fig.~\ref{fig:ring-networks_schem}\textbf{a}, for example, rotating the ring by one step maps every node onto its neighbor and every edge onto the next edge, leaving the graph invariant. However, this rotation maps sources onto sinks, changing the optimization problem. A rotation by two steps, in contrast, leaves the optimization problem itself also invariant. Similarly, in the bipartite network in Fig.~\ref{fig:ring-networks_schem}\textbf{b}, sources and sinks can be relabeled among themselves, but exchanging the two sides would alter the injections. 
Therefore, a relabeling $g$ is a symmetry only if it preserves both the graph and the injection statistics. Since the injections enter the optimization problem only through their second moments $\matr \Sigma$, Eq.~\eqref{eq:D-from-second-moments}, the latter condition reads
\begin{align}
    \matr R_g \matr \Sigma \matr R_g^\top = \matr \Sigma,
\end{align}
where $\matr R_g$ is the permutation matrix that implements the relabeling, with $(\matr R_g)_{nm}=1$ if $g$ maps node $m$ onto node $n$ and zero otherwise.

A node relabeling also permutes the edges and hence the capacities. We first focus on edge-transitive problems, where symmetries map every edge onto every other edge. For such problems, we classify optimal networks into three types, where every edge $e \in \EE$ of the underlying graph is either active, with optimal capacity $k_e^*>0$, or inactive, $k_e^*=0$:
\begin{enumerate}
    \item symmetric: all edges have equal capacities;
    \item weakly symmetry-broken: all edges remain active, but their capacities become unequal;
    \item strongly symmetry-broken: at least one edge becomes inactive, changing the network topology.
\end{enumerate}
A symmetry-broken optimum may retain a subgroup of the original symmetries.

We now formalize these ideas and generalize them beyond edge-transitive networks. The symmetry transformations form a group $\Gamma$. For a connected graph $\mathcal G=(\VV,\EE)$,
\begin{align}
    \Gamma
    =
    \left\{
        g\in\mathrm{Aut}(\mathcal G):
        \matr R_g\matr\Sigma\matr R_g^\top=\matr\Sigma
    \right\},
    \label{eq:symmetry-group}
\end{align}
i.e. $\Gamma$ contains the graph automorphisms that preserve the injection statistics, similar to symmetries of networks with distinct node
types~\cite{stewart2003symmetry}.

Every $g\in\Gamma$ induces a permutation $\pi_g$ of the edges. For an optimal capacity configuration $\vec k^*$, the residual symmetry is
\begin{align}
    \Gamma^*
    =
    \left\{
        g\in\Gamma:
        k^*_{\pi_g(e)}=k_e^*
        \quad\forall e\in\EE
    \right\}.
    \label{eq:residual-symmetry}
\end{align}
Thus, $\Gamma^*=\Gamma$ for a symmetric optimum, whereas $\Gamma^*\subsetneq\Gamma$ signals symmetry breaking. To distinguish weak and strong symmetry breaking, define the support
\begin{align}
    \mathcal S=\{e\in\EE:k_e^*>0\}.
\end{align}
An optimum is weakly symmetry-broken if $\Gamma^*\subsetneq\Gamma$ while $\mathcal S$ remains invariant under $\Gamma$, and strongly symmetry-broken if $\mathcal S$ itself is not invariant. These cases, together with the symmetric case, are mutually exclusive and exhaustive. For edge-transitive problems, this reduces to the simple classification above.

Symmetry-broken optima always occur in families. If an optimum $\vec{k}^*$ is not invariant under a symmetry $g \in \Gamma$, applying $g$   to $\vec{k}^*$ generates a different configuration with the same dissipation, which is also optimal. Applying all symmetries in $g \in \Gamma$ generates the entire family of equivalent optima. Since the symmetries in $\Gamma^*$ leave $\vec{k}^*$ invariant, the family contains $|\Gamma|/|\Gamma^*|$ distinct optima, by the orbit--stabilizer theorem~\cite{Artin2011}. Symmetry breaking is thus necessarily associated with the coexistence of several minima with identical dissipation.

Convexity strongly constrains this picture. Since the optimization problem is convex for $\gamma\geq1$, a unique minimizer must retain every symmetry; otherwise applying a symmetry would generate a distinct solution with the same dissipation. More generally, we obtain the following result.

\begin{thm}
\label{thm:convex}
For $\gamma>1$, the global minimizer of $\bar D$ on $\mathcal K_\gamma$ is unique and symmetric. For $\gamma=1$, the set of global minimizers contains a symmetric configuration. If $\Gamma$ acts transitively on $\EE$, the symmetric minimizer is uniform,
\begin{align}
    k_e=|\EE|^{-1/\gamma}\kappa .
\end{align}
\end{thm}
The proof is given in Supplementary Methods~1. Hence, symmetry breaking is excluded for $\gamma>1$, while at $\gamma=1$ at least one symmetric global optimum exists.

We first consider weak symmetry breaking that arises through a loss of stability of the symmetric configuration. 
To analyze the stability of an optimum, we perturb the capacities by $\vec k\rightarrow\vec k + \eta\vec\delta$, where the vector $\vec \delta$, which we call a mode, determines the direction of the perturbation and $\eta \ll 1$ is its amplitude. An admissible perturbation must preserve the constraint $\sum_e k_e^\gamma = \kappa^\gamma$. Expanding to first order in $\eta$, this requires $\vec\delta$ to belong to the tangent space
\begin{align}
    T_{\vec k}
    =
    \left\{
        \vec\delta:
        \sum_e k_e^{\gamma-1}\delta_e=0
    \right\}.
    \label{eq:tangent-space}
\end{align}
We start with edge-transitive networks and turn to the general case below. For these networks, the symmetric configuration is uniform, $k_e=k_0=|\EE|^{-1/\gamma}\kappa$. Eq.~\eqref{eq:tangent-space} then becomes $\sum_e \delta_e = 0$, i.e., $T_{\vec k}=\vec 1^\perp$. Admissible perturbations redistribute the capacity between edges without changing the total budget. By symmetry, all components of $\vec \nabla \bar D$ are equal, so $\vec \delta \cdot \vec \nabla \bar D=0$ for all $\vec\delta\in T_{\vec k}$. 
Hence, the symmetric configuration is always stationary on the constraint surface.

Weak symmetry breaking occurs when this stationary point loses local stability. It is therefore determined by the Hessian of the Lagrangian $\mathcal L$ (Methods, Eq.~\eqref{eq:lagr-methods}) restricted to $T_{\vec k}$, which we refer to as the constrained Hessian. We define
\begin{align}
    \matr G(\vec k)
    &=
    \matr E^\top\matr L^+\matr E,
    &
    \matr\Psi(\vec k)
    &=
    \matr E^\top\matr L^+\matr\Sigma\matr L^+\matr E ,
    \label{eq:G-Psi}
\end{align}
and use the shorthand $\tilde{\matr G}=\matr G(\vec 1)$ and $\tilde{\matr\Psi}=\matr\Psi(\vec 1)$. Physically, $G_{ef}$ describes the potential difference across edge $e$ induced by a unit current across edge $f$, while $\matr\Psi=\langle\Delta\vec\theta\,\Delta\vec\theta^\top\rangle$ contains the second moments of the edge potential differences. For an edge-transitive network, the diagonal elements $\tilde\Psi_{ee}$ are independent of $e$, and the constrained Hessian takes the form (Methods)
\begin{align}
    \left.
    \nabla^2\mathcal L
    \right|_{T_{\vec k}}
    =
    \frac{1}{k_0^3}
    \left[
        2\,\tilde{\matr G}\circ\tilde{\matr\Psi}
        +
        (\gamma-1)\tilde\Psi_{ee}\matr I
    \right]_{T_{\vec k}},
    \label{eq:scale-free}
\end{align}
where $\circ$ denotes element-wise multiplication. Let $h_{\min}$ denote the smallest eigenvalue of $2\,\tilde{\matr G}\circ\tilde{\matr\Psi}$ on $T_{\vec k}$. The symmetric configuration is locally stable if
$
    h_{\min}
    +
    (\gamma-1)\tilde\Psi_{ee}
    >0 ,
$
and weak symmetry breaking sets in at
\begin{align}
    \gamma_s
    =
    1-\frac{h_{\min}}{\tilde\Psi_{ee}} .
    \label{eq:gamma-s}
\end{align}

The smallest eigenvalue $h_{\min}$, however, does not tell us which mode $\vec \delta$ destabilizes the symmetric configuration. The modes can be classified by how they transform under the symmetries of the problem. For the cube network discussed below, for example, a mode may affect edges in the same way if they point in the same direction, or if they share a source. 
Mathematically, every such symmetry type corresponds to an irreducible representation $\rho$ of $\Gamma$ (see Supplementary Methods~2 for details). The space of perturbations decomposes into independent subspaces $W_\rho$, one for each $\rho$. 
Since symmetry operations do not change the dissipation or the resource constraint, the constrained Hessian $\nabla^2 \mathcal{L} \big|_{T_{\vec k}}$ at a symmetric configuration commutes with every edge permutation $\pi_g$, $g \in \Gamma$. The Hessian thus leaves each subspace invariant and can be written in a block diagonal form, with one block for each nonzero $W_\rho\cap T_{\vec k}$. We refer to both $W_\rho$ and the corresponding diagonal block of the Hessian as a block, and we can check stability for every block separately. 

For edge-transitive networks, the trivial representation, i.e., uniform perturbations, is excluded because it violates the resource constraint, $T_{\vec k}=\bigoplus_{\rho\neq\mathrm{triv}}W_\rho$. 
Let $h_{\min}(\rho)$ be the smallest eigenvalue of the $\rho$th block of $2\,\tilde{\matr G}\circ\tilde{\matr\Psi}$ and define the block-wise thresholds
\begin{align}
    \gamma_s(\rho) = 1-\frac{h_{\min}(\rho)}{\tilde\Psi_{ee}}.
    \label{eq:gamma-s-rho}
\end{align}
The symmetric configuration is stable if $\gamma>\gamma_s(\rho)$ for all $\rho \neq \mathrm{triv}$. As $\gamma$ decreases, the first threshold $\gamma_s$ identifies the block (symmetry type) of the modes that destabilize the symmetric configuration and thereby constrains the patterns of weak symmetry breaking. 
Similar block decompositions are standard in symmetric convex optimization~\cite{gatermann2004symmetry,boyd2009fastest}, cluster synchronization~\cite{pecora2014,sorrentino2016complete} and equivariant bifurcation theory~\cite{golubitsky2002}.

If $\Gamma$ does not act transitively on the edges, the edge set decomposes into distinct orbits, $\mathcal E=\bigcup_{a=1}^r\mathcal E_a$. A symmetric capacity configuration then satisfies $k_e=x_a$ for all $e\in\mathcal E_a$, allowing different capacities on different orbits without breaking symmetry (see Supplementary Note 4 for an example). In this case, Eq.~\eqref{eq:gamma-s-rho} no longer applies, since $\Psi_{ee}$ can differ between orbits. The stability analysis must include an additional block corresponding to symmetric perturbations $T_{\vec k}\cap W_{\mathrm{triv}}$, which redistribute resources between different edge orbits without breaking symmetry.

We now turn to strong symmetry breaking.  Consider a stationary configuration with an inactive edge, $k_e=0$, that is locally stable with respect to variations of its active capacities. Reactivating this edge with a small capacity $\varepsilon$, while rescaling the active capacities to maintain the resource constraint, changes the dissipation by
\begin{align}
    \bar D(\varepsilon)-\bar D(0) = -\Psi_{ee}\,\varepsilon + \nu\,\varepsilon^\gamma + \dots,
    \quad
    \nu = \frac{\bar D(0)}{\gamma\kappa^\gamma} > 0,
    \label{eq:reactivation}
\end{align}
where $\nu$ is the Lagrange multiplier of the resource constraint (Methods, Eq.~\eqref{eq:lagr-methods}), evaluated at the stationary configuration with the inactive edge (Supplementary Methods~3). For $0<\gamma<1$, the second term dominates as $\varepsilon\to0^+$, such that the inactive edge is protected against reactivation. Conversely, for $\gamma>1$ the linear term dominates whenever $\Psi_{ee}>0$, and configurations with inactive edges are generically not local minima. 
Moreover, when $\gamma \rightarrow0^+$, the resource constraint effectively counts active edges, such that the global optimum approaches a minimum-support admissible network. For meshed edge-transitive networks, such a sparse optimum is necessarily strongly symmetry-broken. Detailed derivations are provided in Supplementary Note~3.

These results suggest a generic progression as $\gamma$ is lowered. Strongly symmetry-broken local minima become possible below $\gamma=1$, while sparse configurations become increasingly favorable and eventually one becomes globally optimal. The transition may occur directly from the symmetric configuration or via an intermediate weakly symmetry-broken phase. Further exchanges of global optimality between different sparse configurations are possible before the minimum-support limit is approached as $\gamma\rightarrow0^+$.

The transition to strong symmetry breaking is typically discontinuous~\cite{kaiser2020discontinuous}. This is most transparent for a direct transition from the symmetric configuration in a finite edge-transitive network: the symmetric configuration has $k_e=k_0>0$ for every edge, whereas a strongly symmetry-broken state has $k_{e'}=0$ for at least one edge $e'$. The transition therefore occurs through an exchange of global optimality between competing minima. Denoting their optimized dissipations by $\bar D_{\rm sym}^*(\gamma)$ and $\bar D_{\rm sb}^*(\gamma)$, the transition point $\gamma_c$ satisfies
\begin{align}
    \bar D_{\rm sym}^*(\gamma_c)
    =
    \bar D_{\rm sb}^*(\gamma_c).
\end{align}

\subsection*{Model networks and fluctuating injections}

We illustrate symmetry breaking using three representative network classes. A ring with alternating sources and sinks (Fig.~\ref{fig:ring-networks_schem}) provides the simplest non-trivial example, with a comparatively small symmetry group and cycle rank $\varsigma=1$. Complete bipartite networks represent the opposite limit, with a large symmetry group and rapidly growing cycle rank. Finally, the cube possesses several distinct spatial symmetries and allows us to study strong symmetry breaking with nontrivial residual symmetry. A further example of an extended multilayer network is given in Supplementary Note~4.

We consider sources $\VV_g$ and sinks $\VV_c$, with $N_g=|\VV_g|$ and $N_c=|\VV_c|$. Sink loads are fixed as $S_i=-\mu$, whereas source injections fluctuate according to
\begin{align}
    S_i=\frac{N_c}{N_g}\mu+X_i,
    \qquad i\in\VV_g,
\end{align}
with $\langle X_i\rangle=0$ and $\sum_{i\in\VV_g}X_i=0$. The fluctuation strength is quantified by
\begin{align}
    \beta^2=-\langle X_iX_j\rangle,
    \qquad i\neq j\in\VV_g .
\end{align}
The second moments are then fully determined by $\mu$, $\beta$, and the network size (see Methods).

Only the ratio $\beta/\mu$ affects the stability thresholds, while an overall rescaling of the capacities changes the constrained Hessian only by a positive factor. We can therefore set $\kappa=\mu=1$ and use $(\beta,\gamma)$ as the two control parameters.

\subsection*{Symmetry breaking in networks with ring geometry}

\begin{figure*}[tb]
    \centering
    \includegraphics[width=\linewidth]{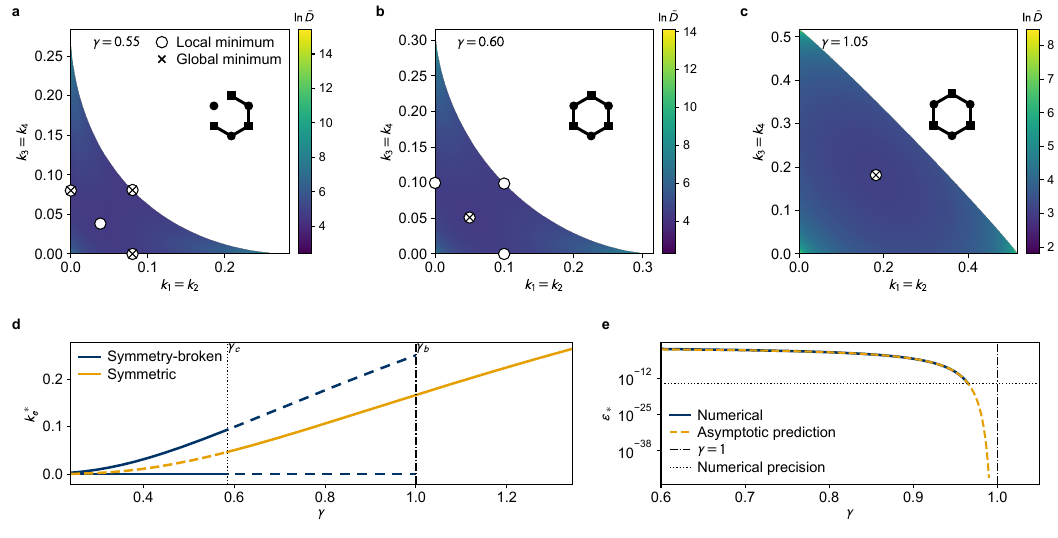}
    \caption{
    \textbf{Symmetry breaking by an exchange of global optimality in a ring network.}
    Ring of $N=3$ sources and $N=3$ sinks with $\mu=0$ and $\kappa=1$. The capacities are equal on the two edges adjacent to each sink, $k_1=k_2$, $k_3=k_4$ and $k_5=k_6$. The resource constraint fixes $k_5$, leaving $k_1$ and $k_3$ as independent variables.
    \textbf{a}--\textbf{c} Logarithm of the expected dissipation, $\ln \bar D(k_1,k_3)$ (color scale), over the feasible domain for $\gamma=0.55$ (\textbf{a}), $\gamma=0.6$ (\textbf{b}) and $\gamma=1.05$ (\textbf{c}). The upper end of the color scale is attained only near the corners of the feasible domain, where $\bar D$ diverges. White circles mark local minima and a black cross marks global minima. Insets show the optimal network, with squares denoting sources and circles denoting sinks. In \textbf{a}, the three symmetry-broken minima on the domain boundary, each with one disconnected sink, are global optima. The symmetric configuration is a local minimum with higher dissipation. In \textbf{b}, the minima have exchanged roles. The symmetric configuration is globally optimal. In \textbf{c}, the problem is convex and the symmetric configuration is the unique minimum.
    \textbf{d} Optimal edge capacities $k_e^*$ as a function of $\gamma$ for the symmetry-broken (dark blue) and symmetric (orange) states. The solid lines correspond to global minima and the dashed lines to local minima. 
    The symmetry breaking transition happens for $\gamma_c=\ln(3/2)/\ln(2)$ where global optimality is exchanged (dotted vertical line). The local minima disappear in a bifurcation at $\gamma_b=1$ (dash-dotted vertical line). 
    \textbf{e} 
    Reactivation capacity $\varepsilon_*$ of the two inactive edges at which the change in dissipation $\bar D(\varepsilon)-\bar D(0)$ changes sign, shown as a function of $\gamma$ on a logarithmic scale. The dark blue solid line shows numerical results, while the orange dashed line shows the leading-order asymptotic prediction. The dash-dotted vertical line marks $\gamma=1$, while the dotted horizontal line denotes the numerical precision. As $\gamma \to 1^-$, $\varepsilon_* \to 0$ rapidly, and the symmetry-broken states become increasingly difficult to resolve numerically.
    }
    \label{fig:results_mu0_N3}
\end{figure*}

We first consider a ring with alternating sources and sinks, illustrated in Fig.~\ref{fig:ring-networks_schem} for $N_c=N_g=3$. The simple geometry allows for a largely analytic characterization of the phase diagram. We label the nodes consecutively, with sources at odd and sinks at even indices, and denote the edge $(e,e+1)$ by $e$.

The symmetries preserving the alternating source--sink statistics form the group $\Gamma\cong D_N$, generated by rotations by two nodes and appropriate reflections. These transformations act transitively on the edges. Hence, the symmetric optimum has equal capacities, weak symmetry breaking corresponds to unequal but positive capacities, and strong symmetry breaking to at least one inactive edge. We investigate which phase is realized as $\gamma$ and $\beta$ are varied.

To identify the basic mechanisms, we first consider the case $\mu=0$, where $S_i=0$ for all sinks $i\in\{2,4,6,\ldots\}$. Kirchhoff's current law and Eq.~\eqref{eq:ke-corson} imply $F_e=F_{e+1}$ and $k_e^*=k_{e+1}^*$ for $e\in\{1,3,5,\ldots\}$, reducing the number of independent optimization variables by half.

For $N=2$, the loss of local stability of the symmetric configuration and the strong transition coincide at $\gamma=1$, without an intermediate weakly broken phase (Supplementary Fig.~2). For $N_c=N_g=3$, the resource constraint gives
$
    k_5
    =
    \left(
        \frac{1}{2}\kappa^\gamma
        -k_1^\gamma-k_3^\gamma
    \right)^{1/\gamma},
$
leaving $k_1$ and $k_3$ as independent variables.
Figure~\ref{fig:results_mu0_N3}\textbf{a}--\textbf{d} shows that the symmetric configuration remains a local minimum for all $\gamma>0$. For $\gamma<1$, three strongly symmetry-broken local minima additionally exist, each with two inactive edges. In each of these states, one sink node is disconnected, which is admissible because $S_i=-\mu=0$. The exchange of global optimality occurs at
\begin{align}
    \gamma_c
    =
    \frac{\ln(3/2)}{\ln 2}
    \approx0.585
\end{align}
(see Supplementary Methods~4). For $\gamma>\gamma_c$ the symmetric configuration is the global minimum, whereas for $\gamma<\gamma_c$ the three symmetry-related broken states are degenerate global minima. These broken states remain local minima up to $\gamma_b=1$, but the reactivation capacity $\varepsilon_*$ at which reactivating the two edges adjacent to the disconnected sink starts to lower the dissipation collapses as $\gamma\to1^-$ (Fig.~2\textbf{e}, see also Supplementary Methods~3 and 4).

\begin{figure*}[tb]
\centering
\includegraphics[width=\linewidth]{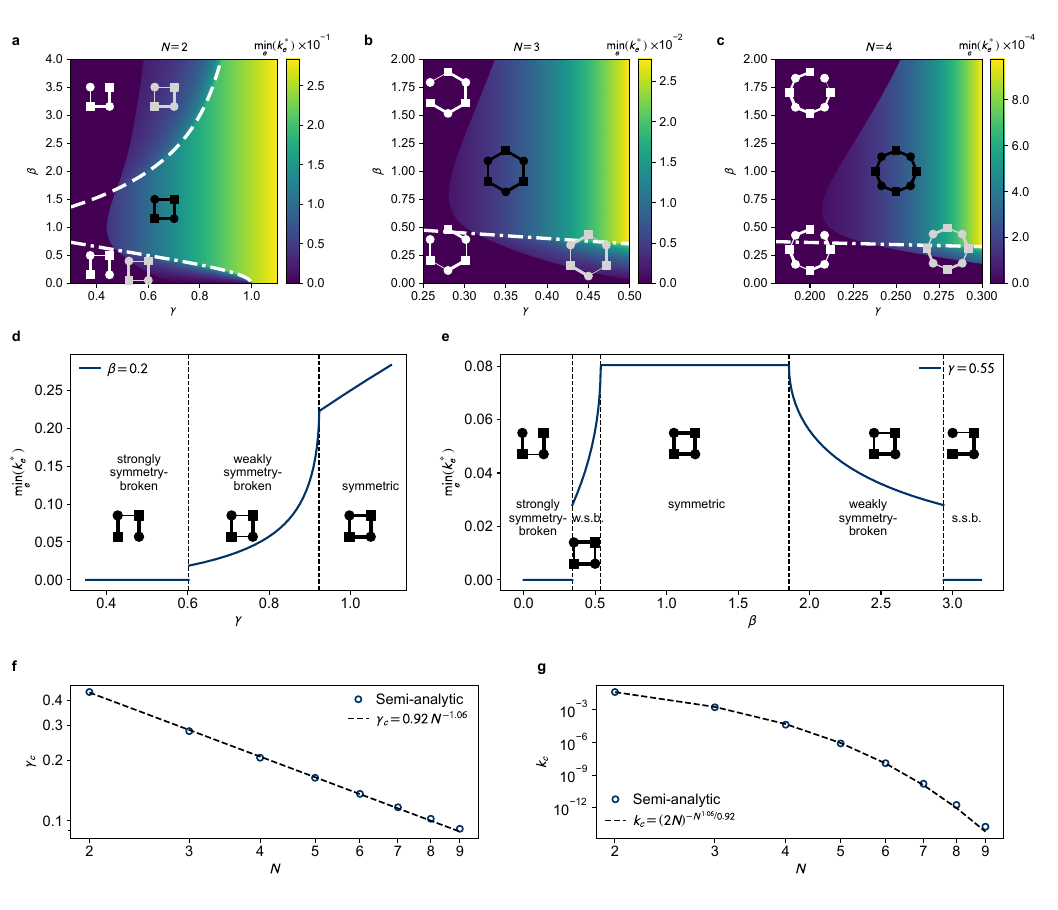}
    \caption{\textbf{Phase diagrams of optimal ring networks.} The rings have $N$ sources and $N$ sinks with alternating injections, with $\mu=1$ and $\kappa=1$ fixed. 
\textbf{a}--\textbf{c} Smallest edge capacity in the optimal network, $\min_{e}k_e^*$ (color scale), as a function of the scaling exponent $\gamma$ and the fluctuation strength $\beta$ for $N=2$ (\textbf{a}), $N=3$ (\textbf{b}) and $N=4$ (\textbf{c}). The dark region with $\min_e k_e^*=0$ indicates strong symmetry breaking, i.e., an open ring with one inactive edge. White dash-dotted lines denote the local stability threshold $\gamma_s$ of the symmetric configuration against the alternating mode. The white dashed line in \textbf{a} denotes the threshold against the sink-paired mode, which destabilizes the symmetric configuration only for $N=2$. Insets show the optimal network in each phase, with squares denoting sources and circles denoting sinks. 
\textbf{d} Smallest edge capacity as a function of $\gamma$ for $N=2$ and $\beta=0.2$. Upon increasing $\gamma$, the optimum changes from strongly symmetry-broken via weakly symmetry-broken to symmetric. Dashed vertical lines denote the transitions.
\textbf{e} Smallest edge capacity as a function of $\beta$ for $N=2$ and $\gamma=0.55$. Upon increasing $\beta$, the optimum changes from strongly symmetry-broken to symmetric and back to a different strongly symmetry-broken phase, each time via a weakly symmetry-broken phase (w.s.b.). The rightmost region, s.s.b., is strongly symmetry-broken.
\textbf{f} Transition point $\gamma_c$, the smallest $\gamma$ at which the optimal network is a closed ring, as a function of $N$. Circles show semi-analytic results and the dashed line shows the fit $\gamma_c=0.92\,N^{-1.06}$.
\textbf{g} Capacity $k_c$ of the symmetric configuration at $\gamma_c$, i.e., the jump of $\min_e k_e^*$ at the transition, as a function of $N$. The dashed line shows the prediction $k_c=(2N)^{-N^{1.06}/0.92}$ obtained from the fit in \textbf{f}.
    }
    \label{fig:phase_diagram_ring}
\end{figure*}

We now turn to the full model with $\mu=1$ and vary the fluctuation strength $\beta$. The competing configurations are
\begin{align}
    &\text{symmetric:}
    && k_e=(2N)^{-1/\gamma}\kappa
    \quad \forall e,
    \nonumber\\
    &\text{weakly broken:}
    && k_e=k_a,\;k_{e+1}=k_b
    \quad \text{alternating},
    \nonumber\\
    &\text{strongly broken:}
    && k_e=0
    \quad \text{for one edge}.
    \label{eq:symm-symm-broken-capacities}
\end{align}
For $\mu>0$ and $\beta>0$, the strongly symmetry-broken configuration is an open ring, since removing two edges would isolate a component with imbalanced injections. For $N=2$, an additional weakly symmetry-broken sink-paired configuration exists, in which the two edges adjacent to a common sink have equal capacities.

The onset of weak symmetry breaking follows from the stability criterion~\eqref{eq:gamma-s}. For the ring geometry, the only generic unstable mode is the alternating pattern, in which neighboring edge pairs acquire different capacities. Its threshold, Eq.~\eqref{eq:gamma-s-rho} is
\begin{align}
    \gamma_s (\text{alt})
    =
    \frac{3\mu^2-(N^2-1)\beta^2}
         {3\mu^2+(N^2-1)\beta^2},
\end{align}
shown as the dash-dotted line in Fig.~\ref{fig:phase_diagram_ring}. Thus, for sufficiently small fluctuations and intermediate values of $\gamma$, the symmetric configuration can lose stability through this alternating mode. For $\gamma>\gamma_s(\text{alt})$ it is locally stable, whereas for $\gamma<\gamma_s(\text{alt})$ the alternating mode is unstable. For $N\geq3$, this is the only route to weak symmetry breaking. The special case $N=2$ admits one additional sink-paired mode, which drives weak symmetry breaking for large $\beta$, shown as the dashed line in Fig.~\ref{fig:phase_diagram_ring}. The details are discussed in Supplementary Methods~6.

To determine the onset of strong symmetry breaking and construct the global phase diagram, we evaluate the expected dissipation of the three competing configurations semi-analytically (Supplementary Methods~5). The resulting phase diagrams in Fig.~\ref{fig:phase_diagram_ring} show that loss of local stability does not necessarily determine the global transition. Starting from large $\gamma$, the optimal network is symmetric. Upon lowering $\gamma$, it either passes through an intermediate weakly symmetry-broken phase or undergoes a direct transition to a strongly symmetry-broken phase. The latter transition is discontinuous and occurs when the competing minima exchange global optimality.

The phase diagram is reentrant in the fluctuation strength $\beta$ (Fig.~\ref{fig:phase_diagram_ring}\textbf{e}). As $\beta$ increases, the optimal network can change from strongly symmetry-broken to symmetric and back again, typically through intermediate weakly symmetry-broken phases. Thus, fluctuations have a non-monotonic effect, with symmetric structures favored only at intermediate noise levels.

The observed behavior can be understood from the competing transport requirements. For small $\beta$, sources and sinks can balance predominantly locally, favoring sparse connections between neighboring source--sink pairs. For large $\beta$, fluctuations between sources dominate and require an extended backbone for global balancing, again favoring asymmetric structures. Intermediate fluctuations require both local and global transport, for which the symmetric network provides the best compromise.

Finally, we examine how the transition scales with $N$. Comparing the dissipation of the symmetric and strongly symmetry-broken states for $N=2,\ldots,9$ yields
\begin{align}
    \gamma_c \sim N^{-\alpha},
    \qquad
    \alpha=1.06\pm0.01,
\end{align}
see Fig.~\ref{fig:phase_diagram_ring}\textbf{f-g}. The range of $\gamma$ in which the open ring is optimal shrinks approximately as $N^{-1}$. Nevertheless, the transition remains discontinuous for every finite $N$: the minimum edge capacity jumps from zero to a finite value at $\gamma_c$, although the magnitude of this jump decreases rapidly with system size. Details of the scaling analysis are given in Supplementary Methods~7. 

\subsection*{Symmetry breaking in a cube network}

\begin{figure*}
    \includegraphics[width = \linewidth]{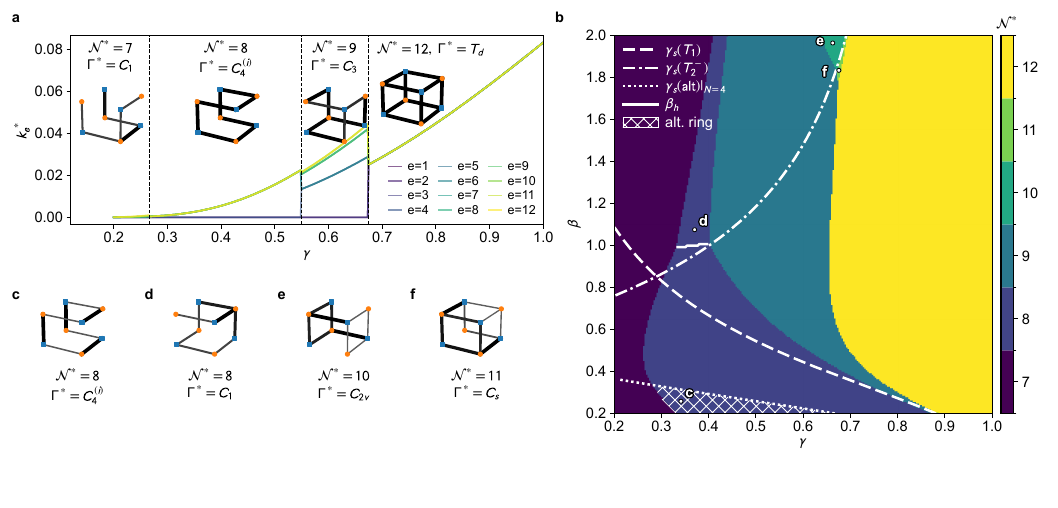}
    \caption{
    \textbf{Phases of the optimal cube network.}
\textbf{a} Optimal edge capacities $k_e^*$ as a function of the scaling exponent $\gamma$ for the fluctuation strength $\beta=0.6$. At each $\gamma$, the twelve capacities are sorted by magnitude and labelled $e=1,\dots,12$. Vertical dashed lines mark changes in the number of active edges $\mathcal{N}^*$. Insets show the optimal network in each phase: the tree ($\mathcal{N}^*=7)$, the eight-edge ring ($\mathcal{N}^*=8)$, the chiral network ($\mathcal{N}^*=9)$ and the symmetric cube ($\mathcal{N}^*=12)$. We include the residual symmetry group $\Gamma^*$ of these configurations below each inset.
\textbf{b} Number of active edges $\mathcal{N}^*$ in the optimal network as a function of $\gamma$ and $\beta$. White dashed and dash-dotted lines show the local stability thresholds $\gamma_s(T_1)$ and $\gamma_s(T_2^-)$ of the symmetric configuration. The dotted line shows $\gamma_s(\mathrm{alt})|_{N=4}$, below which the capacities of the eight-edge ring alternate (hatched region). The solid white line $\beta_h$ separates the two eight-edge phases, the ring and the kite. Circles mark the parameters of panels \textbf{c--f}.
\textbf{c--f} Optimal networks of the phases that do not occur at $\beta=0.6$: the ring with alternating capacities (\textbf{c}, $\gamma=0.34$, $\beta=0.26$), the kite (\textbf{d}, $\gamma=0.37$, $\beta=1.07$), and the networks with ten (\textbf{e}, $\gamma=0.66$, $\beta=1.96$) and eleven active edges (\textbf{f}, $\gamma=0.68$, $\beta=1.83$).
In all network drawings, squares denote sources and circles denote sinks, and line width and darkness increase with $k_e^*$. In all panels, $\mu=1, \kappa=1$.}
    \label{fig:multiplex1}
\end{figure*}

The ring networks considered above have a relatively simple symmetry structure. We now turn to the cube as an example in which different spatial symmetries can be broken independently, allowing for partially symmetry-broken optimal networks.

We consider a cube with four source and four sink nodes, arranged such that every edge connects a source and a sink. The source--sink pattern is preserved by a group of $24$ spatial transformations,
\begin{align}
\Gamma \cong T_d \cong S_4,
\qquad |\Gamma|=24,
\end{align}
corresponding to the symmetry group of the tetrahedron formed by the four source nodes. The group acts transitively on the twelve edges, such that the fully symmetric optimum has equal capacities. In contrast to the ring, however, $\Gamma$ contains several distinct types of transformations, including two- and threefold rotations, reflections, and improper fourfold rotations, which need not be broken simultaneously. The corresponding decomposition into blocks and their stability thresholds is derived in Supplementary Methods~8.

There are only two blocks that can be critical, with thresholds
\begin{align}
    \gamma_s(T_1) = \frac{8\mu^2}{8\mu^2+27\beta^2},
    \quad
    \gamma_s(T_2^-) = \frac{9\sqrt6\,\beta^2-8\mu^2}{8\mu^2+27\beta^2},
\end{align}
such that $T_1$ is critical for $\beta<\beta_\times=\mu\,(16/9\sqrt6)^{1/2}\approx0.85\,\mu$ and $T_2^-$ for $\beta>\beta_\times$.

Figure~\ref{fig:multiplex1}\textbf{a} illustrates the resulting sequence of optimal structures for $\beta=0.6$. Starting from large $\gamma$, the optimum is fully symmetric. Upon lowering $\gamma$, it first enters a chiral nine-edge configuration that retains only a threefold rotational symmetry. Over a broad interval of $\gamma$, the optimum then forms an eight-edge ring embedded in the cube. 
This structure remains highly symmetric, retaining the group $C_4^{(i)}$ generated by an improper fourfold rotation, and thus breaks only part of the original symmetry group. Eventually, the optimum becomes a tree and all spatial symmetries are broken. The residual symmetries of the nine-edge and eight-edge configurations, $C_3$ and $C_4^{(i)}$, are isotropy subgroups of $T_1$ (Supplementary Methods 8). The critical block thus organizes the residual symmetry even across discontinuous transitions.

The full phase diagram in Fig.~\ref{fig:multiplex1}\textbf{b} reveals a non-monotonic influence of fluctuations. At fixed $\gamma$, increasing $\beta$ can induce reentrant transitions from phases with a small number $\mathcal N^*$ of active edges to more highly connected phases and back again.
At weak fluctuations, the capacities of the eight-edge ring alternate below $\gamma_s(\mathrm{alt})|_{N=4}$ (Fig.~\ref{fig:multiplex1}\textbf{c}). At strong fluctuations, the ring is replaced by the kite configuration (Fig.~\ref{fig:multiplex1}\textbf{d}), and networks with ten and eleven active edges appear (Fig.~\ref{fig:multiplex1}\textbf{e},\textbf{f}). 
The cube provides a second example in which fluctuations induce reentrant changes in connectivity and symmetry, now in a network with a substantially richer symmetry structure. Supplementary Methods 9 provides $\bar D$ in closed form when it is possible.

\subsection*{Complete bipartite networks}

Complete bipartite subgraphs are among the most common sources of symmetry in real-world networks~\cite{xiao2008, macarthur2008}. Since every source is connected to every sink, they combine maximal connectivity with a large symmetry group, which makes them a natural setting to explore the interplay between the two. Here, we consider complete bipartite networks $K_{N_g,N_c}$ (Fig.~\ref{fig:ring-networks_schem}\textbf{b}). In contrast to ring networks, these networks contain many alternative paths for redistributing flows. Both the number of independent cycles $\varsigma=(N_g-1)(N_c-1)$ and the size of the symmetry group $|\Gamma|=N_g!\,N_c!$ grow rapidly with network size.

Nevertheless, capacity perturbations fall into only four blocks: (i) uniform changes, (ii) redistribution of capacity between sources, (iii) redistribution between sinks, and (iv) loop redistributions involving pairs of sources and sinks. The latter transfer capacity from two edges $(i,j)$ and $(i',j')$ to the alternative connections $(i,j')$ and $(i',j)$, leaving the total capacity at every node unchanged. The uniform mode is excluded by the resource constraint, while source redistributions never become unstable, leaving only two blocks that can destabilize the symmetric configuration (Supplementary Methods 10). 

Since $\Gamma$ acts transitively on the edges, the symmetric configuration has equal capacities. Its local stability can be analyzed separately for each block according to Eq.~\eqref{eq:gamma-s-rho}. 
The loop and sink redistributions yield the critical thresholds
\begin{align}
  \gamma_s(\mathrm{loop}) &=
  \frac{(N_g-3)\beta^2+\beta^{*2}}{(N_g-1)\beta^2+\beta^{*2}},\\
  \gamma_s(\mathrm{sink}) &=
  \frac{(N_g-1)\beta^2-\beta^{*2}}{(N_g-1)\beta^2+\beta^{*2}},
\end{align}
where $\beta^*=N_c\mu/N_g$ is the mean source injection (Supplementary Methods 10).  
For weak fluctuations $\beta<\beta^*$, we have $\gamma_s(\mathrm{loop})>\gamma_s(\mathrm{sink})$,  such that loop redistributions become unstable first. Conversely, for strong fluctuations $\beta>\beta^*$, we have $\gamma_s(\mathrm{sink}) >\gamma_s(\mathrm{loop})$ and sink redistributions become unstable first.
The two thresholds cross at $\beta=\beta^*$, where the symmetric configuration is most stable with $\gamma_s=1-2/N_g$. In contrast, $\gamma_s=\max_\rho\gamma_s(\rho) \to1$ for both $\beta \to 0$ and $\beta \to \infty $. 
The thresholds depend only on $N_g$ and the relative fluctuation strength $\beta/\beta^*$, despite the increasing size and complexity of the network. The stabilization of the cycle space by fluctuations is consistent with known results~\cite{corson2010fluctuations, katifori2010damage}.

\begin{figure*}[tb]
\centering
\includegraphics[width=\linewidth]{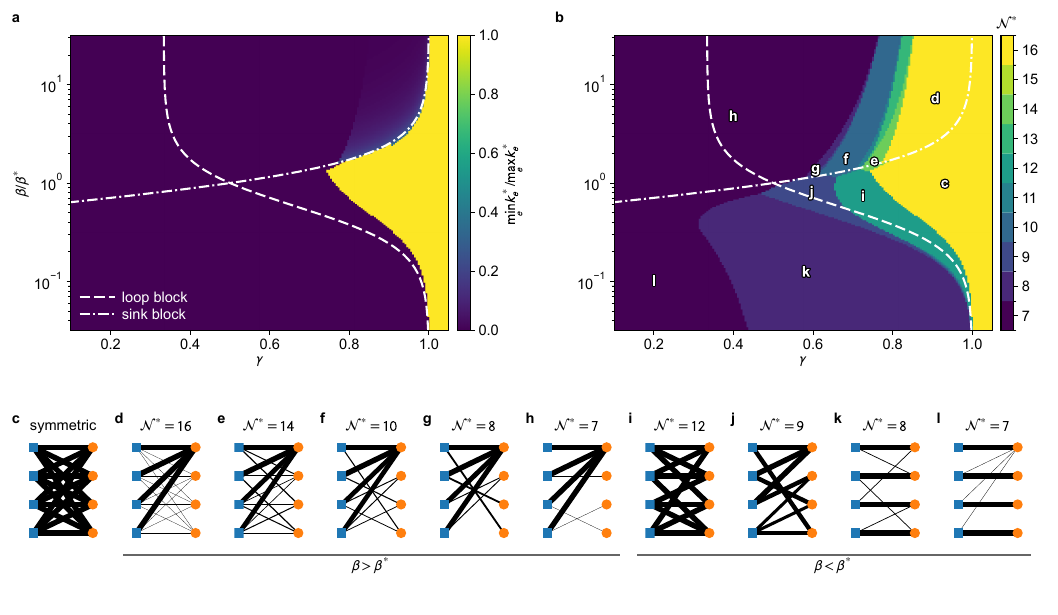}
\caption{\textbf{Phases of the complete bipartite network $K_{4,4}$.}
\textbf{a},\textbf{b}~Ratio of the smallest to the largest optimal edge capacity (\textbf{a}) and number of active edges $\mathcal{N}^*$ (\textbf{b}) as functions of the scaling exponent $\gamma$ and the relative fluctuation strength $\beta/\beta^*$. A ratio of $1$ indicates the symmetric phase, intermediate values weakly symmetry-broken phases, and $0$ strong symmetry breaking. White dashed and dash-dotted lines show the local stability thresholds of the loop block and the sink block, respectively. In \textbf{b}, thin white lines separate distinct configurations with the same $\mathcal{N}^*$, and letters mark the parameters of the networks shown in \textbf{c}--\textbf{l}.
\textbf{c}--\textbf{l}~Optimal networks, with sources (squares) on the left, sinks (circles) on the right, and line widths increasing with $k_e^*/\max_e k_e^*$: the symmetric network (\textbf{c}), phases at strong fluctuations, $\beta>\beta^*$ (\textbf{d}--\textbf{h}), and phases at weak fluctuations, $\beta<\beta^*$ (\textbf{i}--\textbf{l}). Panels \textbf{h} and \textbf{l} show the same phase at different fluctuation strengths. Nodes are ordered such that the sink with the largest total capacity is at the top. Further configurations, which differ from their neighbors by single weak connections, and occupy narrow regions in \textbf{b} and are not shown. Parameters $(\gamma,\beta/\beta^*)$: \textbf{c}~$(0.93, 0.98)$, \textbf{d}~$(0.91, 7.11)$, \textbf{e}~$(0.75, 1.65)$, \textbf{f}~$(0.68, 1.71)$, \textbf{g}~$(0.61, 1.39)$, \textbf{h}~$(0.36, 4.69)$, \textbf{i}~$(0.73, 0.72)$, \textbf{j}~$(0.60, 0.80)$, \textbf{k}~$(0.58, 0.12)$, \textbf{l}~$(0.20, 0.10)$.
In all panels, $\mu = 1, \kappa = 1$.}
    \label{fig:bipartite}
\end{figure*}

Numerical optimization confirms that symmetry is favored near $\beta=\beta^*$. We illustrate this for $N_g=N_c=4$ ($K_{4,4}$) in Fig.~\ref{fig:bipartite}\textbf{a,b}. Here, the symmetric configuration remains globally optimal for values of $\gamma$ well below one, consistent with the enhanced local stability predicted by our theory. This gives rise to reentrant behavior: varying the fluctuation strength at fixed $\gamma$ can drive the optimal network from a symmetry-broken state to a symmetric configuration and back again. At weak and strong fluctuations, the local stability thresholds closely approximate the boundaries of the symmetric phase.

At intermediate fluctuations, strong symmetry breaking determines the transition: a network with fewer active edges becomes globally optimal before the symmetric configuration loses local stability. Consequently, the global transition occurs at $\gamma_c>\gamma_s$. The transition is therefore discontinuous, arising from an exchange of global optimality between competing minima. At strong fluctuations, in contrast, the hub forms continuously out of the symmetric network at $\gamma_s(\mathrm{sink})$, such that the local stability threshold coincides with the phase boundary.

The resulting network structures can be understood from the competing functions of supplying the sinks and balancing fluctuations between sources. 
At intermediate fluctuations, both functions are important, favoring a symmetric network with equally strong connections (Fig.~\ref{fig:bipartite}\textbf{c}). 
For weak fluctuations, each source predominantly supplies one sink through a strong connection, while additional weaker connections balance the small fluctuations. 
For strong fluctuations, the optimum instead develops a hub: all sources strongly connect to a single sink, through which their fluctuations can be balanced, while weaker connections supply the remaining sinks (Fig.~\ref{fig:bipartite}\textbf{d}). With decreasing $\gamma$, the hub network becomes sparser (Fig.~\ref{fig:bipartite}\textbf{e}--\textbf{g}). The final configuration is a tree in which all sources connect to the hub and each remaining sink is attached to a single source (Fig.~\ref{fig:bipartite}\textbf{h}; Supplementary Methods 11). 
For weak fluctuations, we also observe a strongly symmetry-broken configuration that retains considerable residual symmetry: removing one connection per source and sink leaves a network invariant under simultaneous permutations of sources and sinks. For $K_{4,4}$, the configuration shown in Fig.~\ref{fig:bipartite}\textbf{i} is topologically equivalent to the previously discussed cube network. The nine-edge and ring phases of the cube also appear (Fig.~\ref{fig:bipartite}\textbf{j},\textbf{k}; Supplementary Note~5). The tree in Fig.~\ref{fig:bipartite}\textbf{h} remains optimal at small $\gamma$ for all fluctuation strengths, but its capacities change continuously with $\beta$. For $\beta<\beta^*/\sqrt{3}$, the connections supplying the individual sinks dominate, while those to the hub only balance the small fluctuations, cf.\ Fig.~5\textbf{l} and Supplementary Methods~11.

\subsection*{Symmetry breaking in energy grids}

Real transport networks are generally not exactly symmetric, as sources and sinks differ in size and their fluctuations. The strict notion of weak symmetry breaking therefore does not directly apply. However, the characteristic signature of strong symmetry breaking, whereby the optimal network topology changes because one or more edges become inactive, remains meaningful. This is particularly relevant for infrastructure planning, where it determines which of several candidate connections should actually be built.

\begin{figure*}[tb]
    \centering
   \includegraphics[width=\linewidth]{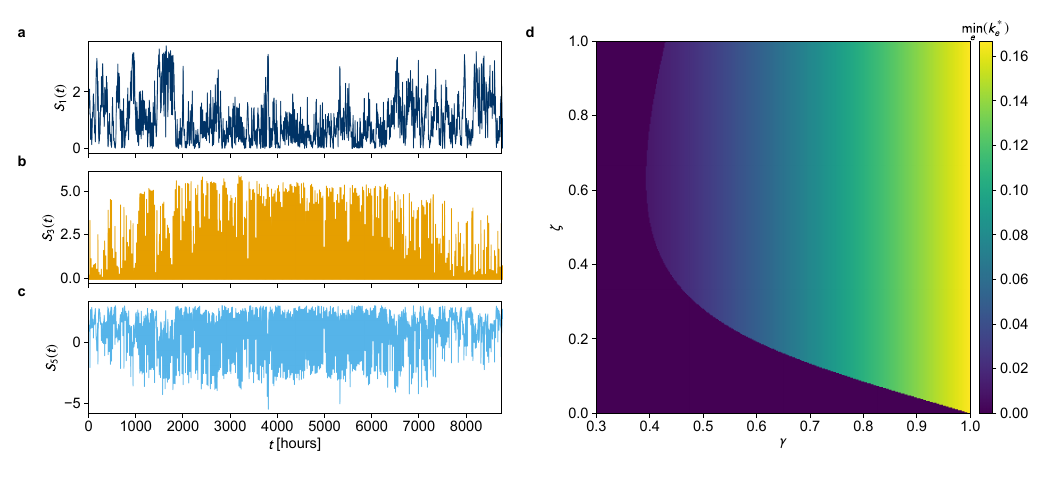}
    \caption{
    \textbf{Strong symmetry breaking in a network with fluctuating generation from renewable power sources.} In a six node ring, nodes 2, 4, 6 have constant demand $S_i=-\mu$, while nodes 1 and 3 host wind and solar generation, $S_i(t)=(1-\zeta)\mu+\zeta\mu Z_i(t)$, and node 5 balances the network, $S_5(t)=3\mu-S_1(t)-S_3(t)$. \textbf{a}-\textbf{c} Time series of the injections $S_1(t)$ (wind), $S_3(t)$ (solar), and $S_5(t)$ (balancing) for $\zeta = 1$ and $\mu=1$. \textbf{d} Smallest optimal edge capacity $\min_e k_e^*$ as a function of the scaling exponent $\gamma$ and the share of fluctuating renewable generation $\zeta$, for $\mu=1$. The open ring, $\min_e k_e^*=0$, appears in the dark region.
    }
    \label{fig:renewables}
\end{figure*}

We illustrate this point for a simple energy system based on the six-node ring introduced above. Three regions have constant demand $S_i=-\mu$, while two regions $i=1,3$ host wind and solar generation,
\begin{align}
    S_i(t)=(1-\zeta)\mu+\zeta\mu Z_i(t),
\end{align}
where $\zeta\in[0,1]$ quantifies the share of fluctuating renewable power generation and thus controls the strength of the injection fluctuations. The remaining region balances the total power injection, modelling a storage infrastructure.
We use wind and solar power time series from renewables.ninja~\cite{pfenninger2016long,staffell2016using} for $Z_i(t)$, with details of the setup provided in Supplementary Note~6. The resulting injections are neither identically distributed nor symmetric, and thus explicitly break the symmetry of the underlying ring geometry. 
Nevertheless, the optimal network displays the same topological transition as the strongly symmetry-broken states studied above. Depending on $\gamma$ and the fluctuation strength $\zeta$, the optimum is either a closed ring with $\min_e k_e^*>0$ or an open ring with $\min_e k_e^*=0$, see Fig.~\ref{fig:renewables}. For intermediate $\gamma$, we again observe a reentrant transition as $\zeta$ is varied. Hence, the mechanism underlying strong symmetry breaking remains relevant even when the primitive data are only approximately symmetric: rather than generating asymmetry spontaneously, the optimization can amplify pre-existing heterogeneity up to a change of network topology.

\section*{Discussion}

We have shown that symmetry breaking is a generic organizing principle of optimal supply networks under a resource constraint. For $\gamma>1$, symmetry breaking is not possible because the optimization problem is convex with a unique symmetric minimizer. For $\gamma<1$, we differentiate between weak and strong symmetry breaking. In weak symmetry breaking, all edges of the underlying network remain active but their capacities become unequal. It occurs when the symmetric configuration loses local stability. The capacity pattern of the emergent symmetry broken configuration is determined by the critical block of the symmetry group of the optimization problem. In contrast, in strong symmetry breaking, edges become inactive and the network topology changes. This transition occurs through an exchange of global optimality between competing local minima, and is generally discontinuous.

Fluctuations play a nontrivial and organizing role. Despite their vastly different connectivity and symmetry groups, rings, the cube, and complete bipartite networks exhibit similar patterns of symmetry breaking: weak and strong fluctuations favor symmetry breaking, while intermediate fluctuations may stabilize the symmetric network. Under increasing noise, the system can exhibit reentrant transitions from strongly symmetry-broken to symmetric and back, implying the existence of an optimal fluctuation level for symmetric network structures. As a control parameter, fluctuations can both suppress and induce symmetry breaking.  Similar noise-induced reentrant phenomena are well known in nonlinear systems~\cite{gammaitoni1998stochastic,pikovsky1997coherence,lindner2004effects,uzuntarla2017inverse,bacic2020paradigmatic, folz2023noise}.

These results offer a perspective on the structural patterns observed in natural and engineered supply networks. 
Real supply networks are never exactly symmetric, but strong symmetry breaking persists as a change of the optimal topology. Our renewable energy example shows that optimization amplifies pre-existing heterogeneity in the injections, sometimes resulting in the removal of links. It also exhibits the reentrant dependence on the share of fluctuating generation.
This highlights the practical relevance of our findings for designing robust and efficient infrastructure, particularly in contexts where variability and intermittency are key factors.

Our work opens several promising avenues for future research. 
On the theoretical side, a systematic study of non-edge-transitive networks would complement our analysis of edge-transitive networks, since the block decomposition framework is still applicable. Symmetries preserve whether an edge connects a source and a sink, meaning that edge transitivity requires a bipartite network with sources and sinks as its two parts. This is the case for the rings, the cube and the complete bipartite graphs studied here. Conversely, in networks such as complete graphs and ring lattices with next nearest neighbor connections, edges between sources, between sinks and between sources and sinks form distinct symmetry classes. The treatment of infinite networks, such as lattices, is another relevant issue. In that case, the block decomposition reduces to a Fourier analysis over a continuous wavevector. 
On the applied side, integrating time-dependent operational constraints and nonlinear flow laws could further enhance the relevance of these models to real-world systems, ranging from vascular transport to smart grids.

\section*{Methods}
\footnotesize

\subsection*{Numerical computation}

We compute locally optimal networks using the self-consistent iteration used by Corson~\cite{corson2010fluctuations}. Stationarity of the Lagrangian requires  $\braket{F_e^2} k_e^{-2} - \nu \gamma k_e^{\gamma-1} = 0$.  The Lagrangian multiplier $\nu$ is fixed by the resource constraint which yields
\begin{align}
    k_e
    =
    \frac{\braket{F_e^2}^{1/(1+\gamma)}}
    {\left(
        \sum_{l\in\EE}
        \braket{F_l^2}^{\gamma/(1+\gamma)}
    \right)^{1/\gamma}}
    \,\kappa .
    \label{eq:ke-corson}
\end{align}
Starting from an initial capacity vector $\vec k$, we first compute the flow variances $\braket{F_e^2}$ for the current network and then update all capacities according to Eq.~\eqref{eq:ke-corson}. This procedure is iterated until the change in $\vec k$ falls below a prescribed tolerance. Since several local minima may exist, the converged solution generally depends on the initial condition.

\subsection*{Second moments}

For the model networks introduced above, the second moments $\Sigma_{ij}=\langle S_iS_j\rangle$ follow directly from
\[
S_i=-\mu \quad (i\in\VV_c),
\qquad
S_i=\frac{N_c}{N_g}\mu+X_i \quad (i\in\VV_g),
\]
with $\langle X_i\rangle=0$, $\langle X_iX_j\rangle=-\beta^2$ for $i\neq j$, and $\sum_{i\in\VV_g}X_i=0$. The latter implies
\[
\langle X_i^2\rangle=(N_g-1)\beta^2.
\]
Hence we obtain
\begin{align}
\Sigma_{ij}
=
\begin{cases}
\displaystyle
\frac{N_c^2}{N_g^2}\mu^2+(N_g-1)\beta^2,
& i=j\in\VV_g,\\[1ex]
\displaystyle
\frac{N_c^2}{N_g^2}\mu^2-\beta^2,
& i\neq j,\quad i,j\in\VV_g,\\[1ex]
\mu^2,
& i,j\in\VV_c,\\[1ex]
\displaystyle
-\frac{N_c}{N_g}\mu^2,
& i\in\VV_g,\ j\in\VV_c .
\end{cases}\label{eq:moments}
\end{align}

\subsection*{Dissipation and effective resistances}

Using Eq.~\eqref{eq:F-L-X}, the dissipation for a given injection vector can be written as
\begin{align}
    D
    &=\vec F^\top\matr K^{-1}\vec F
    =\vec S^\top\matr L^+\matr E\matr K\matr E^\top\matr L^+\vec S
    =\vec S^\top\matr L^+\vec S ,
\end{align}
where we used $\matr L=\matr E\matr K\matr E^\top$ and $\matr L\matr L^+\vec S=\vec S$ for balanced injections. Averaging over the injection statistics yields
\begin{align}
    \bar D
    =\left\langle\vec S^\top\matr L^+\vec S\right\rangle
    =\mathrm{Tr}(\matr L^+\matr\Sigma),
    \qquad
    \matr\Sigma=\langle\vec S\vec S^\top\rangle .
    \label{eq:dissipation-trace-meth}
\end{align}
This expression can be rewritten using the effective resistance~\cite{KleinRandic1993}
\begin{align}
    R_{nm}=L^+_{nn}+L^+_{mm}-2L^+_{nm}.
\end{align}
Since every realization is balanced, we have $\matr\Sigma\vec 1=0$. Hence,
$
    \sum_{n,m\in\VV}\Sigma_{nm}L^+_{nn}
    =
    \sum_{n,m\in\VV}\Sigma_{nm}L^+_{mm}
    =0.
$
We thus obtain, cf.~Ref.~\cite{GhoshBoydSaberi2008},
\begin{align}
    \bar D
    &=\sum_{n,m\in\VV}\Sigma_{nm}L^+_{nm}
     =-\frac{1}{2}
    \sum_{n,m\in\VV}\Sigma_{nm}R_{nm}.
    \label{eq:dissipation_resistance-meth}
\end{align}

For the injection model considered here, the second-moment matrix can be decomposed as
\begin{align}
    \matr\Sigma
    =\mu^2\vec v\vec v^\top+N_g\beta^2\matr Q ,
\end{align}
where $v_n=N_c/N_g$ on sources and $v_n=-1$ on sinks, while
$Q_{nm}=\delta_{nm}-1/N_g$ for $n,m\in\VV_g$ and zero otherwise. The matrix $\matr Q$ is the orthogonal projector onto the vectors that vanish on the sinks and sum to zero over the sources, $\matr Q^2 = \matr Q = \matr Q^\top$. Hence,
\begin{align}
    \bar D
    =
    \mu^2\vec v^\top\matr L^+\vec v
    +
    \beta^2\sum_{i<j\in\VV_g}R_{ij}.
    \label{eq:dissipation_resistance_model-meth}
\end{align}
Thus, the contributions from the mean injection pattern and from source fluctuations separate, with the latter determined by the effective resistances between source nodes.

The average dissipation $\bar D$ is convex in the edge capacities $\vec k$. To see this, recall that for a fixed injection vector $\vec S$,
\begin{align}
    D(\vec k;\vec S)
    =
    \min_{\matr E\vec F=\vec S}
    \sum_{e\in\EE}\frac{F_e^2}{k_e}.
\end{align}
The function $F_e^2/k_e$ is jointly convex in $(F_e,k_e)$ for $k_e>0$. Hence, minimizing over the affine constraint $\matr E\vec F=\vec S$ preserves convexity with respect to $\vec k$. Finally, taking the expectation over the injection statistics preserves convexity, such that
\begin{align}
    \bar D(\vec k)=\langle D(\vec k;\vec S)\rangle
\end{align}
is convex in $\vec k$.

\subsection*{Stability analysis and derivatives of the dissipation}
\label{sec:stability-methods}

We derive the derivatives of the average dissipation
\begin{align}
    \bar D=\operatorname{Tr}(\matr L^+\matr\Sigma),
    \qquad
    \matr L=\matr E\matr K\matr E^\top .
\end{align}
For a connected network, $\matr L^+$ coincides with the inverse of $\matr L$ on the subspace orthogonal to $\vec 1$, which contains all balanced injection vectors. Since $\partial\matr L/\partial k_e$ annihilates $\vec 1$, the kernel of $\matr L$ is not affected, and the additional terms in the general expression for the derivative of a pseudoinverse vanish~\cite{golub1973differentiation}. Hence,
\begin{align}
    \frac{\partial\matr L^+}{\partial k_e}
    =
    -\matr L^+
    \frac{\partial\matr L}{\partial k_e}
    \matr L^+
\end{align}
on this subspace. Since
\begin{align}
    \frac{\partial\matr L}{\partial k_e}
    =
    \vec E_e\vec E_e^\top ,
\end{align}
where $\vec E_e$ denotes column $e$ of the incidence matrix, we obtain
\begin{align}
    \frac{\partial\bar D}{\partial k_e}
    &=
    -\vec E_e^\top
    \matr L^+\matr\Sigma\matr L^+
    \vec E_e .
\end{align}

It is convenient to introduce the edge-space matrices
\begin{align}
    \matr G
    &=
    \matr E^\top\matr L^+\matr E,
    &
    \matr\Psi
    &=
    \matr E^\top
    \matr L^+\matr\Sigma\matr L^+
    \matr E .
    \label{eq:G-Psi-methods}
\end{align}
Here, $G_{ef}$ gives the potential difference across edge $e$ generated by a unit current across edge $f$, while $\matr\Psi$ contains the second moments of the edge potential differences. The first derivative is therefore
\begin{align}
    \frac{\partial\bar D}{\partial k_e}
    =
    -\Psi_{ee}.
    \label{eq:first-derivative-methods}
\end{align}
Differentiating once more yields
\begin{align}
    \frac{\partial^2\bar D}
    {\partial k_e\partial k_f}
    =
    2G_{ef}\Psi_{ef}.
    \label{eq:second-derivative-methods}
\end{align}
These expressions hold whenever the active edges connect all nodes, including the one-sided derivative with respect to an inactive edge at $k_e\to0^+$, where $\Psi_{ee}$ is the mean square potential difference across the open edge. 
Thus, the Hessian of the dissipation is
\begin{align}
    \nabla^2\bar D
    =
    2\,\matr G\circ\matr\Psi ,
\end{align}
where $\circ$ denotes element-wise multiplication. These expressions generalize the gradient and Hessian of the total effective resistance derived in Ref.~\cite{GhoshBoydSaberi2008} to arbitrary injection statistics. Since both $\matr G$ and $\matr\Psi$ are positive semidefinite, the Schur product theorem also provides a direct proof that $\bar D$ is convex in the edge capacities.

To account for the resource constraint, we introduce the Lagrangian
\begin{align}
    \mathcal L
    =
    \bar D
    +
    \nu
    \left(
        \sum_e k_e^\gamma-\kappa^\gamma
    \right). \label{eq:lagr-methods}
\end{align}
At an interior stationary point,
\begin{align}
    \Psi_{ee}
    =
    \nu\gamma k_e^{\gamma-1}.
    \label{eq:stationarity-methods}
\end{align}
The Hessian of the Lagrangian is therefore
\begin{align}
    \nabla^2\mathcal L
    &=
    2\,\matr G\circ\matr\Psi
    +
    \nu\gamma(\gamma-1)
    \operatorname{diag}\!\left(k_e^{\gamma-2}\right)
    \nonumber\\
    &=
    2\,\matr G\circ\matr\Psi
    +
    (\gamma-1)
    \operatorname{diag}\!\left(
        \frac{\Psi_{ee}}{k_e}
    \right).
    \label{eq:lagr-hessian-methods}
\end{align}

At the uniform configuration $k_e=k_0$, we have $\matr L=k_0\tilde{\matr L}$ with $\tilde{\matr L}=\matr E\matr E^\top$, hence $\matr L^+=\tilde{\matr L}^+/k_0$, $\matr G=\tilde{\matr G}/k_0$ and $\matr\Psi=\tilde{\matr\Psi}/k_0^2$. Both terms have the common factor $k_0^{-3}$. Since $\tilde\Psi_{ee}$ is independent of $e$ for edge-transitive networks, we obtain Eq.~\eqref{eq:scale-free}.

\subsection*{AI declaration} 
Generative AI tools, specifically ChatGPT, Claude and GitHub Copilot, were used to assist with language editing and portions of the code development. All outputs were reviewed and validated by the authors, who take full responsibility for the content.

\section*{Data availability}
All numerical results can be reproduced with the code described below. Precomputed results of the computationally expensive parameter scans are provided in the GitHub repository \url{https://github.com/i-bacic/optimal-networks},
archived on Zenodo (\url{https://doi.org/10.5281/zenodo.23729507}). The hourly wind and solar capacity factors used in the energy-system example (Fig.~6) were obtained from renewables.ninja (\url{https://www.renewables.ninja}) for Jülich, Germany, for the year 2019~\cite{pfenninger2016long, staffell2016using}, and are redistributed in the repository under their original licence (CC BY-NC 4.0).

\section*{Code availability}
The code used to generate all results is openly available at \url{https://github.com/i-bacic/optimal-networks} under the MIT License. The version used
here (v1.0.0) is archived on Zenodo
(\url{https://doi.org/10.5281/zenodo.23729507}).

\section*{Acknowledgements}
We thank Boris Stupovski (Institute of Physics Belgrade, Serbia) for helpful suggestions. A.P., A.B., D.W. and I.B. gratefully acknowledge support from the German Federal Ministry of Research, Technology and Space (Bundesministerium für Forschung, Technologie und Raumfahrt) via the project 03SF0751. A.P. was partially supported by the Air Force Office of Scientific Research under grant number FA9550-24-1-0039. The funders had no role in study design, data collection, and analysis, the decision to publish, or any opinions, findings, conclusions, or recommendations expressed in the manuscript.

\section*{Author Contributions}
I.B. and D.W. conceived and designed research. 
I.B. and D.W. developed the theory and the algorithms, with inputs from P.E. for the group theoretic framework.
P.E. and I.B. applied the group theoretic framework to the examples. 
A.P. and I.B. developed and performed the simulations and designed figures. 
A.P. and D.W. prepared the energy grids example with inputs from A.B.
All authors contributed to the writing of the manuscript.

\section*{Competing Interests} 
The authors declare no competing financial or non-financial interests.


\end{document}


\renewcommand{\figurename}{Supplementary Figure}
\renewcommand{\tablename}{Supplementary Table}
\renewcommand{\thefigure}{\arabic{figure}}
\renewcommand{\thetable}{\arabic{table}}

\title{Supplementary Information for Symmetry breaking in minimum dissipation networks}

\author{Aarathi Parameswaran \orcidlink{0009-0005-8255-2152}}
\affiliation{Institute of Climate and Energy Systems: Energy Systems Engineering (ICE-1), Forschungszentrum J\"ulich, 52428 J\"ulich, Germany}
\affiliation{Bonn-Cologne Graduate School of Physics and Astronomy, University of Bonn, Bonn, Germany}
\affiliation{Center for Complex Networks and Systems Research, Luddy School of Informatics, Computing, and Engineering, Indiana University, Bloomington, Indiana 47408, USA}

\author{Pezhman Ebrahimzadeh \orcidlink{0000-0002-9919-3900}}
\affiliation{Neuromorphic Compute Nodes  (PGI-14), Forschungszentrum J\"ulich, 52428 J\"ulich, Germany}

\author{Andrea Benigni \orcidlink{0000-0002-2475-7003}}
\affiliation{Institute of Climate and Energy Systems: Energy Systems Engineering (ICE-1), Forschungszentrum J\"ulich, 52428 J\"ulich, Germany}
\affiliation{RWTH Aachen University, Aachen, Germany}

\author{Dirk Witthaut \orcidlink{0000-0002-3623-5341}}
\email{d.witthaut@fz-juelich.de}
\affiliation{Institute of Climate and Energy Systems: Energy Systems Engineering (ICE-1), Forschungszentrum J\"ulich, 52428 J\"ulich, Germany}
\affiliation{Institute for Theoretical Physics, University of Cologne, 50937 K\"oln, Germany}

\author{Iva Ba\v{c}i\'{c}  \orcidlink{0000-0003-2987-5065}}
\email{i.bacic@fz-juelich.de}
\affiliation{Institute of Climate and Energy Systems: Energy Systems Engineering (ICE-1), Forschungszentrum J\"ulich, 52428 J\"ulich, Germany}
\affiliation{Center of Excellence for Study of Complex Systems, Institute of Physics Belgrade, University of Belgrade, Belgrade, Serbia}
\maketitle

This Supplementary Information contains 
6 Supplementary Notes and 11 Supplementary Methods. Within these, there are 4 Supplementary Figures and 10 Supplementary Tables. 

{\let\origcontentsline\contentsline
 \let\contentsline\nonumcontentsline
 \tableofcontents}
 
\newpage

\section{Supplementary Notes}

\subsection{Supplementary Note 1: Optimal networks with mixed boundary conditions}
\label{sec:mixed-boundary}

The setup considered in the main manuscript assumes that the injection $S_n$ is prescribed at every node $n$ and for every realization of the random injection pattern. In technical transport systems, however, different types of boundary conditions may occur. For instance, the pressure may be fixed at a reservoir in a hydraulic network, or the electric potential may be prescribed at a voltage-controlled node in an electrical network. We therefore briefly discuss the more general case in which the injections are prescribed only at a subset of nodes, while the potentials are prescribed at the remaining nodes.

To this end, we partition the node set as $\VV=\VV_S\cup\VV_\theta$, with $\VV_S\cap\VV_\theta=\emptyset$. At nodes $n\in\VV_S$, the injections $S_n$ are prescribed as in the main text, whereas at nodes $n\in\VV_\theta$ the potentials are fixed to prescribed reference values $\vartheta_n$. For notational simplicity, we order the nodes such that the nodes in $\VV_S$ come first and write
\begin{align}
    \vec\theta =
    \begin{pmatrix}
        \vec\theta_S\\
        \vec\theta_\theta
    \end{pmatrix},
    \qquad
    \vec\theta_\theta=\vec\vartheta,
\end{align}
and partition the network Laplacian accordingly as
\begin{align}
    \matr L=
    \begin{pmatrix}
        \matr L_{SS} & \matr L_{S\theta}\\
        \matr L_{\theta S} & \matr L_{\theta\theta}
    \end{pmatrix}.
\end{align}

The continuity equation is imposed only at the nodes in $\VV_S$, while the equations at the nodes in $\VV_\theta$ are replaced by the prescribed-potential conditions. The resulting linear system reads
\begin{align}
    \begin{pmatrix}
        \vec S_S\\
        \vec\vartheta
    \end{pmatrix}
    =
    \begin{pmatrix}
        \matr L_{SS} & \matr L_{S\theta}\\
        0 & \matr I
    \end{pmatrix}
    \begin{pmatrix}
        \vec\theta_S\\
        \vec{\theta}_\theta
    \end{pmatrix}.
\end{align}
Hence, the unknown potentials are given by
\begin{align}
    \vec\theta_S
    =
    \matr L_{SS}^{-1}
    \left(
        \vec S_S-\matr L_{S\theta}\vec\vartheta     \right).
    \label{eq:mixed-theta}
\end{align}
For a connected network with at least one prescribed-potential node, $\matr L_{SS}$ is a grounded Laplacian and is positive definite, such that the ordinary inverse exists.

The dissipation can be written directly in terms of the nodal potentials as
\begin{align}
    D
    =
    \sum_{e=(m,n)\in\EE}
    k_e(\theta_n-\theta_m)^2
    =
    \vec\theta^\top\matr L\vec\theta .
\end{align}
Inserting Eq.~\eqref{eq:mixed-theta} and $\vec\theta_\theta=\vec\vartheta$ yields
\begin{align}
    D
    =
    \vec S_S^\top\matr L_{SS}^{-1}\vec S_S
    +
    (\vec\vartheta)^\top
    \left(
        \matr L_{\theta\theta}
        -
        \matr L_{\theta S}
        \matr L_{SS}^{-1}
        \matr L_{S\theta}
    \right)
    \vec\vartheta.
\end{align}
The matrix in parentheses is the Schur complement of $\matr L_{SS}$ and corresponds to the Laplacian obtained by Kron reduction onto the potential-controlled nodes~\cite{dorfler2012kron}. If the prescribed potentials are fixed while the injections $\vec S_S$ fluctuate, the average dissipation is therefore
\begin{align}
    \bar D
    =
    \mathrm{Tr}\!\left(
        \matr L_{SS}^{-1}\matr\Sigma_S
    \right)
    +
    \vec\vartheta^\top
    \matr L_{\rm red}
    \vec\vartheta,
    \label{eq:Dbar-mixed}
\end{align}
with $\matr L_{\rm red} =  \matr L_{\theta\theta}  -  \matr L_{\theta S}      \matr L_{SS}^{-1} \matr L_{S\theta}$ and $\matr\Sigma_S  = \left\langle
        \vec S_S\vec S_S^\top \right\rangle$.

An important difference to the setting considered in the main text concerns convexity with respect to the edge capacities. The first term in Eq.~\eqref{eq:Dbar-mixed} is convex in $\vec k$ in full analogy to the main text. 
By contrast, the second term is always concave in $\vec k$. Indeed, by the Dirichlet principle~\cite{lyons2005probability} it admits the variational representation
\begin{align}
   \vec \vartheta^\top
\matr L_{\rm red}
   \vec\vartheta
=
\min_{\vec\theta_S}
\begin{pmatrix}
\vec\theta_S \\
\vec\vartheta
\end{pmatrix}^{\top}
\matr L
\begin{pmatrix}
\vec\theta_S \\
\vec\vartheta
\end{pmatrix},
\end{align}
which is the pointwise minimum of functions that are linear in $\vec k$ and is therefore concave~\cite{boyd2004convex}. Consequently, the dissipation under general mixed boundary conditions is in general neither convex nor concave, and the convexity results derived in the main text do not directly extend to this case. Consistent with this, minimum dissipation networks with prescribed pressures are not necessarily trees~\cite{chang2019minimal}. An important exception occurs if only one potential is prescribed, or more generally if all prescribed potentials are equal. In this case $\matr L_{\rm red}\vec\vartheta=0$, the second term in Eq.~\eqref{eq:Dbar-mixed} vanishes, and the original convexity property is retained, in line with Ref.~\cite{chang2019minimal}, where the classical tree results are recovered when all prescribed pressures are equal.

\newpage

\subsection{Supplementary Note 2: The scaling exponent $\gamma$}
\label{sec:gamma}

The scaling exponent $\gamma$ in the resource constraint $\sum_e k_e^\gamma\leq \kappa^\gamma$ can be related to the geometry and physical realization of an individual edge. Suppose that the edge conductance and the associated resource cost scale with a characteristic transverse size $r$ as
\begin{align}
    k \sim r^b,
    \qquad
    C \sim r^a .
\end{align}
Eliminating $r$ yields $C\sim k^{a/b}$, such that the resource constraint $\sum_e k_e^\gamma\leq \kappa^\gamma$ corresponds to
\begin{align}
    \gamma=\frac{a}{b}.
\end{align}
Here, we assume that all edges have the same length. 
The value of $\gamma$ therefore depends both on the transport law and on how the construction or maintenance cost of an edge scales with its size.

Several simple examples illustrate this relation. For laminar flow of a Newtonian fluid through a cylindrical tube, the Hagen--Poiseuille law gives $k\sim r^4$. If the resource cost is proportional to the cross-sectional area, or equivalently to the enclosed volume for fixed length, one has $C\sim r^2$ and thus $\gamma=1/2$, as in the classical optimization underlying Murray's law \cite{sherman1981connecting}. The same scaling arises naturally for engineered pressure pipes. For a thin-walled cylindrical pressure vessel, the hoop stress is $\sigma_\theta=pr/t$, where $p$ is the internal pressure and $t$ the wall thickness \cite{sinclair2015review}. Hence, at fixed operating pressure and allowable stress, $t\sim r$, and the amount of wall material per unit length scales as $C\sim rt\sim r^2$, again yielding $\gamma=1/2$. If, by contrast, the wall thickness is approximately independent of the pipe radius, the amount of wall material scales with the circumference, $C\sim r$, corresponding to $\gamma=1/4$.
For a bulk electrical conductor in the DC regime, both the conductance and the required amount of conductor material scale with the cross-sectional area, $k \sim r^2$ and $C \sim r^2$, giving $\gamma = 1$. 

The same scaling holds for planar conductors of fixed length and thickness, as in on-chip power-distribution networks~\cite{Boyd2001ISPD}, where the cost enters as a penalty. Since $\bar D$ is homogeneous of degree $-1$, penalty and constraint yield the same optimal configurations up to scale. There, a symmetric mesh required small random perturbations of the edge lengths to break the symmetry~\cite{Boyd2001ISPD}, consistent with degenerate optima at $\gamma=1$ (Theorem~1).

These scaling arguments represent idealized limiting cases. In biological transport networks, in particular, the relation between vessel radius and effective hydraulic conductance can deviate from the Hagen--Poiseuille scaling. Blood exhibits a diameter-dependent apparent viscosity in small vessels, such that the effective exponent $b$ need not be exactly four over all vessel sizes \cite{secomb2013blood}. Likewise, experimentally inferred flow--diameter relations in vascular trees show substantial variation and deviations from idealized predictions such as Murray's law \cite{taylor2024systematic}. More generally, biological transport links may possess additional internal structure: for example, the hydraulic resistance of plant xylem depends not only on conduit diameter but also strongly on end walls and bordered pits \cite{venturas2017plant}. Consequently, when $\gamma=a/b$ is interpreted as an effective scaling exponent for a biological network, it need not coincide exactly with the canonical values $1/4$, $1/2$, or $1$, but may take intermediate values depending on the underlying transport mechanism and network structure.

\clearpage 
\subsection{Supplementary Note 3: The limit $\gamma \rightarrow 0$.}
\label{sec:gamma0}
For fixed injections, optimal networks do not have loops for $\gamma<1$~\cite{banavar2000topology,bohn2007structure,durand2007structure}. With fluctuating injections, tree-like optima were previously observed numerically for small $\gamma$~\cite{corson2010fluctuations}. A related argument for the limit $\gamma\to0$, for single source--sink networks that trade off mixing against dissipation, was given in Ref.~\cite{mentus2021}. In this Supplementary Note, we show that for sufficiently small $\gamma>0$ the optimal network is a spanning tree. For the proof, it is convenient to rescale the resource constraint to
\begin{align}
    \sum_{e\in\EE} k_e^\gamma \le R,
    \qquad R=N-1,
\end{align}
where $R$ is the number of edges in a spanning tree. This rescaling changes all optimal capacities $k_e^*$ only by a common multiplicative factor and therefore leaves the support, and hence the sparsity, of the optimum unchanged for every $\gamma>0$.

The key observation is that
\begin{align}
    \sum_e k_e^\gamma
    \longrightarrow
    \|\vec k\|_0
    \qquad\text{as}\qquad
    \gamma\rightarrow0^+,
\end{align}
for any fixed capacity vector $\vec k$. Thus, in the limiting problem the resource constraint counts the number of active edges and permits at most $R=N-1$ of them. Since a connected feasible network requires at least $N-1$ active edges, the limiting optimum must be a spanning tree. In the following, we formalize this argument and show that the conclusion extends to sufficiently small but finite $\gamma$.

\begin{lem}[Rescaling of the resource constraint]
\label{lem:resource-rescaling}
Consider the optimization problem
\begin{align}
    \min_{\vec k\geq0}\bar D(\vec k)
    \qquad\text{subject to}\qquad
    \sum_{e\in\EE}k_e^\gamma\leq C,
    \label{eq:resource-general}
\end{align}
with $\gamma>0$ and $C>0$. Changing the resource budget from $C$ to any $C'>0$ rescales every optimal capacity by the common factor
$
    a=\left(\frac{C'}{C}\right)^{1/\gamma}.
$
In particular, the support of the optimal network is independent of the normalization of the resource constraint.
\end{lem}

\begin{proof}
The network Laplacian scales linearly with the capacities,
$
    \matr L(a\vec k)=a\matr L(\vec k),
$
and therefore
\begin{align}
    \bar D(a\vec k)
    =
    \operatorname{Tr}\!\left[
        \matr L(a\vec k)^+\matr\Sigma
    \right]
    =
    a^{-1}\bar D(\vec k).
    \label{eq:dissipation-scaling}
\end{align}
At the same time,
\begin{align}
    \sum_e(ak_e)^\gamma
    =
    a^\gamma\sum_e k_e^\gamma.
\end{align}
Hence the map $\vec k\mapsto a\vec k$, with
$a=(C'/C)^{1/\gamma}$, maps the feasible set for budget $C$ bijectively onto the feasible set for budget $C'$, while multiplying the objective by the constant factor $a^{-1}$. It therefore maps global minimizers onto global minimizers. Since $a>0$, the set of active edges is unchanged.
\end{proof}

We may therefore choose the normalization
\begin{align}
    \sum_{e\in\EE}k_e^\gamma\leq R,
    \qquad R=N-1,
    \label{eq:resource-tree-normalized}
\end{align}
without loss of generality. Compared with the original constraint
$\sum_e k_e^\gamma\leq\kappa^\gamma$, the capacities are rescaled by the common factor
\begin{align}
    a=\frac{R^{1/\gamma}}{\kappa}.
\end{align}
This normalization is particularly convenient because a spanning tree with unit capacities, $k_e=1$ on its $R=N-1$ edges and $k_e=0$ otherwise, exhausts the resource budget for every $\gamma>0$.

\begin{lem}[Minimum-support limit]
\label{thm:minimum-support}
Let $\mathcal G=(\VV,\EE)$ be a connected candidate graph with
$N=|\VV|$ nodes and $M=|\EE|$ edges. Assume that the injections are balanced,
$\vec 1^\top\vec S=0$, have finite second moments, and satisfy
\begin{align}
    Q_U
    :=
    \left\langle
        \left(\sum_{i\in U}S_i\right)^2
    \right\rangle
    >0
    \label{eq:cut-nondegenerate}
\end{align}
for every non-empty proper subset $U\subset\VV$. Consider the optimization problem
\begin{align}
    \min_{\vec k\geq0}\bar D(\vec k)
    \qquad\text{subject to}\qquad
    \sum_{e\in\EE}k_e^\gamma\leq R .
    \label{eq:normalized-resource}
\end{align}
Then there exists $\gamma_0>0$ such that, for every
$0<\gamma<\gamma_0$, every global minimizer has exactly $R$ active edges. Its support therefore forms a spanning tree.
\end{lem}

\begin{proof}
First note that the normalization in Eq.~\eqref{eq:normalized-resource} is without loss of generality. Since
\begin{align}
    \bar D(a\vec k)=a^{-1}\bar D(\vec k),
\end{align}
changing the resource budget only rescales all optimal capacities by a common factor and does not change their support or relative values.

Because $\mathcal G$ is connected, it contains a spanning tree $T$. Setting
\begin{align}
    k_e=
    \begin{cases}
        1, & e\in T,\\
        0, & e\notin T,
    \end{cases}
\end{align}
satisfies the resource constraint for every $\gamma>0$. Balanced injections admit a unique flow on this tree, and its expected dissipation $D_T$ is finite. Consequently, every global minimizer $\vec k^{\,*}$ satisfies
\begin{align}
    \bar D(\vec k^{\,*})\leq D_T .
    \label{eq:tree-upper-bound}
\end{align}

We next derive a uniform lower bound on the capacity across every cut. Let
$\delta(U)$ denote the edges connecting $U$ to its complement and let
$s_U=\sum_{i\in U}S_i$ be the net injection of $U$. Kirchhoff's current law gives
\begin{align}
    s_U
    =
    \sum_{e\in\delta(U)}\sigma_e F_e ,
\end{align}
where $\sigma_e=\pm1$ denotes the orientation relative to the cut. By the Cauchy--Schwarz inequality,
\begin{align}
    s_U^2
    &\leq
    \left(
        \sum_{e\in\delta(U)}k_e
    \right)
    \left(
        \sum_{e\in\delta(U)}
        \frac{F_e^2}{k_e}
    \right) \leq
    \left(
        \sum_{e\in\delta(U)}k_e
    \right)D .
    \label{eq:cut-cauchy-schwarz}
\end{align}
Taking the expectation, with $Q_U=\langle s_U^2\rangle$, and using Eq.~\eqref{eq:tree-upper-bound} yields
\begin{align}
    \sum_{e\in\delta(U)}k_e^{\,*}
    \geq
    \frac{Q_U}{\bar D(\vec k^{\,*})}
    \geq
    \frac{Q_U}{D_T}.
\end{align}
Since there are only finitely many cuts and $Q_U>0$, there is a constant
$c>0$, independent of $\gamma$, such that
\begin{align}
    \sum_{e\in\delta(U)}k_e^{\,*}\geq c
    \label{eq:cut-capacity-lower}
\end{align}
for every nontrivial cut.

Choose a constant $0<a<1$ such that $Ma<c$. The subgraph formed by all edges with $k_e^{\,*}\geq a$ must be connected. Otherwise, one of its connected components would define a cut across which every edge has capacity smaller than $a$, giving
\begin{align}
    \sum_{e\in\delta(U)}k_e^{\,*}<Ma<c,
\end{align}
in contradiction to Eq.~\eqref{eq:cut-capacity-lower}. Hence, the edges with capacity at least $a$ contain a spanning tree $T_\gamma$.

Since the $R$ edges of $T_\gamma$ all satisfy $k_e^{\,*}\geq a$, every edge $f\notin T_\gamma$ obeys
\begin{align}
    (k_f^{\,*})^\gamma
    \leq
    R - R \; a^\gamma
    =
    R \, (1-a^\gamma).
    \label{eq:small-extra-edge}
\end{align}
due to the resource constraint. The right-hand side tends to zero as $\gamma\rightarrow0^+$. Hence, all capacities outside $T_\gamma$ tend uniformly to zero; the set of active edges approaches a spanning tree.

We now show that for sufficiently small $\gamma$ such additional edges cannot be present at all. Suppose, to the contrary, that there exists a sequence $\gamma_j\rightarrow0^+$ for which a global minimizer contains an active edge $f_j\notin T_{\gamma_j}$ with capacity
$\varepsilon_j=k_{f_j}^{\,*}>0$. Equation~\eqref{eq:small-extra-edge} implies $\varepsilon_j\rightarrow0$ as $\gamma_j\rightarrow0^+$.

Set this capacity to zero while leaving all other capacities unchanged, and denote the resulting vector by $\widehat{\vec k}^{\,j}$. Since the spanning tree $T_{\gamma_j}$ remains present with all its capacities bounded below by $a$, the network stays connected. Moreover, the edge potential differences remain uniformly bounded: for $f=(m,n)$, the Cauchy--Schwarz inequality along the path from $m$ to $n$ in $T_{\gamma_j}$ gives $(\theta_m-\theta_n)^2\leq (R/a)\,D$, and by Rayleigh monotonicity $\bar D\leq \max_{T'}D_{T'}/a$ for every configuration that contains $T_{\gamma_j}$ with capacities at least $a$. Therefore there exists a constant $C>0$, independent of $j$, such that
\begin{align}
    \bar D(\widehat{\vec k}^{\,j})
    -
    \bar D(\vec k^{\,*}_j)
    \leq
    C\varepsilon_j .
    \label{eq:edge-removal-bound}
\end{align}
This also follows directly by integrating
$\partial\bar D/\partial k_f=-\Psi_{ff}$ from
$k_f=\varepsilon_j$ to $k_f=0$.

The removal of edge $f_j$ frees the resource
$\varepsilon_j^{\gamma_j}$. We redistribute this resource by uniformly rescaling all remaining capacities,
\begin{align}
    \vec k'^{\,j}
    =
    s_j\widehat{\vec k}^{\,j},
    \qquad
    s_j
    =
    \left(
        \frac{R}
        {R-\varepsilon_j^{\gamma_j}}
    \right)^{1/\gamma_j}.
\end{align}
The rescaled configuration again saturates the resource constraint. By homogeneity,
$
    \bar D(\vec k'^{\,j})
    =
    s_j^{-1}\bar D(\widehat{\vec k}^{\,j}).
$ 

Writing $x_j=\varepsilon_j^{\gamma_j}/R$ and using
$0<\gamma_j<1$, we have
$
    s_j^{-1}
    =
    (1-x_j)^{1/\gamma_j}
    \leq
    1-x_j .
$
 
Furthermore, the optimal dissipation is bounded away from zero uniformly for small $\gamma$. Since $x^\gamma\geq 1+\gamma\ln x$ for $x>0$ and $\sum_{e\in T_{\gamma_j}}(k_e^{\,*})^{\gamma_j}\leq R$ with $|T_{\gamma_j}|=R$, we have $\prod_{e\in T_{\gamma_j}}k_e^{\,*}\leq 1$. With $k_e^{\,*}\geq a$ on $T_{\gamma_j}$, this gives $k_e^{\,*}\leq a^{-(R-1)}$ for $e\in T_{\gamma_j}$, while Eq.~\eqref{eq:small-extra-edge} bounds all other capacities by $[R(1-a^{\gamma_j})]^{1/\gamma_j}\to0$. Hence all capacities are uniformly bounded above, and the expectation of Eq.~\eqref{eq:cut-cauchy-schwarz} yields a constant $d_0>0$ such that
$\bar D(\vec k^{\,*}_j)\geq d_0$.
Combining these estimates with
Eq.~\eqref{eq:edge-removal-bound} gives
\begin{align}
    \bar D(\vec k'^{\,j})
    -
    \bar D(\vec k^{\,*}_j)
    \leq
    C\varepsilon_j
    -
    \frac{d_0}{R}\varepsilon_j^{\gamma_j}.
\end{align}
Since $\gamma_j\rightarrow0$ and $\varepsilon_j\rightarrow0$, $
    \varepsilon_j^{\gamma_j}
    \gg
    \varepsilon_j $, 
for sufficiently large $j$. Therefore $
    \bar D(\vec k'^{\,j})
    <
    \bar D(\vec k^{\,*}_j),$
contradicting the global optimality of $\vec k^{\,*}_j$.

Thus, for sufficiently small $\gamma$, no edge outside $T_\gamma$ can be active. Every global minimizer therefore contains at most $R$ active edges. On the other hand, condition~\eqref{eq:cut-nondegenerate} excludes disconnected supports, so every finite-dissipation network must contain at least $R$ active edges. Hence every global minimizer has exactly $R$ active edges and its support is a spanning tree.
\end{proof}

If the network is meshed, $\varsigma\geq1$, a spanning tree leaves at least one edge inactive. If $\Gamma$ acts transitively on the edges, it maps this edge onto an active one, so the optimum is strongly symmetry-broken for $\gamma<\gamma_0$.

\newpage

\subsection{Supplementary Note 4: Symmetry Breaking in Extended Multilayer Networks}
\label{sec:extended-multilayer}

\begin{figure*}[b]
    \includegraphics[width = \linewidth]{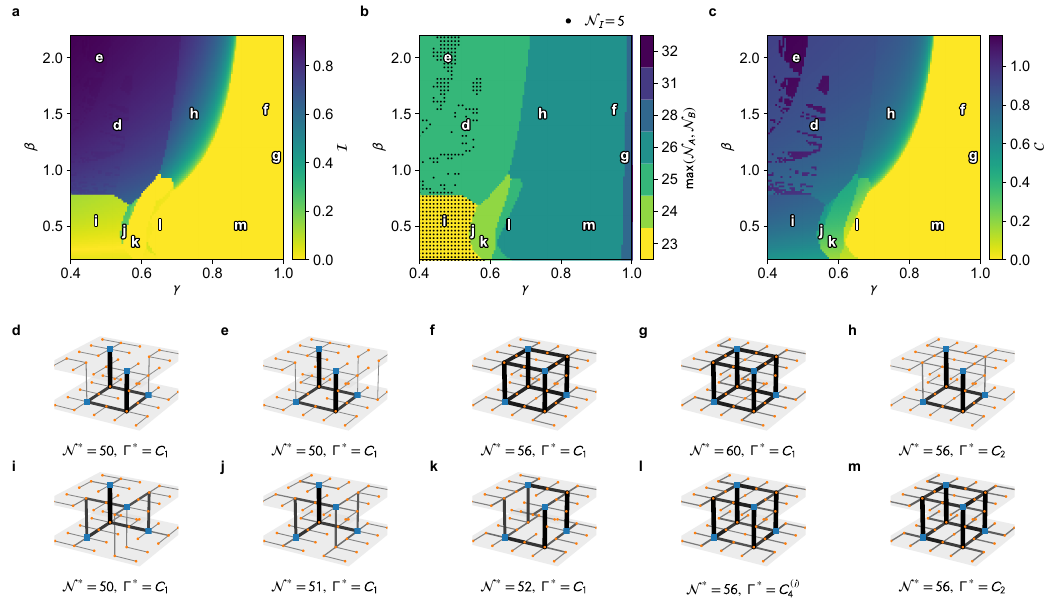}
    \caption{\textbf{Symmetry breaking in a multilayer network.} Two $5\times 5$ lattices with four sources, $\mu = 4/46$. \textbf{a} Layer imbalance $\mathcal{I}$, \textbf{b} number of active edges in the more populated layer, $\max(\mathcal{N}_A,\mathcal{N}_B)$, and \textbf{c} coefficient of variation $\mathcal{C}$ of the active interlayer capacities, as functions of the scaling exponent $\gamma$ and the fluctuation strength $\beta$. In \textbf{b}, dots mark parameters with $\mathcal{N}_I = 5$ active interlayer edges; everywhere else, $\mathcal{N}_I = 4$. Scattered regions in \textbf{b} and \textbf{c} arise from nearly degenerate configurations that differ in single weak edges. Letters mark the parameters of the networks in \textbf{d}--\textbf{m}. \textbf{d}--\textbf{m} Optimal networks, with sources (squares) and sinks (circles); line width and darkness increase with $k_e^*$. Each panel gives the number of active edges $\mathcal{N}^*$ and the residual symmetry $\Gamma^*$. Parameters $(\gamma, \beta)$: \textbf{d} (0.53, 1.39), \textbf{e} (0.48, 2.00), \textbf{f} (0.95, 1.53), \textbf{g} (0.98, 1.12), \textbf{h} (0.75, 1.49), \textbf{i} (0.47, 0.54), \textbf{j} (0.55, 0.45), \textbf{k} (0.58, 0.35), \textbf{l} (0.65, 0.50), \textbf{m} (0.88, 0.50). }
    \label{fig:multiplex-grid}
\end{figure*}

We will now turn to extended multilayer networks. For the sake of simplicity we consider a network with two layers, each consisting of a finite square lattice with two sources (see Supplementary Fig.~\ref{fig:multiplex-grid}). The two layers are connected by edges between all pairs of vertices in the same position in their respective lattices. The four sources are placed in alternating corners of a cube positioned in the center of the whole network. The symmetry group of the optimization problem consists of (i) the identity, (ii) $180^\circ$ rotations around the $x$-, $y$- and $z$-axes, (iii) reflections in the two vertical planes that contain the diagonals, and (iv) improper fourfold rotations around the $z$-axis, such that we have $\Gamma \cong D_{2d}$ with $|\Gamma| = 8$. This is a subgroup of the cube optimization problem symmetry group $T_d$. The rotation around the $z$-axis and the two reflections map each layer onto itself, while the remaining four non-trivial elements exchange the layers.

In this case, $\Gamma$ does not act transitively on the edges. Since every element of $\Gamma$ maps the $z$-axis onto itself, intralayer edges are always mapped onto intralayer edges and interlayer edges onto interlayer edges. Moreover, all the elements involve isometries that fix the center of the network. Thus, edges at different distances from the center will always belong to different orbits.

We set $N_c / N_g \mu = 1$, i.e., $\mu = 4/46 \approx 0.087$.

To explore the optimal network structures across parameter space, we use a two-step parameter scan procedure based on the self-consistent iteration (Methods). In the first step, a coarse scan includes $50\times 50$ points with $\gamma\in[0.4, 1.1]$ and $\beta\in[0.2,2.2]$. At each $\beta$, we lower $\gamma$ from the convex regime $\gamma>1$ by numerical continuation, and additionally start every point from the uniform configuration, which remains $\Gamma$-invariant under the iteration, from a configuration with one layer strengthened, and from four random dense and four random spanning-tree configurations, drawn with a fixed seed. We reduce the resulting configurations to one representative per support class modulo $\Gamma$, distinguishing $\Gamma$-invariant and symmetry-broken representatives. Then, in the second step, we perform a fine scan, on $120\times 120$ points with $\gamma\in[0.4,1.0]$ and $\beta\in[0.2,2.2]$, where we check every point from each of the seed bank representatives and from the uniform configuration. We keep the configuration with the lowest dissipation as the final result. An edge is considered to be active if $k_e > 10^{-5}\max_f k_f$.

To quantify symmetry breaking in the optimal configurations, we introduce several measures. The asymmetry between the two layers, $A$ and $B$, is captured by the normalized capacity imbalance
\begin{align}
\mathcal{I} &= \frac{\left| \sum_{e \in A} k_e \;-\; \sum_{e \in B} k_e \right|}{\sum_{e \in A} k_e \;+\; \sum_{e \in B} k_e}
\end{align}
The group theoretic classification provides an interpretation of the layer imbalance $\mathcal{I}$. If an element that exchanges layers, such as a rotation around the $x$- or $y$-axis, belongs to the residual symmetry $\Gamma^{*}$, both layers have equal total capacities, i.e., $\mathcal{I} = 0$. Thus, $\mathcal{I} > 0$ indicates that all four elements that exchange layers are broken. For strong fluctuations, one layer becomes dominant (Supplementary Fig.~\ref{fig:multiplex-grid}\textbf{d,e}). The strong edges of these configurations are invariant under the rotation around the $z$-axis and the two diagonal reflections, $C_{2v}$, up to relative deviations of $10^{-2}$, while the elements that exchange layers are broken. The weak edges supplying individual sinks break the remaining symmetries, such that $\Gamma^{*} = C_1$, and there are $|\Gamma|/|\Gamma^{*}| = 8$ equally optimal configurations.

For $\gamma > 1$, all links are active and the solution is symmetric. As $\gamma$ decreases below one, links become inactive and symmetry breaking emerges. We therefore track the number of active links in the more populated layer, 
$\max (\mathcal{N}_A,\mathcal{N}_B) $, and the number of active inter-layer links, $\mathcal{N}_I$. To quantify heterogeneity of inter-layer link capacities, we use the coefficient of variation, $\mathcal{C}$:
\begin{align}
\mathcal{C} &= \frac{\sqrt{ \frac{1}{N_\mathcal{I}^{\mathrm{act}}} \sum_{e \in I_{\mathrm{act}}} \left( k_e - \bar{k}_I \right)^2}}{\bar{k}_I }, \\
\bar{k}_I &= \frac{1}{N_\mathcal{I}^{\mathrm{act}}} \sum_{e \in I_{\mathrm{act}}} k_e
\end{align}
evaluated over the $N_\mathcal{I}^{\mathrm{act}}$ active inter-layer links. We plot these results in Supplementary Fig.~\ref{fig:multiplex-grid}.

The resulting phase diagram (Supplementary Fig.~\ref{fig:multiplex-grid}\textbf{a-c}) exhibits several distinct regimes. For $\gamma>1$, the problem is convex and the optimal configuration is symmetric with all edges active (not shown). For $\gamma<1$, edges become inactive. For strong fluctuations, the symmetry between the layers breaks below a $\beta$-dependent threshold, from $\gamma\approx 0.7$ at $\beta\approx 1$ to $\gamma\approx 0.85$ at $\beta\approx 2$, and one layer becomes dominant, $\mathcal{I}>0$. The interlayer edges then divide into a strong and a weak group, $\mathcal{C}\approx 0.8$. For weak fluctuations and small $\gamma$, the optimal network is a spanning tree with one additional edge, $\mathcal{N}^*=50$, with five active interlayer edges (Supplementary Fig.~\ref{fig:multiplex-grid}\textbf{i}). Near $\gamma\approx 1$, both layers remain equivalent, $\mathcal{I}\approx 0$, and the optimum retains a residual symmetry $C_2$ or $C_4^{(i)}$ in parts of the parameter plane (Supplementary Fig.~\ref{fig:multiplex-grid}\textbf{h,l,m}). Many configurations differ only in single weak edges supplying individual sinks and have nearly equal dissipation. The global optimum switches between them under small parameter changes, which produces the scattered regions in Supplementary Fig.~\ref{fig:multiplex-grid}\textbf{b,c}.

In the case of weak fluctuations, a source node typically supplies its neighboring sinks. Hence, the largest squared flows $\langle F_e^2 \rangle$, and thus the largest capacities $k_e$, are generally found for the edges $e$ adjacent to the source nodes. Therefore, there are four strong edges connecting the two layers. For large values of the cost parameter $\gamma$, these four interlayer edges have the same capacity $k_e$ respecting the symmetry of the optimization problem. This symmetry is spontaneously broken when $\gamma$ decreases, and the network approaches a spanning tree.

A different manifestation of symmetry breaking is observed in the case of strong fluctuations ($\beta \gtrsim 1.0$). These fluctuations are balanced by strong flows between the source nodes. Hence, the squared flows $\langle F_e^2\rangle$ and the capacities $k_e$ are large for edges $e$ linking the four source nodes.
For large values of the cost parameter $\gamma$, these strong edges form a cube, respecting the discrete rotational symmetry of the optimization problem.  In particular, a square of strong edges is observed in both layers of the network.
As the cost parameter decreases, redundant connections become less likely~\cite{corson2010fluctuations}. Consequently, the symmetry between the two layers is spontaneously broken below a $\beta$-dependent threshold: only one layer contains a square of strong edges. This ring is connected by two strong edges to sources in the other layer (Supplementary Fig.~\ref{fig:multiplex-grid}\textbf{d,e}). 

The two regimes indicate that the reentrant behavior observed for edge-transitive networks persists in this less symmetric setting. For $\gamma\approx 0.55$--$0.65$, the interlayer capacities are heterogeneous at weak fluctuations (regions j, k in Supplementary Fig.~1), become equal at intermediate fluctuations (region l), and heterogeneous again at strong fluctuations (regions d, h). The layer imbalance $\mathcal I$ follows the same pattern. This reflects a competition between local balancing, which dominates at weak fluctuations, and global balancing between the sources, which dominates at strong fluctuations. We note, however, that $\mathcal I = 0$ and $\mathcal C = 0$ only indicate that the layer-exchanging symmetries and the equivalence of the interlayer edges are preserved, respectively, and do not imply full $\Gamma$-symmetry of the optimum.

\newpage

\subsection{Supplementary Note 5: Inherited phases in nested networks}
\label{snote:nested}

Strong symmetry breaking removes edges, so the optimization effectively reduces to a smaller network, whose optimal configurations can become phases of the original one. Consider a spanning subgraph $G'=(\mathcal V,\mathcal E')$ of $G=(\mathcal V,\mathcal E)$. Since they have the same nodes and injection statistics, a configuration of $G'$ is a configuration of $G$ with $k_e=0$ for all $e\notin\mathcal E'$, with the same dissipation and resource cost. Hence, a global minimizer of $G$ supported on $\mathcal E'$ is also a global minimizer of $G'$. Moreover, for $\gamma<1$, every strict local minimum of $G'$ with a connected support is a local minimum of $G$, since the additional edges are inactive and protected against reactivation (Supplementary Methods~3). The local minima of $G'$ are thus local minima of $G$. They form phases of $G$ wherever they are also globally optimal, which requires comparison with supports not contained in $\mathcal E'$. For $\gamma>1$, configurations with inactive edges are generically not local minima, and no configurations are inherited.

The model networks of the main text are nested as $C_8\subset Q_3\subset K_{4,4}$, with identical injection statistics, Eq.~(20) of the main text with $N_g=N_c=4$. Every cycle through all eight nodes of the cube $Q_3$ alternates between sources and sinks. The eight-edge ring phase of the cube is therefore equivalent to the ring with $N=4$, including the onset of alternating capacities at $\gamma_s(\mathrm{alt})|_{N=4}$ (Supplementary Methods~9). Removing four edges that pair each source with a different sink from $K_{4,4}$ leaves the cube, and there are $4!=24$ such embeddings. At weak fluctuations, the cube configurations are globally optimal in $K_{4,4}$ (Fig.~5i--k of the main text). This removal of edges first reduces $K_{4,4}$ to the cube, which then follows its own sequence of phases until the hub tree becomes optimal at small $\gamma$.

An inherited configuration can retain more symmetry in $K_{4,4}$ than in the cube. The symmetry group of the cube problem is the subgroup of $\Gamma\cong S_4 \times S_4$ that preserves the removed links. If the support of a configuration is contained in several of the 24 cubes embedded in $K_{4,4}$, $\Gamma$ contains permutations that exchange these embeddings. The support of the nine-edge phase is contained in two cubes, which raises $|\Gamma^*|$ from 3 to 6 (Supplementary Table~\ref{tab:bipartite-phases}). In $K_{4,4}$, $\Gamma^* \cong S_3$ contains transpositions, so the nine-edge phase is no longer chiral.

\newpage

\subsection{Supplementary Note 6: Energy-system example with renewable generation}
\label{sec:renewables}

We consider the six-node ring introduced in the main text, where every node represents a geographical region. Regions $i=2,4,6$ have constant demand,
\begin{align}
    S_i=-\mu .
\end{align}
Regions $i=1$ and $i=3$ host wind and solar generation, respectively,
\begin{align}
    S_1(t) &= (1-\zeta)\mu+\zeta\mu Z_1(t),\\
    S_3(t) &= (1-\zeta)\mu+\zeta\mu Z_3(t),
\end{align}
where $Z_1(t)$ and $Z_3(t)$ denote normalized wind and solar capacity factors. The parameter $\zeta\in[0,1]$ quantifies the share of fluctuating renewable power generation and therefore controls the strength of the injection fluctuations. We rescale the capacity factors such that
\begin{align}
    \langle Z_1\rangle=\langle Z_3\rangle=1 .
\end{align}
The remaining node $i=5$ models a storage or balancing infrastructure and compensates the total mismatch,
\begin{align}
    S_5(t)=3\mu-S_1(t)-S_3(t),
\end{align}
such that the network remains balanced at every time step.

For the numerical experiment, we use wind and solar generation time series obtained from renewables.ninja~\cite{pfenninger2016long,staffell2016using} for the year 2019 at the location Jülich. The second moments entering the optimization are computed from the empirical time average
\begin{align}
    \langle S_iS_j\rangle
    =
    \frac{1}{T}\sum_{t=1}^{T}S_i(t)S_j(t),
\end{align}
where $T$ denotes the number of time steps. Wind and solar generation are anti-correlated on seasonal time scales, with solar generation being stronger in summer and wind generation stronger in winter~\cite{heide2010seasonal}. The balancing injection $S_5$ is anti-correlated with the renewable generation by construction, as the storage or balancing infrastructure compensates variations in the renewable supply.

The injection variables $S_1$, $S_3$, and $S_5$ are not identically distributed, such that the full stochastic problem is not strictly symmetric. However, the network remains symmetric on average: all three generation nodes satisfy $\langle S_i\rangle=\mu$, while the three demand nodes have $S_i=-\mu$. Thus, the underlying ring and the mean injection pattern retain the alternating source--sink symmetry, whereas the fluctuations explicitly break it. Consequently, the transition discussed in the main manuscript is not a spontaneous symmetry-breaking transition in the strict group-theoretic sense. Nevertheless, the topological signature associated with strong symmetry breaking remains well defined: the support of the optimal network changes when one or more edge capacities vanish.

\newpage 

\section{Supplementary Methods}

\subsection{Supplementary Methods 1: Proof of Theorem 1}

\begin{thm}
For $\gamma>1$, the global minimizer of $\bar D$ on $\mathcal K_\gamma$ is unique and symmetric. For $\gamma=1$, the set of global minimizers contains a symmetric configuration. If $\Gamma$ acts transitively on $\EE$, the symmetric minimizer is uniform,
\begin{align}
    k_e=|\EE|^{-1/\gamma}\kappa .
\end{align}
\end{thm}

\begin{proof}
To ensure that $\bar D$ remains convex when edges are inactive, we use the following convention: if $F_e = k_e = 0$, then $F_e^2/k_e = 0$. If $k_e = 0, F_e \neq 0$, we formally set $F_e^2/k_e = + \infty$. Then $\bar D(\vec k) = +\infty$ if some connected component of the active subgraph has unbalanced injections, and $\bar D(\vec k)=\operatorname{Tr} (\matr L^+ \matr \Sigma)$ otherwise (Eq.~(6) of the main text; $^+$ denotes the Moore-Penrose pseudoinverse). With this convention, $\bar D$ is convex on $\vec k \geq 0$ (Methods and Ref.~\cite{boyd2004convex}). This holds for every $\gamma$, since $\bar D$ does not depend on $\gamma$. Moreover, for $\gamma \geq1$, the feasible set $\mathcal K_\gamma =
    \left\{
        \vec k\geq0:
        \sum_{e\in\EE}k_e^\gamma\leq\kappa^\gamma
    \right\}$
is convex, since $k_e \mapsto k_e^\gamma$ is convex on $k_e \geq 0$, and the constraint function is a sum of convex functions.

We now prove $\Gamma$-invariance of $\bar D$. For $g\in\Gamma$, let $g \cdot \vec k$ denote the relabeled capacities, $(g\cdot\vec k)_{\pi_g(e)}=k_e$. The node permutation maps the incidence column $\matr E_e$ onto $\pm\matr E_{\pi_g(e)}$, where the sign accounts for the edge orientation. Thus
\begin{align}
    \matr L(g\cdot\vec k)
    = \sum_{e} k_e\, \matr E_{\pi_g(e)}\matr E_{\pi_g(e)}^\top
    = \sum_{e} k_e\, \matr R_g\matr E_e\matr E_e^\top\matr R_g^\top
    = \matr R_g\matr L(\vec k)\matr R_g^\top .\label{eq:L_relabel}
\end{align}
Since $\matr R_g$ is orthogonal, $\bigl(\matr R_g\matr L\matr R_g^\top\bigr)^+=\matr R_g\matr L^+\matr R_g^\top$. Using cyclicity of the trace and $\matr R_g^\top\matr\Sigma\matr R_g=\matr\Sigma$, equivalent to Eq.~(8) of the main text, we obtain
\begin{align}
    \bar D(g\cdot\vec k)
    = \operatorname{Tr}\bigl(\matr R_g\matr L^+(\vec k)\matr R_g^\top\matr\Sigma\bigr)
    = \operatorname{Tr}\bigl(\matr L^+(\vec k)\matr R_g^\top\matr\Sigma\matr R_g\bigr)
    = \operatorname{Tr}\bigl(\matr L^+(\vec k)\matr\Sigma\bigr)
    = \bar D(\vec k).\label{eq:D_invariance}
\end{align}
The identity also holds where $\bar D=+\infty$. A component $C$ of the active subgraph of $\vec k$ is mapped onto the component $g(C)$ of the active subgraph of $g\cdot\vec k$, and $\matr\Sigma\vec 1_{g(C)}=\matr R_g\matr\Sigma\vec 1_C$, i.e., unbalanced components remain unbalanced.

The feasible set $\mathcal K_\gamma$ is also invariant under $\Gamma$. Since $g$ only permutes the capacities, $g\cdot\vec k\geq 0$ and $\sum_e (g\cdot\vec k)_e^\gamma=\sum_e k_e^\gamma$.

For uniqueness, we first show that every global minimizer exhausts the resource budget. 
Since $\matr L$ is linear in $\vec k$, $\matr L(\lambda\vec k)=\lambda\matr L(\vec k)$ for $\lambda>0$, and $\matr L(\lambda\vec k)^+=\lambda^{-1}\matr L(\vec k)^+$. Therefore $\bar D(\lambda\vec k)=\operatorname{Tr}\left(\lambda^{-1}\matr L^+\matr\Sigma\right)=\lambda^{-1}\bar D(\vec k)$. 
Moreover, for $\matr \Sigma \neq0$, we have $\bar D>0$. If all edges are active, $\bar D$ is finite, since the graph is connected and has balanced injections. Therefore, the optimal dissipation is finite and positive. A configuration with finite $\bar D$ that satisfies $\sum_e k_e^\gamma<\kappa^\gamma$ can be rescaled by some $\lambda>1$ to a configuration that exhausts the budget and has strictly lower dissipation. Thus every global minimizer must satisfy $\sum_e k_e^\gamma = \kappa^\gamma$.

We now prove uniqueness for $\gamma>1$ by contradiction. Suppose that $\vec k^{(1)} \neq \vec k^{(2)}$ are both global minimizers. Their midpoint is in $\mathcal K_\gamma$ by convexity of $\mathcal K_\gamma$, and it is a global minimizer by convexity of $\bar D$. By strict convexity of $\vec k\mapsto\sum_e k_e^\gamma$, however, it satisfies $\sum_e k_e^\gamma<\kappa^\gamma$, contradicting the exhaustion of the budget shown above. The global minimizer $\vec k^*$ is therefore unique.
For every $g \in \Gamma$, $g\cdot\vec k^*$ belongs to $\mathcal K_\gamma$ and, by invariance, has the same dissipation, so it is also a global minimizer. Uniqueness yields $g \cdot \vec k^* = \vec k^*$, i.e.\ $\vec k^*$ is symmetric~\cite{GhoshBoydSaberi2008}.

At $\gamma=1$, the argument above cannot be applied, since $\sum_e k_e$ is not strictly convex and the minimizer may not be unique. We can nevertheless construct a symmetric minimizer. Let $\vec k^*$ be a global minimizer and
$
    \bar{\vec k}=\frac{1}{|\Gamma|}\sum_{g\in\Gamma} g\cdot\vec k^*
$
its orbit average, i.e., the average over the orbit $\{g\cdot\vec k^*: g\in\Gamma\}$ of configurations obtained by relabeling the network in all possible ways. It is $\Gamma$-invariant, since relabeling $\bar{\vec k}$ only permutes the terms of the sum. It is in $\mathcal K_\gamma$ by convexity and invariance of $\mathcal K_\gamma$. 
By convexity (Jensen's inequality) and invariance of $\bar D$, we conclude that $\bar{\vec k}$ is a symmetric global minimizer~\cite{Guo2012, gatermann2004symmetry,boyd2009fastest},
\begin{align}
    \bar D(\bar{\vec k})\leq \frac{1}{|\Gamma|}\sum_{g\in\Gamma}\bar D(g\cdot\vec k^*)=\bar D(\vec k^*).
\end{align}

Finally, let $\Gamma$ act transitively on $\EE$. A symmetric configuration satisfies $k_{\pi_g(e)}=k_e$ for all $e\in\EE$ and $g\in\Gamma$. By transitivity, for any two edges $e,f$ there is a $g\in\Gamma$ with $\pi_g(e)=f$. Then, $k_f=k_e$, and all capacities share the same value $k$. Since the symmetric minimizer exhausts the budget, $|\EE|\,k^\gamma=\kappa^\gamma$, i.e., $k=|\EE|^{-1/\gamma}\kappa$.
\end{proof}

For $\gamma=1$ and $\matr\Sigma\propto\matr I-\vec 1\vec 1^\top/|\mathcal V|$, $\bar D$ is proportional to the total effective resistance of the network. For this case, convexity of the optimization problem and uniform optimal capacities on edge-transitive graphs were established in Ref.~\cite{GhoshBoydSaberi2008}. There, uniqueness follows from strict convexity of the objective. 
We generalize this result to arbitrary injection statistics $\matr\Sigma$ and to symmetry groups that preserve them. For $\gamma>1$ we prove uniqueness based on the  strict convexity of the resource constraint. For $\gamma=1$ we obtain a symmetric minimizer by orbit averaging, a standard argument for convex problems invariant under a finite group~\cite{gatermann2004symmetry,boyd2009fastest}.

\newpage
\subsection{Supplementary Methods 2: Block decomposition of the constrained Hessian}

Here we derive the commutation of the Hessian with all symmetry operations, the decomposition of the tangent space into blocks, the block-wise thresholds of Eq.~(17) for edge-transitive networks, and the additional symmetric block for networks that are not edge-transitive. We also include a general result on residual symmetries, which we use for the cube in Supplementary Methods~8.

Consider a stationary point $\vec k$ of $\bar D$ on the constraint surface with $k_e>0$ for all $e$. Its local stability is determined by the Hessian of the Lagrangian (Eq.~(51) of the Methods), restricted to the tangent space $T_{\vec k}$ defined in Eq.~(13) of the main text. If this Hessian is positive definite, $\vec k$ is a strict local minimum. If it has a negative eigenvalue, $\vec k$ is not a local minimum.

\paragraph{Commutation with symmetry operations.} We first show the invariance of the Lagrangian $\mathcal{L}$ under relabeling. 
For $g \in \Gamma$, the relabeled capacities are $g \cdot \vec k=\matr \Pi_g \vec k$, where $\matr \Pi_g$ is the permutation matrix of $\pi_g$ on edge space. Since both $\bar D$ (Supplementary Methods~1) and the constraint function $\sum_e k_e^\gamma$ are invariant under relabeling, the Lagrangian $\mathcal L$ is also invariant under relabeling,
\begin{align}
    \mathcal L(\matr\Pi_g\vec k)=\mathcal L(\vec k) \qquad \text{for all }\vec k .
\end{align}
Differentiating with respect to $\vec k$ yields
\begin{align}
    \matr\Pi_g^\top\,\nabla\mathcal L(\matr\Pi_g\vec k)=\nabla\mathcal L(\vec k),\qquad
    \matr\Pi_g^\top\,\nabla^2\mathcal L(\matr\Pi_g\vec k)\,\matr\Pi_g=\nabla^2\mathcal L(\vec k).
\end{align}
At a symmetric configuration, $\matr \Pi_g \vec k = \vec k$, the second relation shows that the Hessian commutes with every $\matr\Pi_g$
\begin{align}
    \matr\Pi_g^\top\,\nabla^2\mathcal L(\vec k)\,\matr\Pi_g=\nabla^2\mathcal L(\vec k)
    \quad\Longleftrightarrow\quad
    \nabla^2\mathcal L(\vec k)\,\matr\Pi_g=\matr\Pi_g\,\nabla^2\mathcal L(\vec k)
    \qquad \text{for all } g\in\Gamma.
    \label{eq:commutation}
\end{align}
That is, a mode $\vec \delta$ and the relabeled mode $\matr \Pi_g \vec \delta$ affect the Lagrangian equally to second order.

Since $\bar D$ is itself invariant under relabeling, $\bar D(\matr\Pi_g\vec k)=\bar D(\vec k)$ holds for all $\vec k$ (Supplementary Methods~1). Differentiating this once with respect to $\vec k$ gives $
    \matr\Pi_g^\top\nabla\bar D(\matr\Pi_g\vec k)=\nabla\bar D(\vec k).$ 
At a symmetric configuration, $\matr\Pi_g\vec k=\vec k$, we get $\matr\Pi_g^\top\nabla\bar D(\vec k)=\nabla\bar D(\vec k)$. Multiplying by $\matr\Pi_g$ from the left, we obtain
\begin{align}
    \matr\Pi_g\nabla\bar D(\vec k)=\nabla\bar D(\vec k)\qquad\text{for all } g\in\Gamma.
    \label{eq:gradient-invariance}
\end{align}
With $\partial\bar D/\partial k_e=-\Psi_{ee}$, Eq.~(46) of the Methods, $\Psi_{\pi_g(e)\pi_g(e)}=\Psi_{ee}$ holds for all $g\in\Gamma$, i.e., $\Psi_{ee}$ takes the same value on equivalent edges. In other words, $\Psi_{ee}$ is constant on edge orbits, where an edge orbit of $e$ is the set $\{\pi_g(e):g\in\Gamma\}$ of edges onto which $e$ can be mapped by a symmetry. The edge set decomposes into $r$ disjoint orbits, $\mathcal E=\bigcup_{a=1}^r\mathcal E_a$, with $r=1$ if and only if $\Gamma$ acts transitively on the edges.

\paragraph{Block decomposition.} 
The action of $\Gamma$ decomposes edge space into mutually orthogonal blocks $W_\rho$, one for each irreducible representation $\rho$ of $\Gamma$. Each block $W_\rho$ consists of the perturbations transforming according to $\rho$~\cite{Serre1977}. The orthogonal projector onto $W_\rho$ is a linear combination of the permutation matrices $\matr\Pi_g$~\cite{Serre1977, cotton1990chemical,pecora2014},
\begin{align}
\matr P_\rho = \frac{d_\rho}{|\Gamma|} \sum_{g \in \Gamma} \overline{\chi_\rho(g)}\, \matr \Pi_g
= \frac{d_\rho}{|\Gamma|} \sum_{g \in \Gamma} \chi_\rho(g)\, \matr \Pi_g,\label{eq:projector}
\end{align}
where $d_\rho=\dim\rho$ and $\chi_\rho(g)$ is the character of $\rho$, i.e., the trace of the matrix representing $g$ in $\rho$. The second equality holds for all symmetry groups we consider ($D_N$, $S_4$ and $S_{N_g}\times S_{N_c}$), since they have real characters. All their irreducible representations are real representations, thus all blocks are real subspaces of edge space. Note that for general symmetry groups, where irreducible representations may not be real, the block-diagonal structure below is not affected, but eigenvalue degeneracies will increase~\cite{Serre1977}.
By Eq.~\eqref{eq:commutation}, $\nabla^2\mathcal L$ commutes with every $\matr P_\rho$ and is therefore block diagonal. By Schur's lemma, $\nabla^2\mathcal L$ restricted to $W_\rho$ reduces to an $m_\rho\times m_\rho$ matrix, where $m_\rho$ is the number of copies of $\rho$ in edge space, with each eigenvalue having multiplicity $d_\rho = \dim\rho$~\cite{Serre1977, gatermann2004symmetry,boyd2009fastest}. The trivial block $W_{\mathrm{triv}}$ consists of the perturbations that are constant on edge orbits. Its dimension is equal to the number of edge orbits $r$, cf.\ Ref.~\cite{boyd2009fastest,pecora2014}.

\paragraph{Resource constraint.}Admissible perturbations are orthogonal to the normal vector of the constraint surface, $n_e=\gamma k_e^{\gamma-1}$, i.e., $T_{\vec k}=\vec n^\perp$, Eq.~(13) of the main text. At a symmetric configuration, $\vec n$ is constant on edge orbits, i.e., $\vec n \in W_{\mathrm{triv}}$. Thus all nontrivial blocks are orthogonal to $\vec n$ and in $T_{\vec k}$, i.e., symmetry-breaking perturbations preserve the resource budget automatically. Then only $W_{\mathrm{triv}} \cap T_{\vec k}$ remains of the trivial block. This consists of symmetric perturbations that redistribute resources between edge orbits, and
\begin{align}
    T_{\vec k}=\left(T_{\vec k}\cap W_{\mathrm{triv}}\right)\oplus\bigoplus_{\rho\neq\mathrm{triv}}W_\rho,
    \qquad \dim\left(T_{\vec k}\cap W_{\mathrm{triv}}\right)=r-1.
    \label{eq:tangent-decomposition}
\end{align}
Since $\matr \Pi_g \vec n = \vec n$, the projector onto $T_{\vec k}$ commutes with every $\matr\Pi_g$, and the constrained Hessian remains block diagonal, with one block for each nonzero $W_\rho\cap T_{\vec k}$. This is the decomposition used in the main text. 

By Eq.~\eqref{eq:gradient-invariance}, also $\nabla \bar D\in W_{\mathrm{triv}}$ at a symmetric configuration, so its projection onto $T_{\vec k}$ lies in $T_{\vec k}\cap W_{\mathrm{triv}}$. A symmetric configuration is thus stationary when it is stationary with respect to symmetric perturbations.

A loss of local stability results in symmetry breaking only if it occurs in a nontrivial block, i.e., when the smallest eigenvalue of that block crosses zero. This is weak symmetry breaking. An instability in $T_{\vec k}\cap W_{\mathrm{triv}}$ leads to another symmetric configuration. This is analogous to cluster synchronization, where perturbations within the synchronization manifold and those transverse to the synchronization manifold belong to separate blocks~\cite{sorrentino2016complete, pecora2014}. Note that the Hessian cannot determine strong symmetry breaking, which occurs through an exchange of global optimality (Eq.~(19) of the main text).

\paragraph{Residual symmetry.}
When a nontrivial block $W_{\rho^*}$ loses stability, the unstable modes  transform according to $\rho^*$. The residual symmetry of a configuration $\vec k+\eta\vec\delta$ branching off the symmetric configuration along an unstable mode $\vec\delta\in W_{\rho^*}$ is $\{g\in\Gamma:\matr\Pi_g\vec\delta=\vec\delta\}$, i.e., an isotropy subgroup of the action of $\Gamma$ on $W_{\rho^*}$, as in equivariant bifurcation theory~\cite{golubitsky2002}. The Hessian cannot determine whether the resulting transition is continuous. We use this result for the cube in Supplementary Methods~8.

\paragraph{Edge-transitive networks.}
If $\Gamma$ acts transitively on $\EE$, then $r=1$, $W_{\mathrm{triv}}=\mathrm{span}(\vec 1)$, and at the uniform configuration $T_{\vec k}=\vec 1^\perp=\bigoplus_{\rho\neq\mathrm{triv}}W_\rho$. Since $\Psi_{ee}$ is constant on edge orbits, it has equal values for all edges, so the uniform configuration is stationary, and the constrained Hessian takes the form of Eq.~(15) of the main text (derived in the Methods). Since the identity term shifts all eigenvalues equally, an eigenvector of $2\tilde{\matr G}\circ\tilde{\matr\Psi}$ in $W_\rho$ with eigenvalue $h$ is a stable direction if and only if
\begin{align}
    \gamma>1-\frac{h}{\tilde\Psi_{ee}}.
\end{align}
Dividing by $\tilde\Psi_{ee}$ preserves the inequality, since $\tilde\Psi_{ee}>0$: as a mean-square potential difference it is nonnegative, and since it is equal on all edges, $\tilde\Psi_{ee}=0$ only if all injections are zero. Applied to the smallest eigenvalue $h_{\min}(\rho)$ of each block, this yields the block-wise thresholds of Eq.~(17) of the main text. The overall threshold of Eq.~(16) of the main text is 
\begin{align}
\gamma_s=\max_\rho\gamma_s(\rho).
\end{align}
At $\gamma=1$, the constrained Hessian reduces to $2\,\tilde{\matr{G}}\circ\tilde{\matr{\Psi}}/k_0^3$, which is positive semidefinite. Hence $\bar D$ is flat to second order along block $\rho$ if and only if $h_{\min}(\rho)=0$, i.e., $\gamma_s(\rho)=1$. At discontinuous transitions with $\gamma_c<1$, $\bar D$ is generically not flat, since the competing configurations are isolated minima with equal values of the dissipation.

\paragraph{Non-edge-transitive networks.}
If $r>1$, $\Psi_{ee}$ can differ between orbits, and the term $(\gamma-1)\,\mathrm{diag}(\Psi_{ee}/k_e)$ in Eq.~(51) of the Methods is no longer proportional to the identity. It still commutes with every $\matr \Pi_g$, so the block structure is still applicable, but the thresholds must be obtained from the blocks of the full constrained Hessian rather than from Eq.~(17), and the block $T_{\vec k}\cap W_{\mathrm{triv}}$ must be included.

\newpage
\subsection{Supplementary Methods 3: Dissipation expansion for reactivated edges}
\label{sec:dissipation-reactivation}
Consider a configuration with an inactive edge $e$, such that $k_e=0$, whose active edges $\mathcal A$ connect all nodes. Assume that the resource constraint is exhausted, i.e., $\sum_f k_f^\gamma = \kappa^\gamma$, and that the stationarity condition $\Psi_{ff} = \nu \gamma k_f^{\gamma-1}$ (Eq.~(50) of the Methods) is satisfied on $\mathcal A$, where $\nu$ is the Lagrange multiplier of the resource constraint.

For $\gamma<1$, the derivative of the constraint function $\gamma k_e^{\gamma-1}$ diverges as $k_e \to 0^+$, so an inactive edge remains consistent with the stationarity condition, however large $\Psi_{ee}$. To determine whether reactivation lowers the dissipation we use the following expansion.

We expand $\bar D$ for feasible configurations in which edge $e$ is reactivated with a small capacity $k_e=\varepsilon$, while the active capacities change by $\delta k_f$ to keep the budget exhausted, $\sum_{f\in\mathcal A}\gamma k_f^{\gamma-1}\delta k_f=-\varepsilon^\gamma$ to leading order. Since the two capacity changes scale as $\varepsilon$ and $\varepsilon^\gamma$, the expansion contains terms of both orders. Using $\partial\bar D/\partial k_e=-\Psi_{ee}$, Eq.~(46) of the Methods, and $\Psi_{ff}=\nu\gamma k_f^{\gamma-1}$, we find
\begin{align}
    \bar D(\varepsilon)-\bar D(0) = -\Psi_{ee}\,\varepsilon - \sum_{f\in\mathcal A}\Psi_{ff}\,\delta k_f + \dots
    = -\Psi_{ee}\,\varepsilon + \nu\,\varepsilon^\gamma + \dots
\end{align}
To leading order, this result does not depend on how the compensation is distributed over the active edges. To identify the coefficient $\nu$, we differentiate $\bar D(\lambda\vec k)=\lambda^{-1}\bar D(\vec k)$ at $\lambda=1$, which yields $\sum_f k_f\Psi_{ff}=\bar D$. With Eq.~(50) of the Methods and the exhausted budget, this gives $\nu=\bar D(0)/(\gamma\kappa^\gamma)>0$, resulting in Eq.~(18) of the main text. The neglected terms are of higher order than $\nu\varepsilon^\gamma$. They are small in the vicinity of $\varepsilon_*$, if $\varepsilon_* \ll \kappa$.

For $\gamma<1$, the term $\nu\varepsilon^\gamma$ dominates as $\varepsilon\to0^+$, so reactivation raises the dissipation. A configuration that is a strict local minimum with respect to its active capacities, cf.\ Eq.~(51) restricted to $\mathcal A$, therefore remains a local minimum when its inactive edges are reactivated, as shown for trees in Ref.~\cite{kaiser2020discontinuous}. For $\gamma>1$, the linear term dominates whenever $\Psi_{ee}>0$, so the configuration is not a local minimum. At $\gamma=1$, Eq.~(50) gives $\Psi_{ff}=\nu$ on all active edges, and the inactive edge stays inactive if $\Psi_{ee}<\nu$, i.e., if the mean-square potential difference across it is smaller than across the active edges. An equivalent criterion appears in the optimization of total effective resistance and of power-distribution networks~\cite{GhoshBoydSaberi2008,Boyd2001ISPD}.

For $\gamma<1$, the two terms are balanced at the crossover scale
\begin{align}
    \varepsilon_*=\left(\frac{\nu}{\Psi_{ee}}\right)^{\frac{1}{1-\gamma}}
    =\left(\frac{\bar D(0)}{\gamma\kappa^\gamma\Psi_{ee}}\right)^{\frac{1}{1-\gamma}} .
\end{align}
Reactivation raises the dissipation for $\varepsilon\ll\varepsilon_*$, and the expansion indicates a decrease beyond $\varepsilon_*$. Therefore $\varepsilon_*$ estimates the extent of the basin of the configuration along the reactivation direction. For a configuration with $\Psi_{ee}>\nu$, the crossover scale $\varepsilon_*$ collapses as $\gamma \to 1^-$, because the exponent $1/(1-\gamma)$ diverges. The broken configuration then remains a local minimum for all $\gamma<1$, but its basin along this direction becomes vanishingly small, and numerical searches may lose it as $\gamma \to 1^-$. Fig.~2\textbf{e} of the main text shows this for the ring example.

If the active edges do not connect all nodes, e.g., if edge $e$ joins two balanced components of the active subgraph, edges must be reactivated together, as in the example in Supplementary Methods 4.

\newpage
\subsection{Supplementary Methods 4: Dissipation of hexagon with $\mu=0$}
\label{sec:hexagon-triangle}

For $N=3$ and $\mu=0$, Kirchhoff's current law implies that the flows on the two edges adjacent to each sink are equal,
\begin{align*}
    F_1=F_2 \equiv f_1,\qquad
    F_3=F_4 \equiv f_2,\qquad
    F_5=F_6 \equiv f_3.
\end{align*}
At a stationary point, the two edges of each pair therefore have equal capacities, Eq.~(31) of the Methods. In the following, we denote the common capacities of the three pairs by $k_1$ (edges 1 and 2), $k_2$ (edges 3 and 4) and $k_3$ (edges 5 and 6). Furthermore, we have Kirchhoff's current law at the source nodes
\begin{align*}
    f_1-f_3=X_1,\qquad
    f_2-f_1=X_2,\qquad
    f_3-f_2=X_3,
\end{align*}
with $X_1+X_2+X_3=0$ and Kirchhoff's voltage law
\begin{align}
    2\sum_{j=1}^3 \frac{f_j}{k_j} = 0.
\end{align}
Solving this linear system of equations yields
\begin{align}
    f_3 &= -\frac{k_3\big((k_1+k_2)X_1+ k_1 X_2\big)}{k_1 k_2+k_1 k_3+k_2 k_3}, 
    \\
    f_1 &= X_1 + f_3,
    \\
    f_2 &= X_1+X_2+f_3.
\end{align}
Hence, the expected dissipation reads 
\begin{align*}
    \bar D &= 2\sum_{j=1}^3 \frac{\langle f_j^2 \rangle}{k_j} = \frac{2\left[ (k_1 + k_2) \langle X_1^2 \rangle +2 k_1 \langle X_1 X_2 \rangle  + (k_1 + k_3) \langle X_2^2 \rangle
    \right] }{k_1 k_2+k_1 k_3+k_2 k_3} .
\end{align*}
Using the second moments given in Eq.~(32) of the Methods, the expected dissipation
can be expressed directly in terms of the edge pair capacities as
\begin{align}
    \bar D(k_1,k_2,k_3)=4\beta^2\frac{k_1 + k_2 + k_3}{k_1 k_2+k_1 k_3+k_2 k_3}.\label{eq:diss-hexagon}
\end{align}

We now compare the dissipation of the symmetric configuration $k_1 = k_2 = k_3 = 6^{-1/\gamma}\kappa $ and that of the strongly symmetry-broken state in which one sink is disconnected. We set $k_1 = 0$, $k_2 = k_3 = 4^{-1/\gamma}\kappa$ without loss of generality. The dissipations become equal at the transition point
\begin{align*}
    \gamma_c = \frac{ \ln (3/2)}{\ln 2} \approx 0.585.
\end{align*}

To analyze the stability of this symmetry-broken state, we apply the reactivation expansion of Supplementary Methods~3. In this case, the inactive pair must be reactivated together, since flow can only pass through the sink if both of its adjacent edges are active. Setting $k_1=\varepsilon$, the budget cost is $2\varepsilon^\gamma$, and Eq.~\eqref{eq:diss-hexagon} yields
\begin{align}
    \bar D(\varepsilon)-\bar D(0) = -C_1\varepsilon + C_2\varepsilon^\gamma + \dots,
    \qquad
    C_1=-\frac{\partial\bar D}{\partial k_1}\bigg|_{\varepsilon=0}=\frac{12\beta^2}{k_0^2},
    \qquad
    C_2=2\nu=\frac{16\beta^2}{\gamma k_0\kappa^\gamma},
\end{align}
with $k_0=4^{-1/\gamma}\kappa$ and $\nu=\bar D(0)/(\gamma\kappa^\gamma)$. For $\gamma<1$, the $\varepsilon^\gamma$ term dominates and the broken state is a local minimum. At $\gamma=1$, $C_2/C_1=1/3<1$, so the minimum disappears at $\gamma_b=1$. The crossover scale, $\varepsilon_*=(C_2/C_1)^{1/(1-\gamma)}$, collapses as $\gamma\to1^-$ (Fig.~2\textbf{e} of the main text).

\newpage
\subsection{Supplementary Methods 5: Expected dissipation of the ring configurations}
\label{sec:dissipation-as}

Here we compute the expected dissipation, using Eq.~(38) of the Methods,
\begin{align}
    \bar{D} = \mu^2 \vec{v}^\top \matr{L}^{+} \vec{v} + \beta^2 \sum_{i<j\in\mathcal{V}_g} R_{ij},
    \label{eq:ring_D_split}
\end{align}
for the symmetric, the alternating, the sink-paired ($N=2$ only) and the open ring configuration.
The rings have $N$ sources and $N$ sinks, labeled consecutively, with sources at odd and sinks at even indices. Edge $e$ connects nodes $e$ and $e+1$ (mod $2N$). For $N_g = N_c = N$, the mean injection pattern $\vec v$ has $v_n = +1$ on sources and $v_n = -1$ on sinks.

\paragraph{Alternating configuration.} 
Here, we have alternating capacities $k_a$ and $k_b$ on consecutive edges, subject to $N(k_a^\gamma + k_b^\gamma) = \kappa^\gamma$. Every node has one neighbor of each capacity, both with opposite sign of $v_n$. Then $\vec v$ is an eigenvector of the Laplacian $\matr L$ with eigenvalue $2(k_a+k_b)$. For $N_g = N_c$, we obtain
\begin{align}
\mu^2\vec v^\top\matr L^+\vec v = \frac{\mu^2 |\vec{v}|^2}{2(k_a+k_b)} = \frac{\mu^2N}{k_a+k_b}. 
\end{align}
For the fluctuating part, we only need the effective resistances between sources. Since sinks don't have fluctuating injections, the two edges adjacent to a sink are connected in series. Therefore, the $2N$ ring is equivalent to a ring of $N$ sources connected by edges with uniform effective capacities $k_{\mathrm{eff}}=k_a k_b/(k_a+k_b)$. Two sources at distance $d$ along this ring divide the ring into two arcs, with $d$ and $N-d$ edges each, respectively. Each arc is a chain of edges connected in series, with resistances $d/k_{\mathrm{eff}}$ and $(N-d)/k_{\mathrm{eff}}$. The two arcs are connected in parallel, such that~\cite{Klein1995}
$
    R_{ij} = \frac{1}{k_{\mathrm{eff}}} \frac{d(N-d)}{N}.
$
Going around the ring in one direction, each source is connected to every other source at distances $d = 1, \dots, N-1$. Summing over all sources counts every pair twice, therefore~\cite{Lukovits1999}
\begin{align}
    \sum_{i<j\in\mathcal{V}_g} R_{ij} = \frac{1}{2 k_{\mathrm{eff}}} \sum_{d=1}^{N-1} d(N-d) = \frac{N(N^2-1)}{12\, k_{\mathrm{eff}}}.
\end{align}
Equivalently, summing the definition of the effective resistance, Eq.~(35) of the Methods, over all pairs gives $\sum_{i<j} R_{ij} = N\,\mathrm{Tr}(\tilde{\matr{L}}_N^{+})/k_{\mathrm{eff}}$, where $\tilde{\matr{L}}$ is the Laplacian of the unit-capacity ring with $N$ nodes. Using the diagonal entries $(N^2-1)/(12N)$ of $\tilde{\matr{L}}_N^{+}$~\cite{Bendito2010} yields the same result. Combining the two terms, we obtain the dissipation
\begin{align}
    \bar{D}_{\mathrm{alt}}(k_a, k_b) = \frac{\mu^2 N}{k_a + k_b} + \frac{\beta^2 N (N^2 - 1)}{12} \, \frac{k_a + k_b}{k_a k_b}.
    \label{eq:D_alt}
\end{align}

\paragraph{Symmetric configuration.} This configuration is a special case of Eq.~\eqref{eq:D_alt} with $k_a=k_b=(2N)^{-1/\gamma}\kappa$, i.e.,
\begin{align}
    \bar{D}_{\mathrm{sym}} = \frac{(2N)^{1/\gamma}}{\kappa} \left[ \frac{N\mu^2}{2} + \frac{N(N^2-1)\beta^2}{6} \right].
    \label{eq:D_sym}
\end{align}

\paragraph{Sink-paired configuration ($N=2$).}
Here, the capacities of edges $1,\dots,4$ are $(k_a, k_a, k_b, k_b)$, subject to $2(k_a^\gamma + k_b^\gamma) = \kappa^\gamma$. The two edges adjacent to each sink are equal. Kirchhoff's current law determines the mean flows up to one loop flow. Kirchhoff's voltage law around the ring then implies that each source provides $\mu/2$ for each of its two neighboring sinks, independently of $k_a$ and $k_b$.  The two sources are connected by two parallel arcs with resistances $2/k_a$ and $2/k_b$, such that $R_{13} = 2/(k_a + k_b)$. The dissipation is then given by
\begin{align}
    \bar{D}_{\mathrm{pair}}(k_a, k_b) = \frac{\mu^2}{2} \left( \frac{1}{k_a} + \frac{1}{k_b} \right) + \frac{2\beta^2}{k_a + k_b}.
    \label{eq:D_pair}
\end{align}

\begin{figure}[t]
  \centering
  \includegraphics[width=\linewidth]{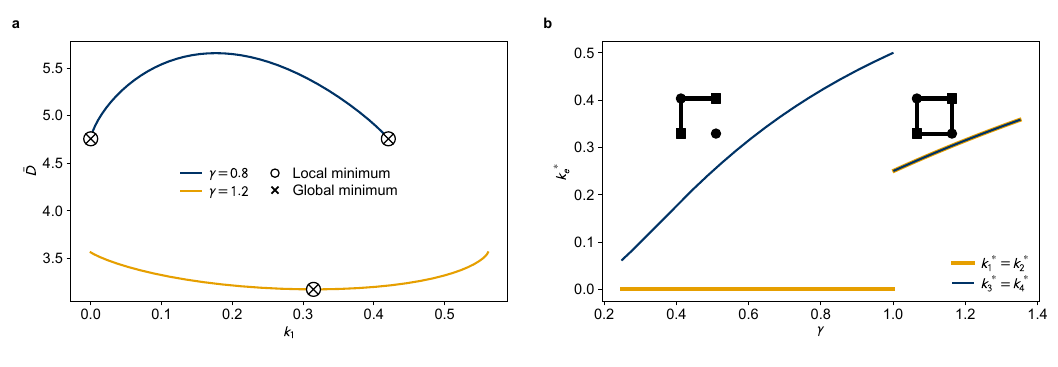}
  \caption{Symmetry breaking via bifurcation in an elementary ring network with $N=2$ and $\mu = 0$. \textbf{a} Expected dissipation $\bar D$ as a function of $k_1$ along the resource constraint for $\gamma = 0.8$ and $\gamma = 1.2$. \textbf{b} Optimal capacities as a function of $\gamma$.}
  \label{fig:ring_N2_mu0}
\end{figure}

For $\mu = 0$, the sinks don't have injections, and the two edges adjacent to each sink act in series. Eq.~\eqref{eq:D_pair} then reduces to
\begin{align}
\bar D^{\mathrm{pair}}(k_a, k_b) = \frac{2\beta^2}{k_a + k_b},
\end{align}
i.e., the fluctuation $X_1 = -X_3$, with variance $\beta^2$, flows through two parallel paths of conductance $k_a/2$ and $k_b/2$. Minimizing $\bar D$ thus amounts to maximizing $k_a + k_b$ under $2(k_a^\gamma + k_b^\gamma) = \kappa^\gamma$. For $\gamma > 1$ the maximum is symmetric. For $\gamma < 1$ it is on the boundary, where one pair of edges is inactive. At $\gamma = 1$, all feasible configurations are optimal. The symmetric configuration thus loses stability at $\gamma_s = 1$ (Supplementary Methods~6), which coincides with the strong transition, without an intermediate weakly broken phase (Supplementary Fig.~\ref{fig:ring_N2_mu0}).

\paragraph{Open ring.}
For the strongly symmetry-broken state, one edge is inactive. We set $k_{2N} = 0$ without loss of generality. Since the open ring is a tree, Kirchhoff's current law determines the edge flows independently of the edge capacities. The cumulative injection of the nodes $1, \dots, e$, flows through edge $e$. There are $n_g = \lceil e/2 \rceil$ sources and $n_c = \lfloor e/2 \rfloor$ sinks, so the flow through edge $e$ is given by
\begin{align*}
    F_e = \sum_{i=1}^{e} S_i = \mu\,(n_g - n_c) + \sum_{\substack{i \le e \\ i \in \mathcal{V}_g}} X_i .
\end{align*}
Using $\braket{X_i} = 0$, $\braket{X_i^2} = (N-1)\beta^2$ and $\braket{X_i X_j} = -\beta^2$ for $i \neq j$, we obtain
\begin{equation}
\left\langle F_e^2\right\rangle = \mu^2\,\delta_{e \bmod 2,\,1} + n_g(N-n_g)\,\beta^2,
\qquad n_g=\lceil e/2\rceil, \qquad e=1,\dots,2N-1,
\end{equation}
where $\delta_{e \bmod 2,\,1}$ is the Kronecker delta. Since the flows do not depend on the capacities, the optimal capacities are given by Eq.~(31) of the Methods,
\begin{equation}
k_e^* = \frac{\left\langle F_e^2\right\rangle^{1/(1+\gamma)}}{\left(\sum_{l=1}^{2N-1}\left\langle F_l^2\right\rangle^{\gamma/(1+\gamma)}\right)^{1/\gamma}}\,\kappa .
\end{equation}
Inserting them into $\bar D=\sum_e \langle F_e^2\rangle/k_e$, Eq.~(6) of the main text, yields
\begin{equation}
\bar D^{\mathrm{open}} = \frac{1}{\kappa}\left[\sum_{e=1}^{2N-1}\left\langle F_e^2\right\rangle^{\frac{\gamma}{1+\gamma}}\right]^{\frac{1+\gamma}{\gamma}} .\label{eq:D_open}
\end{equation}

\newpage
\subsection{Supplementary Methods 6: Group-theoretic stability analysis of the ring network}
\label{sec:ring-stability-thm2}
Here, we explicitly evaluate the stability criterion for the ring networks. We identify the symmetry group $\Gamma$, derive the decomposition of the space of capacity perturbations into irreducible representations, and obtain all of the stability thresholds.

We keep the labeling from Supplementary Methods~5, orienting edge $e$ from node $e$ to node $e+1$ (mod $2N$). It is convenient to group the edges into the $N$ pairs $P_j = \{2j-1,\, 2j\}$, for $j = 1,\ldots,N$, consisting of the two edges incident to the sink $2j$.

\paragraph{The symmetry group of the optimization problem.} 
The automorphism group of the $2N$-ring is the dihedral group $D_{2N}$. Preserving the alternating source/sink pattern excludes the edge-centered reflections, which map sources to sinks, and leaves
\begin{align}
    \Gamma \cong D_N, \qquad |\Gamma| = 2N ,
    \label{eq:ring-group}
\end{align}
generated by the rotation $r$ by two nodes, $i \mapsto i+2$, together with the $N$ node-centered reflections. The reflection about node $1$ acts as $i \mapsto 2-i$ on nodes and $e \mapsto 1-e$ on edges. It preserves the parity of the node indices, mapping sources onto sources, and exchanges odd and even edges.

The types of elements and their invariant nodes and edges are listed in Supplementary Table~\ref{tab:ring-classes}. The identity element leaves everything invariant. A rotation $r^k$ maps $e \mapsto e+2k$ and leaves an edge invariant only for $k \equiv 0$, while a node-centered reflection $\sigma$ leaves only the two nodes on its axis invariant. The symmetries act on the capacities and capacity perturbations through the permutation matrices $\matr \Pi_g$ on edge space (Supplementary Methods~2). The character of this permutation representation, $\chi(g) = \mathrm{Tr}\,\matr\Pi_g$, is the number of edges left invariant by $g$, i.e.,
\begin{align}
    \chi(g) = 2N\, \delta_{g,\mathrm{id}}.
    \label{eq:ring-character}
\end{align}
The rotations map every odd edge onto every other odd edge, and the reflections exchange odd and even edges, so the action of $\Gamma$ on the edge set $\EE$ is transitive.

\begin{table}[!b]
\caption{Element types of the symmetry group $\Gamma \cong D_N$ of the ring network. In the last two columns we list the number of nodes and edges left invariant. The latter is equal to the character $\chi$ of $\matr \Pi_g$.}
\label{tab:ring-classes}
\begin{ruledtabular}
\begin{tabular}{llcc}
class & geometric action & invariant nodes & invariant edges (character $\chi$) \\
\hline
$\mathrm{id}$              & identity                              & $2N$ & $2N$ \\
$r^k$, $k \neq 0$ & rotation by $2k$ nodes                & $0$  & $0$ \\
$\sigma$          & reflection about a node-centered axis & $2$  & $0$ \\
\end{tabular}
\end{ruledtabular}
\end{table}

\paragraph{Decomposition of the space of capacity perturbations.}
\label{sec:ring-irreps}
A perturbation of the optimal network is a vector $\vec \delta = (\delta_1,\ldots,\delta_{2N})$ that determines the direction in which the symmetric configuration $\vec k_0$ is perturbed, $k_e = k_0 + \eta \delta_e$, where $\eta$ is the amplitude of the perturbation. Since $\chi(g)$ is zero for every $g$ in Supplementary Table~\ref{tab:ring-classes} except the identity, the standard multiplicity formula~\cite{cotton1990chemical} yields
\begin{align}
    m_\rho = \frac{1}{|\Gamma|} \sum_{g \in \Gamma} \chi(g)\, \chi_\rho(g) = \frac{1}{2N}\, 2N \chi_\rho(\mathrm{id}) = d_\rho ,
    \label{eq:ring-mult}
\end{align}
where we used the fact that the character of the identity equals the dimension of the representation, $\chi_\rho(\mathrm{id}) = d_\rho$. Therefore, every irreducible representation occurs in edge space with multiplicity equal to its dimension. Using the standard Mulliken labels~\cite{cotton1990chemical}, the edge space $\mathbb{R}^{2N}$ is decomposed as~\cite{Serre1977}, for even $N$, 
\begin{align}
    \mathbb{R}^{2N} = A_1 \oplus A_2 \oplus B_1 \oplus B_2
        \oplus \bigoplus_{m=1}^{N/2-1} 2\, E_m ,
    \label{eq:ring-decomp-even}
\end{align}
while for odd $N$ the two $B$ blocks are absent,
\begin{align}
    \mathbb{R}^{2N} = A_1 \oplus A_2
        \oplus \bigoplus_{m=1}^{(N-1)/2} 2\, E_m.
    \label{eq:ring-decomp-odd}
\end{align}

We label the one-dimensional blocks by the characters $(\chi(r),\chi(\sigma))$ of the rotation and of a source-centered reflection, with $A_1 = (+,+)$, $A_2 = (+,-)$, $B_1 = (-,+)$ and $B_2 = (-,-)$. Since each one-dimensional irreducible representation occurs once, $\dim W_\rho=1$. Therefore these irreducible representations span an eigendirection of the constrained Hessian. On the two-dimensional block $E_m$, $\chi_{E_m}(r^k) = 2\cos(2\pi m k/N)$ and $\chi_{E_m}(\sigma) = 0$~\cite{Serre1977}. Since $m_{E_m} = 2$, Schur's lemma yields a $2 \times 2$ matrix rather than a scalar on each $E_m$.

It remains to identify which pattern of capacity perturbations spans each block. We collect the $2N$ perturbations $\delta_e$ in an $N \times 2$ array, with one row per pair $P_j$ and one column for each of its two edges,
\begin{align}
    \begin{pmatrix}
        \delta_1 & \delta_2 \\
        \vdots & \vdots \\
        \delta_{2j-1} & \delta_{2j} \\
        \vdots & \vdots \\
        \delta_{2N-1} & \delta_{2N}
    \end{pmatrix}
    \nonumber
\end{align}
Every symmetry permutes the rows, since $D_N$ acts on the $N$ sinks. Node-centered reflections also exchange the two columns, because the reflection about node $1$ maps $P_j \to P_{1-j}$ and exchanges the odd and the even edge within each pair. Thus, the column-symmetric and column-antisymmetric patterns are $N$-dimensional invariant subspaces,
\begin{align}
    \mathbb{R}^{2N} = \mathcal{U}_+ \oplus \mathcal{U}_- ,
    \qquad
    \mathcal{U}_+: \ \delta_{2j-1} = \delta_{2j} ,
    \qquad
    \mathcal{U}_-: \ \delta_{2j-1} = -\delta_{2j} ,
    \label{eq:ring-SA}
\end{align}
where $\mathcal{U}_+$ is the permutation representation of $D_N$ on the $N$ sinks and $\mathcal{U}_- \cong \mathcal{U}_+ \otimes \varepsilon$. The sign character is $\varepsilon = +1$ on rotations and $\varepsilon = -1$ on reflections. This is an orthogonal decomposition. The rotation $r$ acts within each subspace by cyclically permuting the rows. Its eigenvalues are $e^{2\pi i m /N}$, where $m$ labels the block. In each subspace, $m=0$ and, for even $N$, $m=N/2$,  give real one-dimensional blocks with $\chi(r) = +1$ and $-1$, while the conjugate pairs $\{m, N-m\}$ span the real planes $E_m$. 
A source-centered reflection permutes the rows and also swaps the two columns. Since the swap acts as $+1$ on $\mathcal{U}_+$ and $-1$ on $\mathcal{U}_-$, the two subspaces have opposite reflection characters for the same $m$. Thus, the correspondence between the blocks and perturbation patterns is as follows:
\begin{align}
    A_1 &: \ \vec \delta = \vec 1, \ \text{the uniform perturbation (excluded by the resource constraint)};
        \nonumber \\
    A_2 &: \ \delta^{\rm alt}_e = (-1)^{e+1}, \ \text{the } r\text{-invariant direction in } \mathcal{U}_- ;
        \nonumber \\
    B_2 &: \ \delta^{\rm sink}_{2j-1} = \delta^{\rm sink}_{2j} = (-1)^{j+1}, \ \text{the direction in } \mathcal{U}_+ \text{ on which } r \text{ acts as } -1 ;
        \nonumber \\
    B_1 &: \ \delta^{\rm src}_{2j-1} = -\delta^{\rm src}_{2j} = (-1)^{j+1}, \ \text{the direction in } \mathcal{U}_- \text{ on which } r \text{ acts as } -1 ;
        \nonumber \\
    E_m &: \ \text{the planes on which } r \text{ rotates by } 2\pi m/N, \ \text{one in } \mathcal{U}_+ \text{ and one in } \mathcal{U}_- .
    \label{eq:ring-blocks}
\end{align}
The mode $\vec \delta^{\rm alt}$ corresponds to the alternating configuration, while the sink-paired mode $\vec \delta^{\rm sink}$ and the source-paired mode $\vec \delta^{\rm src}$ satisfy $\delta_{e+2} = -\delta_e$ and group the edges adjacent to a common sink and to a common source, respectively. At $N=2$, $\vec\delta^{\rm sink}$ yields the sink-paired configuration of Supplementary Methods~5. The blocks $B_1$ and $B_2$ exist only for even $N$. Each $E_m$ has one copy in $\mathcal{U}_+$ and one in $\mathcal{U}_-$, so its multiplicity space is spanned by two perturbations: one in which the two edges of every pair are perturbed equally, and one in which they are perturbed by equal and opposite amounts.

\paragraph{Stability thresholds.}
\label{sec:ring-thresholds}

At unit capacities $\vec K = \matr I$, $\tilde G_{ef}$ is the potential difference across edge $e$ induced by a unit current across edge $f$ (Eq.~(14) of the main text). With all edges oriented consistently around the ring, this current flows partly through edge $f$, with a share of $1 - 1/(2N)$, and partly through the remaining $2N-1$ edges in series, with the current $-1/(2N)$. Hence
\begin{align}
    \tilde{\matr G} = \matr I - \frac{1}{2N}\, \matr J,
    \label{eq:Gtilde-ring}
\end{align}
where $\matr J$ is the $2N \times 2N$ all-ones matrix. Since $\tilde{\matr \Psi}_{ee}$ does not depend on $e$ (edge-transitivity), $\matr H = 2\tilde{\matr G}\circ\tilde{\matr\Psi}$ becomes
\begin{align}
    \matr H = 2\, \tilde \Psi_{ee}\, \matr I - \frac{1}{N}\, \tilde{\matr \Psi}.
    \label{eq:H-ring}
\end{align}
Therefore, $\matr H$ and $\tilde{\matr \Psi}$ share eigenvectors. Since $\matr H$ commutes with every $\matr \Pi_g$ (Supplementary Methods~2), so does $\tilde{\matr \Psi}$. Then, by Schur's lemma, both act as scalars on every block of multiplicity one, and as $2 \times 2$ matrices on the two copies of each $E_m$. Writing $\psi_\rho$ for the eigenvalue of $\tilde{\matr \Psi}$ on block $\rho$, we obtain $h_\rho = 2\tilde \Psi_{ee} - \psi_\rho/N$. Therefore, the thresholds for the ring reduce to
\begin{align}
    \gamma_s(\rho) = 1 - \frac{h_\rho}{\tilde \Psi_{ee}} = -1 + \frac{\psi_\rho}{N\, \tilde \Psi_{ee}} ,
    \qquad
    \psi_\rho = \frac{\vec \delta^{\rho\top} \tilde{\matr \Psi}\, \vec \delta^\rho}{|\vec \delta^\rho|^2} .
    \label{eq:gammas-master}
\end{align}
It remains to find $\tilde \Psi_{ee}$ and the eigenvalues $\psi_\rho$. 

We first obtain the diagonal element from $\sum_e k_e \Psi_{ee} = \bar D$ (Supplementary Methods~3), i.e., $\tilde \Psi_{ee} = \bar D(\vec 1)/(2N)$ at unit capacities. Setting $k_a = k_b = 1$ in Eq.~\eqref{eq:D_alt} yields $\bar D(\vec 1) = \mu^2 N/2 + \beta^2 N(N^2-1)/6$, and
\begin{align}
    \tilde \Psi_{ee} = \frac{\mu^2}{4} + \frac{(N^2-1)\beta^2}{12}
    = \frac{3\mu^2 + (N^2-1)\beta^2}{12} .
    \label{eq:Psi-ee-ring}
\end{align}

The nodal injections decompose into a deterministic mean load and a fluctuating part, $S_n = \mu v_n + X_n$, where $v_n$ is equal to $+1$ on sources and $-1$ on sinks, while $X_n$ vanish on the sinks and sum to zero over the sources. Since $\langle X_n \rangle = 0$, the cross terms are zero in $\matr \Sigma = \langle \vec S \vec S^\top \rangle$, and the matrix $\tilde{\matr \Psi}$ also decomposes into a deterministic and fluctuating part $\tilde{\matr \Psi} = \tilde{\matr \Psi}^{(\mu)} + \tilde{\matr \Psi}^{(\beta)}$.

Each of the two parts acts only within one of the subspaces of Eq.~\eqref{eq:ring-SA}. For the deterministic part, $\vec v$ is an eigenvector of $\tilde{\matr L}$ with eigenvalue $4$ (Supplementary Methods~5). Then $\tilde{\matr L}^+ \vec v = \vec v/4$ yields the potential differences $\Delta\theta^{(\mu)}_e = \frac{1}{4} \mu (v_e - v_{e+1}) = \frac12 \mu\, \delta^{\rm alt}_e$, such that
\begin{align}
    \tilde{\matr \Psi}^{(\mu)}
    = \frac{\mu^2}{4}\, \vec \delta^{\rm alt} \vec \delta^{{\rm alt}\top} ,
    \qquad
    \Delta \vec \theta^{(\mu)} \in \mathcal{U}_- .
    \label{eq:Psi-mu}
\end{align}
For the fluctuating part, recall that the injections vanish on the sinks. Then, Kirchhoff's current law at sink $2j$ implies equal potential differences on the two edges that meet at that sink,
\begin{align}
    \Delta\theta_{2j-1} = \Delta\theta_{2j} \equiv \varphi_j ,
    \qquad
    \Delta \vec \theta^{(\beta)} \in \mathcal{U}_+ .
    \label{eq:pair-flow}
\end{align}
Since $\tilde{\matr \Psi}^{(\mu)}$ vanishes on $\mathcal{U}_+$ and $\tilde{\matr \Psi}^{(\beta)}$ on $\mathcal{U}_-$, the matrix $\tilde{\matr \Psi}$ is block-diagonal with respect to $\mathcal{U}_+ \oplus \mathcal{U}_-$. Each of the eigenvalues $\psi_\rho$ depends on either $\mu$ or $\beta$, but not on both at the same time.

Eq.~\eqref{eq:Psi-mu} acts on $\mathcal{U}_-$. It is rank one, so
\begin{align}
    \psi_{\rm alt} = \frac{\mu^2}{4}\, |\vec \delta^{\rm alt}|^2 = \frac{N\mu^2}{2} ,
    \qquad
    \psi_\rho = 0 \quad \text{on } \mathcal{U}_- \ominus \operatorname{span}(\vec \delta^{\rm alt}) .
    \label{eq:R-antisym}
\end{align}
In particular, $\tilde{\matr \Psi}\, \vec \delta^{\rm src} = 0$, so $\psi_{\rm src} = 0$.

We now turn to $\mathcal{U}_+$. Besides the uniform block $A_1$ and, for even $N$, the sink-paired block $B_2$, it contains one copy of each two-dimensional block $E_m$, whose second copy lies in $\mathcal{U}_-$.
On $\mathcal{U}_+$, we denote with $\varphi_j$ the common potential difference on the two edges of the pair $P_j$. Kirchhoff's current law at the source $2j-1$ then reads $\varphi_j - \varphi_{j-1} = X_j$. Each pair acts as a single effective edge, and the ring reduces to a cycle of $N$ sources with balanced injections $\vec X$.

Every block in $\mathcal{U}_+$ except $A_1$ has perturbations summing to zero over the pairs, so they can be written as $\delta_{2j-1} = \delta_{2j} = g_j - g_{j+1}$ with $\vec g \perp \vec 1$ unique. Abel summation yields the scalar product,
\begin{align}
    \sum_e \delta_e \Delta\theta_e = 2 \sum_j (g_j - g_{j+1}) \varphi_j
    = 2 \sum_j g_j \bigl( \varphi_j - \varphi_{j-1} \bigr)
    = 2\, \vec g \cdot \vec X ,
    \label{eq:drops-scalar}
\end{align}
whose variance is $\langle ( \vec g \cdot \vec X )^2 \rangle = N\beta^2 |\vec g|^2$ because $\vec g \perp \vec 1$. With $|\vec \delta|^2 = 2 \sum_j (g_j - g_{j+1})^2$, Eq.~\eqref{eq:gammas-master} yields
\begin{align}
    \psi = 2N\beta^2\, \frac{|\vec g|^2}{\vec g^\top \matr L_N \vec g} ,
    \label{eq:R-sym-general}
\end{align}
where $\matr L_N$ is the Laplacian of the source cycle, $\vec g^\top \matr L_N \vec g = \sum_j (g_j - g_{j+1})^2$. Since $\matr L_N$ commutes with the rotation $r$, its eigenvectors are the blocks of Eq.~\eqref{eq:ring-blocks} contained in $\mathcal{U}_+$. For $m = 1, \ldots, \lfloor N/2 \rfloor$, the copy of $E_m$ in $\mathcal{U}_+$ consists of the perturbations with
\begin{align}
    g_j = a \cos\left(\frac{2\pi m j}{N}\right) + b \sin\left(\frac{2\pi m j}{N}\right), \qquad a, b \in \mathbb{R},
\end{align}
where, for even $N$ and $m = N/2$, the sine term vanishes and $\vec g$ spans the block $B_2$. Using $\cos(x-y) + \cos(x+y) = 2\cos x \cos y$ and $\sin(x-y) + \sin(x+y) = 2\sin x \cos y$ with $x = 2\pi m j/N$ and $y = 2\pi m/N$, we obtain
\begin{align}
    (\matr L_N \vec g)_j = 2g_j - g_{j-1} - g_{j+1}
    = \left[ 2 - 2\cos\left(\frac{2\pi m}{N}\right) \right] g_j
    = \lambda_m\, g_j, \qquad
    \lambda_m = 4 \sin^2\left(\frac{\pi m}{N}\right),
    \label{eq:cycle-eigenvalues}
\end{align}
for all $a$ and $b$. Thus, $\matr L_N$ is a multiple of the identity on each two-dimensional block $E_m$ and the Rayleigh quotient is constant there. Hence
\begin{align}
    \psi_m = \frac{2N\beta^2}{\lambda_m} = \frac{N\beta^2}{2\sin^2(\pi m/N)} ,
    \qquad m = 1,\ldots,\lfloor N/2 \rfloor .
    \label{eq:R-sym}
\end{align}
The uniform block $A_1$, corresponding to $m=0$, is excluded by the resource constraint. For even $N$,
$m = N/2$ is the block $B_2$ spanned by $\vec \delta^{\rm sink}$, giving
$\psi_{\rm sink} = N\beta^2/2$. Together with Eq.~\eqref{eq:R-antisym}, the two copies
of each $E_m$ carry $\psi_m$ and $0$, so the multiplicity block is already diagonal.

Inserting Eqs.~\eqref{eq:R-antisym} and~\eqref{eq:R-sym} together with Eq.~\eqref{eq:Psi-ee-ring} into the master formula Eq.~\eqref{eq:gammas-master}, we obtain stability boundaries
\begin{align}
    \gamma_s({\rm alt}) &= \frac{3\mu^2 - (N^2-1)\beta^2}{3\mu^2 + (N^2-1)\beta^2} ,
    \label{eq:gs-alt}\\
    \gamma_s(m) &= \frac{6\beta^2 - \sin^2(\pi m/N) \bigl[3\mu^2 + (N^2-1)\beta^2\bigr]}{\sin^2(\pi m/N) \bigl[3\mu^2 + (N^2-1)\beta^2\bigr]} ,
    \qquad m = 1,\ldots, \lfloor N/2\rfloor ,
    \label{eq:gs-m}\\
        \gamma_s &= -1 \qquad \text{on every block of } \mathcal{U}_- \text{ except } A_2 .
    \label{eq:gs-B}
\end{align}
The block $B_1$, spanned by $\vec \delta^{\rm src}$, is one of these, so $\gamma_s({\rm src}) = -1$. Since we consider only $\gamma > 0$, $A_2$ is the only block of $\mathcal{U}_-$ that can destabilize the symmetric configuration.

For even $N$, Eq.~\eqref{eq:gs-m} with $m = N/2$ describes the sink-paired mode,
\begin{align}
    \gamma_s({\rm sink}) = \frac{(7-N^2)\beta^2 - 3\mu^2}{3\mu^2 + (N^2-1)\beta^2} .
    \label{eq:gs-A}
\end{align}

Among the blocks with $m = 1, \ldots, \lfloor N/2 \rfloor$, $\lambda_m$ is smallest for $m=1$, so the largest threshold on $\mathcal{U}_+$ belongs to $E_1$ rather than to the sink-paired mode $B_2$. It is positive if and only if
\begin{align}
    \beta^2 \bigl[ 6 - (N^2-1)\, \sin^2(\pi/N) \bigr] > 3 \mu^2 \sin^2(\pi/N) ,
    \label{eq:m1-condition}
\end{align}
and $(N^2-1)\sin^2(\pi/N)$ equals $3$ at $N=2$, exactly $6$ at $N=3$, and grows
monotonically to $\pi^2$ for larger $N$.

The two relevant boundaries can be inverted for the fluctuation strength,
\begin{align}
    \beta &= \mu \sqrt{\frac{3(1-\gamma)}{(N^2-1)(1+\gamma)}} ,
    \label{eq:beta-alt}\\
    \beta &= \mu \sqrt{\frac{3 \sin^2(\pi/N) (1+\gamma)}{6 - \sin^2(\pi/N) (N^2-1)(1+\gamma)}} ,
    \label{eq:beta-m}
\end{align}
the latter is real only where its denominator is positive, i.e.\ under the same
condition as Eq.~\eqref{eq:m1-condition}.

The sign of the bracket in Eq.~\eqref{eq:m1-condition} determines whether perturbations in $\mathcal{U}_+$ can destabilize the symmetric configuration for $\gamma > 0$:
\begin{itemize}
    \item For $N=2$, $m=1$ coincides with the sink-paired mode, and Eq.~\eqref{eq:m1-condition} reduces to $\beta > \mu$. Equation~\eqref{eq:beta-m} then gives $\beta = \mu \sqrt{(1+\gamma)/(1-\gamma)}$.
    \item For $N=3$, the bracket vanishes and $\gamma_s(1) = -3\mu^2/(3\mu^2 + 8\beta^2) \le 0$, with equality only at $\mu = 0$.
    \item For $N \ge 4$, the bracket is negative and $\gamma_s(m) < 0$ for every $m$ and all $\beta, \mu$. For odd $N$, the sink- and source-paired modes do not exist in the first place.
\end{itemize}
Thus, for $N \ge 3$ the alternating configuration is the only route to weak symmetry breaking. At $N=2$ and $\mu = 0$ we obtain $\gamma_s({\rm alt}) = -1$ and $\gamma_s(1) = +1$, so the symmetric configuration is unstable for every $\gamma < 1$. At $N=3$ and $\mu = 0$, the largest threshold vanishes, so the symmetric configuration remains a local minimum for every $\gamma > 0$ even where it is no longer the global optimum.

\newpage

\subsection{Supplementary Methods 7: Scaling of the transition point and capacity jump with ring size}

We analyze how the strong symmetry-breaking transition of the ring scales with system size. We set $\mu = \kappa = 1$. We define the transition point $\gamma_c(N)$ as the leftmost point of the region with $\min_e k_e^* > 0$ in Fig.~3\textbf{a}--\textbf{c} of the main text. For $\gamma < \gamma_c$, the optimal network is an open ring for every fluctuation strength. At the leftmost point, the open ring and the symmetric configuration exchange global optimality, such that $\gamma_c$ is the minimum over $\beta$ of the boundary $\bar{D}_{\mathrm{sym}}(\gamma) = \bar{D}_{\mathrm{open}}(\gamma)$, with the dissipations given in Eqs.~\eqref{eq:D_sym} and \eqref{eq:D_open} of Supplementary Methods~5.

Fitting $\gamma_c(N)$ to the power law $\gamma_c = c\, N^{-\alpha}$, for $N = 2, \dots, 9$, we obtain
\begin{align}
     c = 0.92 \pm 0.01, \qquad \alpha = 1.06 \pm 0.01,
    \label{eq:gamma_c_fit}
\end{align}
shown in Fig.~3\textbf{f} of the main text.

At the transition, the minimum edge capacity jumps from zero to the capacity of the symmetric configuration,
\begin{align}
    k_c = (2N)^{-1/\gamma_c}.\label{eq:k_c}
\end{align}
Inserting the power law into \eqref{eq:k_c} yields
\begin{align}
    k_c \simeq (2N)^{-N^{\alpha}/c},
    \label{eq:k_c-2}
\end{align}
i.e., the jump decays super-exponentially with $N$. Figure~3\textbf{g} of the main text compares $k_c$ evaluated at the computed $\gamma_c(N)$ with Eq.~\eqref{eq:k_c-2}. Although the transition remains discontinuous for finite $N$, the magnitude of the discontinuity, quantified by $k_c$, decreases rapidly with increasing system size.

\newpage 
\subsection{Supplementary Methods 8: Group-theoretic stability analysis of the cube}
\label{app:cube_symmetry}

Here we identify the symmetry group $\Gamma$ of the cube network, decompose the space of capacity perturbations into blocks, obtain the stability thresholds of all blocks, and find the residual symmetries of the symmetry-broken configurations.

\paragraph{The symmetry group of the optimization problem.}
\label{app:cube_group}
The graph automorphisms of the cube coincide with its geometric symmetries, $\mathrm{Aut}(Q_3) \cong O_h$, $|O_h| = 48$. The symmetries of the optimization problem are the elements of the point group $O_h$ that map sources to sources and sinks to sinks. The four sources are positioned in alternating corners of the cube and span a regular tetrahedron, whose symmetry is $T_d$, as shown in Supplementary Fig.~\ref{fig:cube_symmetry}\textbf{a}. The sinks span a complementary tetrahedron, and the remaining 24 operations of $O_h$, among them the inversion through the center, exchange the two tetrahedra and are therefore excluded. Therefore, $\Gamma$ is given by the full symmetry group of the tetrahedron formed by the sources,
\begin{equation}
  \Gamma \;\cong\; T_d \;\cong\; S_4 , \qquad |\Gamma| = 24 ,
  \label{eq:cube_gamma_app}
\end{equation}
where the isomorphism to $S_4$ is the action on the four sources. We can fix the sources as a reference, because each sink is the unique node adjacent to a given triple of sources. An element that leaves every source invariant thus leaves every node invariant.

The five conjugacy classes of $\Gamma$ are listed in Supplementary Table~\ref{tab:cube_classes}\footnote{The identity class is typically denoted with $E$ in the Schoenflies notation~\cite{cotton1990chemical}. However, we will denote it with $\mathrm{id}$ to avoid confusion with the two-dimensional irreducible representation $E$.}, three of which are illustrated in Supplementary Fig.~\ref{fig:cube_symmetry}: a threefold rotation around a space diagonal (Supplementary Fig.~\ref{fig:cube_symmetry}\textbf{b}), an improper fourfold rotation around a face axis (Supplementary Fig.~\ref{fig:cube_symmetry}\textbf{c}), and a reflection in one of the six mirror planes (Supplementary Fig.~\ref{fig:cube_symmetry}\textbf{d}).

Since every $g \in \Gamma$ maps sources to sources, it preserves the source to sink orientation of every edge. 
The permutation matrices $\matr{\Pi}_g$ on edge space (Supplementary Methods~2) define a representation of $\Gamma$ with character $\chi(g) = \mathrm{Tr}\,\matr{\Pi}_g$, equal to the number of edges left invariant by $g$, cf. the last column of Supplementary Table~2.

Finally, $\Gamma$ acts transitively on the twelve edges: it permutes the four sources transitively, and the threefold rotation about the space diagonal through a source cycles the three edges at that source. By the orbit-stabilizer theorem, the stabilizer of an edge thus has order $|\Gamma|/12 = 2$. The stabilizer consists of the identity together with the reflection in the $\sigma_d$ plane that contains that edge. 

\begin{table}
\caption{Conjugacy classes of the symmetry group $\Gamma \cong T_d \cong S_4$ of the cube network. Each class is determined by the axis or plane it acts about. In the last two columns, we list the number of nodes and edges it leaves invariant. The edge counts are the character $\chi$ of the permutation representation on edge space.}
\label{tab:cube_classes}
\begin{ruledtabular}
\begin{tabular}{lllcc}
class & axis or plane & geometric action & invariant nodes & invariant edges \\
\hline
$\mathrm{id}$         & ---                              & identity                                & 8 & 12 \\
$8C_3$      & space diagonal                   & cycles the three edges at each source   & 2 & 0  \\
$3C_2$      & face axis                        & rotation by $180^\circ$, maps every edge to a parallel one                 & 0 & 0  \\
$6S_4$      & face axis                        & improper rotation, interchanges two normal directions    & 0 & 0  \\
$6\sigma_d$ & plane through two opp. edges & reflection fixing the two edges in the plane   & 4 & 2  \\
\end{tabular}
\end{ruledtabular}
\end{table}

\begin{figure*}
\centering
\begin{tikzpicture}[cubeview]
 
\begin{scope}
  \cubecoords
  \cubeedges
  \drawset{tet}{000/011,000/101,000/110,011/101,011/110,101/110}
  \cubenodes
  \panlab{a}
\end{scope}
 
\begin{scope}[xshift=2.9cm]
  \cubecoords
  \cubeedges
  \drawset{hi}{\srcAedges}
  \draw[ax] (-0.24,-0.24,-0.24) -- (1.24,1.24,1.24);
  \cubenodes
  \panlab{b}
\end{scope}
 
\begin{scope}[xshift=5.8cm]
  \cubecoords
  \cubeedges
  \draw[ax] (0.5,0.5,-0.32) -- (0.5,0.5,1.32);
  \fill[black] (0.5,0.5,0) circle (1.1pt) (0.5,0.5,1) circle (1.1pt);
  \cubenodes
  \panlab{c}
\end{scope}
 
\begin{scope}[xshift=8.7cm]
  \cubecoords
  \fill[mir] (-0.12,-0.12,-0.12) -- (-0.12,-0.12,1.12)
          -- (1.12,1.12,1.12) -- (1.12,1.12,-0.12) -- cycle;
  \cubeedges
  \drawset{hi}{000/001,110/111}
  \cubenodes
  \panlab{d}
\end{scope}
 
\begin{scope}[xshift=11.6cm]
  \cubecoords
  \drawset{ed}{\xedges}
  \drawset{ed}{\yedges}
  \drawset{hi}{\zedges}
  \cubenodes
  \panlab{e}
\end{scope}
 
\begin{scope}[xshift=14.5cm]
  \cubecoords
  \drawset{ed}{\srcBedges}
  \drawset{ed}{\srcCedges}
  \drawset{ed}{\srcDedges}
  \drawset{hi}{\srcAedges}
  \cubenodes
  \panlab{f}
\end{scope}
 
\end{tikzpicture}
\caption{Symmetry of the cube network. Squares mark sources, circles sinks.
\textbf{a} Every edge joins a source to a sink. The four sources span a regular tetrahedron, and $\Gamma$ is its full symmetry group $T_d\cong S_4$.
\textbf{b} A threefold rotation around a space diagonal cyclically permutes the three edges at each source.
\textbf{c} An improper fourfold rotation around a face axis, whose square is a twofold rotation. 
\textbf{d} One of the six mirror planes. The two edges lying in the plane are fixed. 
\textbf{e, f} Two families of capacity perturbations: thick edges have $\delta_e>0$, thin edges $\delta_e<0$. In \textbf{e} the perturbation depends only on the spatial direction of the edge ($E$), while in \textbf{f} it depends only on the source it touches ($T_2$).}
\label{fig:cube_symmetry}
\end{figure*}

\paragraph{Decomposition of the space of capacity perturbations.}
\label{app:cube_decomposition}
A perturbation of the symmetric configuration is a vector $\vec{\delta} \in \mathbb{R}^{12}$ that assigns a capacity change $\delta_e$ to each edge. Every $g \in \Gamma$ acts on $\vec{\delta}$ by permuting its entries, $\vec{\delta} \mapsto \matr{\Pi}_g \vec{\delta}$, which decomposes $\mathbb{R}^{12}$ into blocks $W_\rho$, one for each irreducible representation $\rho$ of $\Gamma$ (Supplementary Methods~2). Since the constrained Hessian commutes with this action, it is block diagonal, and the stability thresholds can be obtained block by block. As a first step, we determine the blocks.

From Supplementary Table~\ref{tab:cube_classes}, the character of the permutation representation is
\begin{equation}
  \chi = (12, 0, 0, 0, 2) \quad \text{on} \quad (\mathrm{id}, 8C_3, 3C_2, 6S_4, 6\sigma_d).
\end{equation}

We use the standard Mulliken labels for the irreducible representations of $T_d$~\cite{cotton1990chemical}, whereby $A$ denotes a one-dimensional irreducible representation, $E$ a two-dimensional one and $T$ a three-dimensional one. The subscript distinguishes irreducible representations of equal dimension by their sign under the improper operations $S_4$ and $\sigma_d$. Their characters can be found in Ref.~\cite{cotton1990chemical}. We write $\chi(c)$ and $\chi_\rho(c)$ for the values of $\chi$ and $\chi_\rho$ on class $c$,
\begin{equation}
  \begin{array}{l|rrrrr}
     \chi_\rho(c) & \mathrm{id} & 8C_3 & 3C_2 & 6S_4 & 6\sigma_d \\ \hline
    A_1 & 1 & 1 & 1 & 1 & 1 \\
    A_2 & 1 & 1 & 1 & -1 & -1 \\
    E   & 2 & -1 & 2 & 0 & 0 \\
    T_1 & 3 & 0 & -1 & 1 & -1 \\
    T_2 & 3 & 0 & -1 & -1 & 1
  \end{array}
  \label{eq:td_character_table}
\end{equation}
These characters are orthonormal when summed over the group. The multiplicity $m_\rho$ of the irreducible representation $\rho$ in $\mathbb{R}^{12}$ is~\cite{cotton1990chemical}
\begin{equation}
  m_\rho \;=\; \frac{1}{|\Gamma|}\sum_{g\in\Gamma}\chi(g)\,\chi_\rho(g)
  \;=\; \frac{1}{24}\sum_{\text{classes}} n_c\,\chi (c) \,\chi_{\rho} (c) ,
  \label{eq:multiplicity}
\end{equation}
where $n_c$ is the number of elements in class $c$. Taking $\rho = T_2$ as an example and using $\chi_{T_2}(\mathrm{id}) = d_{T_2} = 3$, we find
\begin{equation}
  m_{T_2} = \frac{1}{24}\Big[\underbrace{1\cdot 12\cdot 3}_{\mathrm{id}} + \underbrace{6\cdot 2\cdot 1}_{6\sigma_d}\Big] = 2.
\end{equation}
For the remaining blocks we obtain accordingly $m_{A_1}=1$, $m_{A_2}=0$, $m_{E}=1$ and $m_{T_1}=1$. Hence, the decomposition of edge space into irreducible representations is
\begin{equation}
  \mathbb{R}^{12} \;=\; A_1 \oplus E \oplus T_1 \oplus 2\,T_2 .
  \label{eq:cube_decomposition_app}
\end{equation}
The dimensions are consistent, since $1 + 2 + 3 + 2\times 3 = 12$. 

To identify the perturbations in each block, note that every edge has exactly one source endpoint and points along one of the three axis directions $x$, $y$, $z$, and every source has exactly one edge along each direction. We therefore collect the twelve perturbations in a $4 \times 3$ matrix, with one row per source and one column per direction,
\begin{equation*}
  \begin{pmatrix}
    \delta_{1x} & \delta_{1y} & \delta_{1z} \\
    \delta_{2x} & \delta_{2y} & \delta_{2z} \\
    \delta_{3x} & \delta_{3y} & \delta_{3z} \\
    \delta_{4x} & \delta_{4y} & \delta_{4z}
  \end{pmatrix}.
\end{equation*}
This arrangement suggests three types of perturbations: those that depend only on the source (row), those that depend only on the direction (column), and those whose rows and columns all sum to zero. A similar approach is standard in the statistical analysis of tables~\cite{Diaconis1988}. A source-only or direction-only perturbation generally contains a uniform contribution. We remove it by requiring these perturbations to sum to zero, such that the first two types do not overlap with the uniform perturbation. The third type sums to zero automatically, since its rows already sum to zero. Together with the uniform perturbation, the four types are orthogonal to each other, and their dimensions add up to $1 + 3 + 2 + 6 = 12$, thus accounting for all possible perturbations. Since every symmetry only permutes rows and columns, a perturbation of a given type remains of that type. We can therefore identify the irreducible representations within each type separately.

Comparing with Eq.~\eqref{eq:cube_decomposition_app}, the uniform perturbation transforms as $A_1$, and the direction-only perturbations transform as $E$, the only two-dimensional irreducible representation. For the source-only perturbations, the character is the number of invariant sources minus one. Indeed, the number of invariant sources is the character of all perturbations that depend only on the source, including those with nonzero sum, and the uniform perturbation among them contributes one. By comparison with Eq.~\eqref{eq:td_character_table}, the source-only perturbations thus transform as $T_2$, and the perturbations of the third type as $T_1 \oplus T_2$. To summarize:
\begin{align}
  A_1 &: \quad \delta_e \text{ constant, the uniform perturbation (excluded by the resource constraint);} \nonumber\\
  E &: \quad \delta_e \text{ depends only on the direction of the edge, with the three values summing to zero;} \nonumber\\
  T_2 &: \quad \delta_e \text{ depends only on the source, with the four values summing to zero;} \nonumber\\
  T_1\oplus T_2 &: \quad \text{all row and column sums of the matrix are zero.}\label{eq:cube_patterns}
\end{align}
The block $E$ distinguishes the three cube axes, cf.\ Supplementary Fig.~\ref{fig:cube_symmetry}\textbf{e}, and the first copy of $T_2$ distinguishes the four sources, cf.\ Supplementary Fig.~\ref{fig:cube_symmetry}\textbf{f}. The second copy of $T_2$ depends on both the source and the direction. Since $T_2$ occurs twice, the cube has four thresholds in total: one threshold each for $E$ and $T_1$, and two thresholds from $W_{T_2}$. Since the Hessian couples the two copies, its eigenvectors in $W_{T_2}$ are linear combinations of the two copies, which we denote by $T_2^\pm$ below.

\paragraph{Stability thresholds.}
\label{app:cube_thresholds}
We first evaluate $\tilde{\Psi}_{ee}$. Since $\Gamma$ acts transitively on the edges, $\tilde{\Psi}_{ee}$ is the same for all twelve edges, and $\sum_e k_e \Psi_{ee} = \bar D$ (Supplementary Methods~3) yields
\begin{equation}
  \tilde{\Psi}_{ee} = \frac{\bar D(\vec{1})}{12}.\label{eq:psi_from_dbar}
\end{equation}
By Eq.~(38) of the Methods, $\bar D$ has separate contributions for the mean injections and the fluctuations. Like for the alternating ring (Supplementary Methods~5), every neighbor of a node $n$ has the opposite sign of $v_n$. Since every node has three neighbors, $\vec{v}$ is an eigenvector of $\tilde{\matr{L}}$ with eigenvalue 6, and we have
\begin{align}
  \mu^2 \vec{v}^\top \tilde{\matr{L}}^+ \vec{v} = \frac{\mu^2 |\vec{v}|^2}{6} = \frac{4}{3}\mu^2 .\label{eq:cube_mean_part}
\end{align}
For the fluctuating part, we notice that there are no adjacent sources, and any two sources are in the opposite corners of a shared face. All $\binom{4}{2}=6$ source pairs are thus connected by a face diagonal. Evaluating Eq.~(31) of Ref.~\cite{pippenger2010hypercube} for the $n=3$-dimensional hypercube and distance two yields the effective resistance $R_2 = 3/4$ across a face diagonal of the unit-capacity cube. Then,
\begin{align}
  \beta^2 \sum_{i<j\in\mathcal{V}_g} R_{ij} = 6\beta^2 R_2 = \frac{9}{2}\beta^2 .
  \label{eq:cube_fluct_part}
\end{align}
Combining Eqs.~\eqref{eq:cube_mean_part} and \eqref{eq:cube_fluct_part}, we obtain
\begin{align}
  \bar D(\vec{1}) = \frac{4}{3}\mu^2 + \frac{9}{2}\beta^2
  = \frac{8\mu^2 + 27\beta^2}{6} ,\label{eq:cube_D1}
\end{align}
and Eq.~\eqref{eq:psi_from_dbar} then yields
\begin{align}
  \tilde\Psi_{ee} = \frac{\mu^2}{9} + \frac{3\beta^2}{8}
  = \frac{8\mu^2+27\beta^2}{72} .
  \label{eq:cube_psi_ee}
\end{align}

The thresholds are obtained from the eigenvalues of $\matr{H} = 2\tilde{\matr{G}} \circ \tilde{\matr{\Psi}}$ on each block, where $\tilde{\matr{\Psi}} = \langle \Delta\vec{\theta}\,\Delta\vec{\theta}^\top \rangle$ contains the second moments of the edge potential differences at unit capacities. As for the ring (Supplementary Methods~6), the injections $S_n = \mu v_n + X_n$ divide $\Delta\vec{\theta}$ into a deterministic and a fluctuating part. Due to $\langle X_n \rangle = 0$, the cross terms vanish and $\tilde{\matr{\Psi}} = \tilde{\matr{\Psi}}^{(\mu)} + \tilde{\matr{\Psi}}^{(\beta)}$. Since $\tilde{\matr{L}}^+\vec{v} = \vec{v}/6$, the deterministic potential difference across every edge, oriented from its source, is $\Delta\theta^{(\mu)}_e = \mu/3$. Therefore,
\begin{align}
  \tilde{\matr{\Psi}}^{(\mu)} &= \frac{\mu^2}{9}\matr{J}, \label{eq:cube_Psimu}\\
  \matr{H}^{(\mu)} &= 2\tilde{\matr{G}} \circ \tilde{\matr{\Psi}}^{(\mu)} = \frac{2\mu^2}{9}\tilde{\matr{G}}. \label{eq:cube_Hmu}
\end{align}

\begin{figure*}
  \centering
  \includegraphics[width=\textwidth]{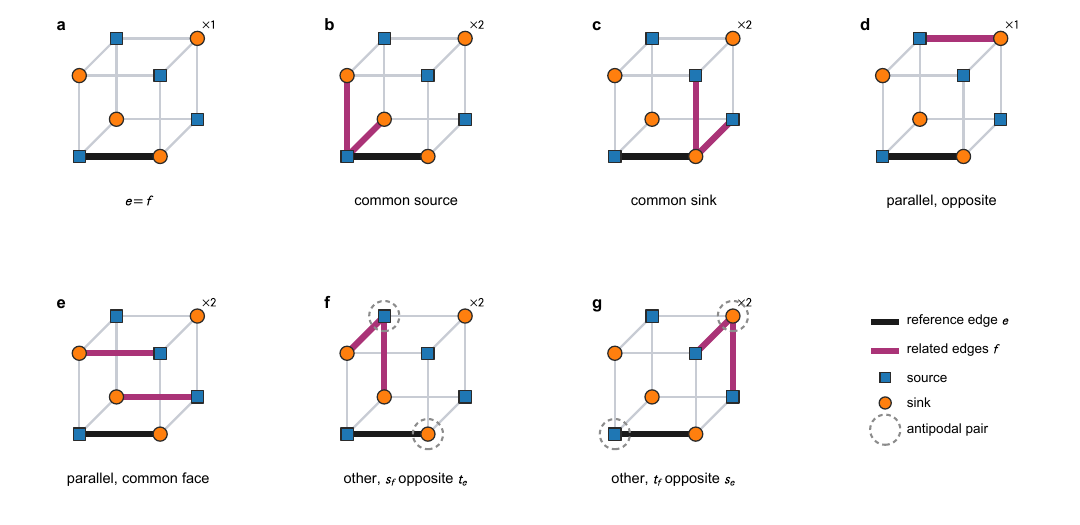}
  \caption{\label{fig:cube_orbits}
    Classes of edge pairs on the cube. Squares denote sources, circles sinks. In each panel, the reference edge $e$ is black and the edges $f$ related to it are colored in magenta. The label $\times n$ gives the number of such $f$ for fixed $e$. \textbf{a}--\textbf{e} The five relations fixed by transposition. \textbf{f}, \textbf{g} The residual relation, which splits into the two orbits that transposition exchanges, distinguished by which endpoint of $e$ is antipodal to which endpoint of $f$ (dashed circles).}
\end{figure*}

To evaluate $\tilde{\matr G}$ and $\tilde{\matr\Psi}^{(\beta)}$, we use their symmetry. Since every $g$ preserves the edge orientations (paragraph a), relabeling the nodes permutes the edges without sign changes, $\matr R_g \matr E = \matr E \matr \Pi_g$. Together with $\matr R_g \tilde{\matr L}^+ \matr R_g^\top = \tilde{\matr L}^+$, Eq.~\eqref{eq:L_relabel}, and $\matr R_g^\top \matr\Sigma \matr R_g = \matr\Sigma$, cf.\ Eq.~\eqref{eq:D_invariance}, this shows that $\tilde{\matr{G}}$ and $\tilde{\matr{\Psi}}$ are invariant under relabeling, $\tilde G_{ef} = \tilde G_{\pi_g(e)\pi_g(f)}$ and $\tilde\Psi_{ef} = \tilde\Psi_{\pi_g(e)\pi_g(f)}$. Then $\matr{H}$ is also invariant under relabeling. These matrices thus take the same value on any two pairs of edges that are related by a symmetry. We call a set of mutually related pairs an orbit of $\Gamma$ acting on ordered pairs of edges. Since $\chi(g)^2$ counts the ordered pairs of edges left invariant by $g$, the number of such orbits is $|\Gamma|^{-1}\sum_g \chi(g)^2 = \sum_\rho m_\rho^2 = 7$~\cite{Serre1977}, with the multiplicities from Eq.~\eqref{eq:cube_decomposition_app}. 
Six of the orbits are mapped onto themselves by the transposition $(e,f)\mapsto(f,e)$. The seventh consists of the pairs in which $e$ and $f$ are neither parallel nor share an endpoint, and consists of two orbits that the transposition exchanges (Supplementary Fig.~\ref{fig:cube_orbits}\textbf{f,g}). Since $\tilde{\matr G}$, $\tilde{\matr\Psi}$ and $\matr H$ are symmetric, these two orbits have the same value, and each matrix is determined by six values. They correspond to the following classes $\mathcal C_i$ of edge pairs $e$ and $f$:
\begin{enumerate}
  \item $\mathcal{C}_0: \quad$ $e=f$;
  \item $\mathcal{C}_1: \quad$ $e$ and $f$ share their source;
  \item $\mathcal{C}_2: \quad$ $e$ and $f$ share their sink;
  \item $\mathcal{C}_3: \quad$ $e \parallel f$ and the two don't share a common face (opposite edges);
  \item $\mathcal{C}_4: \quad$ $e \parallel f$ and the two share a common face;
  \item $\mathcal{C}_5: \quad$ other.
\end{enumerate}
Counting from a fixed $e$, the six classes contain 1, 2, 2, 1, 2 and 4 edges, respectively: the source and the sink of $e$ each have two further edges, three edges are parallel to $e$, one opposite and two sharing a face with it, and the remaining four do not touch an endpoint of $e$ (Supplementary Fig.~\ref{fig:cube_orbits}).

To evaluate $\tilde{\matr G}$, we identify the effective resistances for each of the classes. Writing ${\rm source}(e)$ and ${\rm sink}(e)$ for the source and sink of $e$, the column of the node-edge incidence matrix $\matr E$ for edge $e$ is $\vec 1_{{\rm source}(e)} - \vec 1_{{\rm sink}(e)}$, so that, using Eq.~(35) of the Methods,
\begin{equation}
  \tilde G_{ef}
  = \frac{1}{2}\bigl[ R({\rm source}(e),{\rm sink}(f)) + R({\rm sink}(e),{\rm source}(f)) - R({\rm source}(e),{\rm source}(f)) - R({\rm sink}(e),{\rm sink}(f)) \bigr] ,
  \label{eq:Gtilde_resistances}
\end{equation}
where $R$ is the effective resistance of the cube with unit capacity. Two sources, or two sinks, are always at distance two, yielding $R_2$, or $0$ if the nodes coincide. A source and a sink are either neighbors, at distance one, or antipodal, at distance three. Thus, only three values can appear, $R_1 = 7/12$, $R_2 = 3/4$ and $R_3 = 5/6$~\cite{pippenger2010hypercube}. Supplementary Table~\ref{tab:cube_Gtilde} lists the corresponding values, together with the final value of $\tilde G_{ef}$ for each of the six classes.

\begin{table}
\caption{\label{tab:cube_Gtilde}
Evaluation of Eq.~\eqref{eq:Gtilde_resistances} for the six classes.}
\begin{ruledtabular}
\begin{tabular}{lccccc}
$\mathcal{C}_i$ & $R({\rm source}(e),{\rm sink}(f))$ & $R({\rm sink}(e),{\rm source}(f))$ & $R({\rm source}(e),{\rm source}(f))$ & $R({\rm sink}(e),{\rm sink}(f))$ & $\tilde G_{ef}$ \\
\hline
$e=f$                  & $R_1$ & $R_1$ & $0$   & $0$   & $7/12$  \\
common source          & $R_1$ & $R_1$ & $0$   & $R_2$ & $5/24$  \\
common sink            & $R_1$ & $R_1$ & $R_2$ & $0$   & $5/24$  \\
parallel, opposite     & $R_3$ & $R_3$ & $R_2$ & $R_2$ & $1/12$  \\
parallel, common face  & $R_1$ & $R_1$ & $R_2$ & $R_2$ & $-1/6$  \\
other                  & $R_1$ & $R_3$ & $R_2$ & $R_2$ & $-1/24$ \\
\end{tabular}
\end{ruledtabular}
\end{table}

We proceed to $\tilde{\matr \Psi}^{(\beta)}$, determined by the sources with fluctuating injections. Let
\begin{align}
  \vec w_e = \tilde{\matr L}^{+}\bigl(\vec 1_{{\rm source}(e)} - \vec 1_{{\rm sink}(e)}\bigr)
\end{align}
be the potential generated by a unit current across $e$. The fluctuating potential difference across $e$ is $\Delta\theta^{(\beta)}_e = \vec w_e \cdot \vec X$. With $\langle \vec X \vec X^\top\rangle = 4\beta^2\matr Q$, Eq.~(37) of the Methods with $N_g = 4$, and since $\matr Q$ is an orthogonal projector, $\matr Q^2 = \matr Q = \matr Q^\top$, we obtain
\begin{equation}
  \tilde\Psi^{(\beta)}_{ef}
  = \bigl\langle \Delta\theta^{(\beta)}_e \Delta\theta^{(\beta)}_f \bigr\rangle
  = 4\beta^2\, \vec w_e^\top \matr{Q}\, \vec w_f
  = 4\beta^2\, \vec u_e \cdot \vec u_f ,
  \label{eq:Psitilde_overlap}
\end{equation}
where $\vec u_e = \matr Q \vec w_e$ restricts $\vec w_e$ to the sources and subtracts their mean. It is independent of the additive constant in $\vec w_e$. Every entry is thus determined by the four-component vectors $\vec u_e$ and $\vec u_f$, which we obtain from the same resistances as $\tilde{\matr G}$.

Measuring $\vec w_e$ from its value at ${\rm source}(e)$, we need $\bigl(\vec 1_a - \vec 1_{{\rm source}(e)}\bigr)^\top \tilde{\matr L}^+ \bigl(\vec 1_{{\rm source}(e)} - \vec 1_{{\rm sink}(e)}\bigr)$. This has the same form as $\tilde G_{ef}$, and Eq.~\eqref{eq:Gtilde_resistances} applies with ${\rm source}(f) = a$ and ${\rm sink}(f) = {\rm source}(e)$, since it holds for any pair of nodes, not only for edges. This gives
\begin{align}
  w_e(a) - w_e\bigl({\rm source}(e)\bigr)
  = \frac{1}{2}\Bigl[ R\bigl(a,{\rm sink}(e)\bigr) - R\bigl(a,{\rm source}(e)\bigr) - R_1 \Bigr] .
  \label{eq:w_offsets}
\end{align}
Among the four sources, there are three scenarios. For $a = {\rm source}(e)$, the offset vanishes. We denote the one source antipodal to ${\rm sink}(e)$ as ${\rm anti}(e)$. It lies at distance two from ${\rm source}(e)$, yielding $\frac{1}{2}[R_3 - R_2 - R_1] = -\frac{1}{4}$. The remaining two sources are adjacent to ${\rm sink}(e)$ and again at distance two from ${\rm source}(e)$. They are exchanged by the reflection that stabilizes $e$ (paragraph a), and both yield $\frac{1}{2}[R_1 - R_2 - R_1] = -\frac{3}{8}$. The mean of
$\bigl(0, -\frac{1}{4}, -\frac{3}{8}, -\frac{3}{8}\bigr)$ is $-\frac{1}{4}$. After subtracting it
\begin{align}
  u_e\bigl({\rm source}(e)\bigr) = \frac{1}{4} , \qquad
  u_e\bigl({\rm anti}(e)\bigr) = 0 , \qquad
  u_e = -\frac{1}{8} \ \text{ on the remaining two sources.}
  \label{eq:u_pattern}
\end{align}

Each $\vec u_e$ is thus fixed by the position of two sources, ${\rm source}(e)$ and ${\rm anti}(e)$. Ordering the four sources as $\bigl({\rm source}(e),\ {\rm anti}(e),\ c_1,\ c_2\bigr)$, so that $\vec u_e = (\frac{1}{4}, 0, -\frac{1}{8}, -\frac{1}{8})$ throughout, evaluating Eq.~\eqref{eq:Psitilde_overlap} on a class amounts to locating ${\rm source}(f)$ and ${\rm anti}(f)$ within that ordering, as in Supplementary Table~\ref{tab:cube_Psitilde}.

\begin{table}
\caption{\label{tab:cube_Psitilde}
Evaluation of Eq.~\eqref{eq:Psitilde_overlap} on the six classes.}
\begin{ruledtabular}
\begin{tabular}{lcccc}
$\mathcal{C}_i$ & ${\rm source}(f)$ & ${\rm anti}(f)$ & $\vec u_f$ & $\tilde\Psi^{(\beta)}_{ef}$ \\
\hline
$e=f$                 & ${\rm source}(e)$ & ${\rm anti}(e)$   & $(\frac14,0,-\frac18,-\frac18)$ & $\frac{3}{8}\beta^2$   \\
common source         & ${\rm source}(e)$ & $c_1$             & $(\frac14,-\frac18,0,-\frac18)$ & $\frac{5}{16}\beta^2$  \\
common sink           & $c_1$             & ${\rm anti}(e)$   & $(-\frac18,0,\frac14,-\frac18)$ & $-\frac{3}{16}\beta^2$ \\
parallel, opposite    & ${\rm anti}(e)$   & ${\rm source}(e)$ & $(0,\frac14,-\frac18,-\frac18)$ & $\frac{1}{8}\beta^2$   \\
parallel, common face & $c_1$             & $c_2$             & $(-\frac18,-\frac18,\frac14,0)$ & $-\frac{1}{4}\beta^2$  \\
other                 & ${\rm anti}(e)$   & $c_2$             & $(-\frac18,\frac14,-\frac18,0)$ & $-\frac{1}{16}\beta^2$ \\
\end{tabular}
\end{ruledtabular}
\end{table}

We can finally assemble $\matr H$. Adding Eq.~\eqref{eq:cube_Psimu} to the values of $\tilde{\matr{\Psi}}^{(\beta)}$ yields $\tilde\Psi_{ef} = \mu^{2}/9 + \tilde\Psi^{(\beta)}_{ef}$, and
\begin{align}
  H_{ef} = 2 \tilde G_{ef}\Bigl(\frac{\mu^{2}}{9} + \tilde\Psi^{(\beta)}_{ef}\Bigr) ,
  \label{eq:cube_H_assembled}
\end{align}
evaluated class by class in Supplementary Table~\ref{tab:cube_pair_values}. 

\begin{table}
\caption{Values of $\tilde{\matr{G}}$, $\tilde{\matr{\Psi}}^{(\beta)}$ and
$\matr{H}$ on the six classes of edge pairs. The entries of $\matr{H}$ follow
from the two preceding columns by Eq.~\eqref{eq:cube_H_assembled}. The last
column gives the number of edges $f$ in each class for fixed $e$.}
\label{tab:cube_pair_values}
\begin{ruledtabular}
\begin{tabular}{lcccc}
$\mathcal{C}_i$ & $\tilde G_{ef}$ & $\tilde\Psi^{(\beta)}_{ef}$ & $H_{ef}$ & count \\
\hline
$e=f$                 & $7/12$  & $\frac{3}{8}\beta^{2}$   & $\frac{7}{54}\mu^{2}   + \frac{7}{16}\beta^{2}$   & 1 \\
common source         & $5/24$  & $\frac{5}{16}\beta^{2}$  & $\frac{5}{108}\mu^{2}  + \frac{25}{192}\beta^{2}$ & 2 \\
common sink           & $5/24$  & $-\frac{3}{16}\beta^{2}$ & $\frac{5}{108}\mu^{2}  - \frac{5}{64}\beta^{2}$   & 2 \\
parallel, opposite    & $1/12$  & $\frac{1}{8}\beta^{2}$   & $\frac{1}{54}\mu^{2}   + \frac{1}{48}\beta^{2}$   & 1 \\
parallel, common face & $-1/6$  & $-\frac{1}{4}\beta^{2}$  & $-\frac{1}{27}\mu^{2}  + \frac{1}{12}\beta^{2}$   & 2 \\
other                 & $-1/24$ & $-\frac{1}{16}\beta^{2}$ & $-\frac{1}{108}\mu^{2} + \frac{1}{192}\beta^{2}$  & 4 \\
\end{tabular}
\end{ruledtabular}
\end{table}

The thresholds require the eigenvalues of $\matr H$ on each block. They can be obtained using the six values of Supplementary Table~\ref{tab:cube_pair_values} and the character table, Eq.~\eqref{eq:td_character_table}.

For each of the six classes of edge pairs $\mathcal{C}_i$, we can define its $12 \times12$ adjacency matrix on edge space $\matr A_i$, such that $(\matr A_i)_{ef}=1$ if $(e,f) \in \mathcal{C}_i$ and zero otherwise.
Matrices that are constant on the orbits of $\Gamma$ acting on ordered pairs of edges commute with every $\matr \Pi_g$, i.e.\ they form the commutant, or centralizer ring, of the permutation representation~\cite{bai2009truss}. Its dimension is equal to the number of these orbits, $\sum_\rho m_\rho^2 = 7$. Since two of them are exchanged by transposition, the six matrices $\matr A_i$ span its symmetric elements, including $\matr H$.
The general theory of these relation matrices and the algebra they generate is developed in Ref.~\cite{higman1975coherent}.
Since all pairs in a class share the same entry of $\matr H$, it is a linear combination of these matrices, where the coefficient $H_i$ is the value of $H_{ef}$ on class $i$, provided in Supplementary Table~\ref{tab:cube_pair_values},
\begin{equation}
  \matr{H} = \sum_{i=0}^{5} H_i \matr{A}_i .
  \label{eq:H_as_scheme}
\end{equation}

We now determine the eigenvalues of the matrices $\matr A_i$. First, consider the case when $m_\rho=1$ (everything except $T_2$). Then, $W_\rho$ is a single copy of the irreducible representation $\rho$.
According to Schur's lemma, $\matr A_i$ acts on $W_\rho$ as a scalar $p_i(\rho)$, i.e., $\matr A_i \matr P_\rho = p_i(\rho) \matr{P}_\rho$, where $\matr P_\rho$ is the orthogonal projector onto $W_\rho$. It satisfies $\operatorname{Tr} \matr{P}_\rho = \operatorname{dim} W_\rho = d_\rho m_\rho = d_\rho$. Taking the trace then yields $p_i(\rho) = \frac{1}{d_\rho} \mathrm{Tr} \bigl(\matr A_i\matr{P}_\rho \bigr)$.
The projector is given by Eq.~\eqref{eq:projector}, cf.\ Supplementary Methods~2. In general, $\chi_\rho(g)$ appears complex conjugated in this formula. The conjugation can be omitted, since all characters of $T_d$ are real.
Since $(\matr A_i \matr \Pi_g)_{ee} = (\matr A_i)_{e,\pi_g(e)}$, the trace counts the edges $e$ for which $(e, \pi_g(e))$ belongs to class $\mathcal{C}_i$,
\begin{equation}
  a_i(g) := \operatorname{Tr}\bigl(\matr{A}_i\matr{\Pi}_g\bigr)
  = \sum_{e \in \EE} (\matr{A}_i)_{e,\pi_g(e)}.
  \label{eq:a_i_def}
\end{equation}
Hence,
\begin{equation}
  p_i(\rho) = \frac{1}{|\Gamma|}\sum_{g\in\Gamma}\chi_\rho(g)\,a_i(g)
  = \frac{1}{24}\sum_{\text{classes } c} n_c\,\chi_\rho(c)\, a_i(c),
  \label{eq:p_i}
\end{equation}
where $n_c$ is the size of class $c$, and $\chi_\rho(c)$ is provided by the character table, Eq.~\eqref{eq:td_character_table}. This has the same form as the multiplicity
formula~\eqref{eq:multiplicity}. It reduces to Eq.~\eqref{eq:multiplicity} for $\matr A_0 = \matr{I}$.

\begin{table}
\caption{\label{tab:cube_p_i}
The block traces $p_i(\rho)$ of Eq.~\eqref{eq:p_i}. For the blocks of multiplicity one, these are the eigenvalues of $\matr A_i$. For $T_2$, of multiplicity two, the entry is the trace of the $2\times2$ matrix by which $\matr A_i$ acts on $W_{T_2}$.}
\begin{ruledtabular}
\begin{tabular}{lcccccc}
 $\rho$ & $e=f$ & common\ source & common sink & parallel, opposite & parallel, common face & other \\
\hline
$A_1$ & 1 & 2  & 2  & 1  & 2  & 4  \\
$E$   & 1 & -1 & -1 & 1  & 2  & -2 \\
$T_1$ & 1 & -1 & -1 & -1 & 0  & 2  \\
$T_2$ & 2 & 1  & 1  & 0  & -2 & -2 \\
\end{tabular}
\end{ruledtabular}
\end{table}

\begin{table}
\caption{\label{tab:cube_a_i}
The counts $a_i(g)$ of Eq.~\eqref{eq:a_i_def}: for a representative $g$ of each
conjugacy class, the number of edges standing in relation $i$ to their image
$\pi_g(e)$.}
\begin{ruledtabular}
\begin{tabular}{lcccccc}
 & $e=f$ & common\ source & common sink & parallel, opposite & parallel, common face & other \\
\hline
$\mathrm{id}$          & 12 & 0 & 0 & 0 & 0 & 0 \\
$8C_3$       & 0  & 3 & 3 & 0 & 0 & 6 \\
$3C_2$       & 0  & 0 & 0 & 4 & 8 & 0 \\
$6S_4$       & 0  & 0 & 0 & 0 & 4 & 8 \\
$6\sigma_d$  & 2  & 4 & 4 & 2 & 0 & 0 \\
\end{tabular}
\end{ruledtabular}
\end{table}

To evaluate Eq.~\eqref{eq:a_i_def} for a representative $g$ of each conjugacy class, we determine, for each of the twelve edges $e$, the class $\mathcal{C}_i$ to which the pair $(e, \pi_g(e))$ belongs, and count the edges in each class. For example, a reflection $\sigma_d$ fixes the two edges in its mirror plane, so that $a_0(\sigma_d) = 2$. This yields the counts $a_i(g)$ in Supplementary Table~\ref{tab:cube_a_i}. Contracting them with the characters~\eqref{eq:td_character_table} via Eq.~\eqref{eq:p_i} gives the $p_i(\rho)$ in Supplementary Table~\ref{tab:cube_p_i}.

Since $\matr H$ is a linear combination of the matrices $\matr A_i$, cf.\ Eq.~\eqref{eq:H_as_scheme}, and each of them acts on a block of multiplicity one as the scalar $p_i(\rho)$, the eigenvalue of $\matr H$ on $W_\rho$ is
\begin{equation}
  h_\rho = \sum_i H_i\, p_i(\rho) , \qquad m_\rho = 1 .
  \label{eq:h_rho}
\end{equation}
With the values of Supplementary Table~\ref{tab:cube_pair_values},
\begin{align}
  h_{A_1} &= H_0 + 2H_1 + 2H_2 + H_3 + 2H_4 + 4H_5
    = \frac{2\mu^{2}}{9} + \frac{3\beta^{2}}{4} , \\
  h_{E}   &= H_0 - H_1 - H_2 + H_3 + 2H_4 - 2H_5
    = \frac{9\beta^{2}}{16} , \\
  h_{T_1} &= H_0 - H_1 - H_2 - H_3 \phantom{{}+2H_4} + 2H_5
    = \frac{3\beta^{2}}{8} .
\end{align}

The $\mu^2$ terms are absent from $h_E$ and $h_{T_1}$, because $\tilde{\matr G}$, and hence $\matr H^{(\mu)}$, Eq.~\eqref{eq:cube_Hmu}, vanishes on these blocks. Indeed, $\tilde{\matr G} = \matr E^\top(\matr E\matr E^\top)^+\matr E$ is the orthogonal projector onto the row space of $\matr E$, whose orthogonal complement is the cycle space. Hence $W_E \oplus W_{T_1}$ is the cycle space of the cube, of dimension $\varsigma = 5$.

The remaining block $W_{T_2}$ cannot be treated this way, because $m_{T_2} = 2$. By Schur's lemma, $\matr{H}$ acts on $W_{T_2}$ as $\matr{I}_3 \otimes \matr{H}_{T_2}$, with a symmetric $2\times 2$ matrix $\matr{H}_{T_2}$ rather than a scalar~\cite{bai2009truss}. Each of its two eigenvalues $h_{T_2^\pm}$ therefore occurs $d_{T_2} = 3$ times in the spectrum of $\matr{H}$. The eigenvectors of $\matr{H}_{T_2}$ select the two copies of $T_2$ within $W_{T_2}$ that are invariant under $\matr{H}$, which we denote by $T_2^\pm$. We obtain $\operatorname{Tr} \matr H_{T_2}$ and $\det \matr H_{T_2}$ from the first two moments of $\matr H$.

Every eigenvalue on a block of multiplicity one occurs $d_\rho$ times. Since
$\operatorname{Tr} \matr{H} = 12 H_0$, and these eigenvalues are already known,
\begin{equation}
  \operatorname{Tr} \matr{H}_{T_2} = h_{T_2^{+}} + h_{T_2^{-}}
  = \frac{1}{3}\Bigl( \operatorname{Tr} \matr{H} - h_{A_1} - 2h_{E} - 3h_{T_1} \Bigr)
  = \frac{4\mu^{2}}{9} + \frac{3\beta^{2}}{4} .
  \label{eq:trM}
\end{equation}
The second moment is available from Supplementary Table~\ref{tab:cube_pair_values}. Because $\matr{H}$ is
symmetric and takes a single value on each class, we have
\begin{equation}
  \operatorname{Tr} \matr{H}^{2} = \sum_{e,f} H_{ef}^{2} = 12 \sum_i |\mathcal C_i| H_i^{2}
  = \frac{28\mu^{4}}{81} + \frac{4\mu^{2}\beta^{2}}{3} + \frac{387\beta^{4}}{128} .
  \label{eq:cube_trH2}
\end{equation}
Subtracting the contributions of the other blocks in the same way gives
$h_{T_2^{+}}^{2} + h_{T_2^{-}}^{2} = \frac13(\operatorname{Tr}\matr{H}^2 - h_{A_1}^2 - 2h_E^2 - 3h_{T_1}^2)$,
so that
\begin{equation}
  \det \matr H_{T_2} = h_{T_2^{+}} h_{T_2^{-}}
  =   \frac{1}{2} \Bigl[(\operatorname{Tr}\matr H_{T_2})^{2} - \bigl(h_{T_2^{+}}^{2} + h_{T_2^{-}}^{2}\bigr)\Bigr]
  = \frac{4\mu^{4}}{81} + \frac{\mu^{2}\beta^{2}}{6} + \frac{3\beta^{4}}{64} .
  \label{eq:detM}
\end{equation}
The eigenvalues of a symmetric $2\times2$ matrix can be obtained from its trace and determinant,
\begin{equation}
  h_{T_2^{\pm}} = \frac{\operatorname{Tr} \matr H_{T_2}
    \pm \sqrt{(\operatorname{Tr} \matr H_{T_2})^{2} - 4\det \matr H_{T_2}}}{2}
  = \frac{2\mu^{2}}{9} + \frac{3 \pm \sqrt{6}}{8}\,\beta^{2} .
  \label{eq:cube_hT2}
\end{equation}
The two eigenvalues $h_{T_2^\pm}$ coincide at $\beta = 0$ and split only in the presence of fluctuations.

Inserting these eigenvalues into Eq.~(17) of the main text, $\gamma_s(\rho) = 1 - h_\rho/\tilde\Psi_{ee}$, with $\tilde\Psi_{ee}$ from Eq.~\eqref{eq:cube_psi_ee}, yields
\begin{align}
  \gamma_s(T_1) &= \frac{8\mu^{2}}{8\mu^{2}+27\beta^{2}},
  \label{eq:cube_gs_T1}\\
  \gamma_s(E) &= \frac{16\mu^{2}-27\beta^{2}}{16\mu^{2}+54\beta^{2}},
  \label{eq:cube_gs_E}\\
  \gamma_s(T_2^{\pm}) &= -\,\frac{8\mu^{2} \pm 9\sqrt{6}\,\beta^{2}}{8\mu^{2}+27\beta^{2}} .
  \label{eq:cube_gs_T2}
\end{align}

Since $\gamma_s(T_2^{+}) < 0$ and $\gamma_s(E) < \gamma_s(T_1)$ for all $\beta > 0$, only $T_1$ and $T_2^{-}$ can be critical. In particular, the $E$ block, whose perturbations distinguish the three spatial directions, never determines $\gamma_s$.

\paragraph{Residual symmetries}
\label{app:cube_residual}
The isotropy subgroups of the critical block constrain which symmetries remain in the symmetry-broken configuration (Supplementary Methods~2). As representations of $T_d$, $T_2$ is equivalent to the action on ordinary vectors in $\mathbb{R}^3$, while $T_1 \cong T_2 \otimes A_2$ is equivalent to the action on ordinary vectors with an additional sign change under the improper operations $S_4$ and $\sigma_d$. Since $T_2^+$ and $T_2^-$ are single copies of $T_2$, they share the same isotropy subgroups. The maximal isotropy subgroups are:
\begin{itemize}
  \item $T_1$, space diagonal: $C_3$. A reflection in a plane containing the axis reverses a $T_1$ vector along it because of the additional sign. Reflections cannot remain.
  \item $T_1$, face axis: $C_4^{(i)}$\footnote{We write $C_4^{(i)}$ rather than the Schoenflies symbol $S_4$ to avoid confusion with $\Gamma \cong S_4$.}, the cyclic group of order four generated by an improper fourfold rotation.
  \item $T_2$, space diagonal: $C_{3v}$. The reflections in planes containing the axis fix a vector along the axis.
  \item $T_2$, face axis: $C_{2v}$. The improper fourfold rotation reverses a vector along its axis. Only the twofold rotation and the two reflections remain.
\end{itemize}
As discussed in the main text, an optimum with residual symmetry $\Gamma^*$ belongs to a family of $|\Gamma|/|\Gamma^*|$ equivalent optima, i.e., $8$ for $C_3$, $6$ for $C_4^{(i)}$, $4$ for $C_{3v}$, and $6$ for $C_{2v}$. The residual symmetries of the nine-edge and eight-edge configurations, $C_3$ and $C_4^{(i)}$, are exactly the two isotropy subgroups of $T_1$, the critical block for $\beta < \beta_\times$.

For $\beta > \beta_\times$, where $T_2^-$ is critical, the ten-edge configuration has the residual symmetry $\Gamma^* \cong C_{2v}$, the isotropy subgroup of $T_2$ for a face axis. The eleven-edge configuration retains only a single reflection, $\Gamma^* \cong C_s$, the isotropy subgroup of a $T_2$ vector lying in one mirror plane. Its order is two, so it is not a maximal isotropy subgroup. The kite and the tree only have the trivial symmetry, $\Gamma^* = C_1$.

\newpage

\subsection{Supplementary Methods 9: Expected dissipation of the cube phases}
\label{app:cube_dissipations}

Here, we provide the expected dissipation of the symmetric, tree and ring configurations of the cube in closed form. We summarize the remaining optimal configurations in Supplementary Table~\ref{tab:cube_configs}, whose $\bar D$ we evaluate numerically with the self-consistent iteration of the Methods, from coarse and fine parameter scans with multiple initial conditions.

\paragraph{Symmetric configuration.} All twelve capacities are equal to $k=\kappa\,12^{-1/\gamma}$. Using Eq.~\eqref{eq:cube_D1} we obtain
\begin{align}
  \bar D_{\mathrm{sym}}
  = \frac{12^{1/\gamma}}{\kappa}\Big(\frac{4}{3}\mu^{2}+\frac{9}{2}\beta^{2}\Big).
  \label{eq:cube_D12}
\end{align}

\paragraph{Tree configuration.} On a tree, the flows are determined by Kirchhoff's current law. Cutting the tree at edge $e$ leaves $n_g$ sources and $n_c$ sinks on one side, and for $N_g = N_c$
\begin{equation}
  \langle F_e^2\rangle = (n_g-n_c)^{2}\mu^{2} + n_g\,(N_g-n_g)\,\beta^{2}.
  \label{eq:tree_flow_constants}
\end{equation}
Since the flows do not depend on the capacities, we use Eq.~(31) of the Methods, and the formula we derived for the open ring, Eq.~\eqref{eq:D_open} in Supplementary Methods~5. 
The optimal tree is found by comparing the $384$ spanning trees of the cube. Accounting for symmetry, there are only 16 distinct trees. Over the relevant range of $\gamma$, only one tree is always optimal (depicted in Fig.~4\textbf{a} of the main text, cf. inset at $\gamma = 0.2$). There are $24$ equivalent optima.
Equation~\eqref{eq:tree_flow_constants} evaluates on its seven cuts to
$\langle F_e^2\rangle = \mu^{2}+3\beta^{2}$, $\mu^{2}+4\beta^{2}$, $\mu^{2}$ twice and
$3\beta^{2}$ three times. We thus obtain
\begin{align}
  \bar D_{\mathrm{tree}} = \frac{1}{\kappa}
  \Big[(\mu^{2}+3\beta^{2})^{\frac{\gamma}{1+\gamma}}
     + (\mu^{2}+4\beta^{2})^{\frac{\gamma}{1+\gamma}}
     + 2\,\mu^{\frac{2\gamma}{1+\gamma}}
     + 3\,(3\beta^{2})^{\frac{\gamma}{1+\gamma}}\Big]^{\frac{1+\gamma}{\gamma}}.
  \label{eq:cube_D7}
\end{align}

\paragraph{Ring configuration.} The eight active edges form a cycle which alternates between sources and sinks, i.e., the elementary ring with $N=4$ (Supplementary Note~5). Its optimum is uniform as long as the alternating mode is stable, cf.\ Eq.~\eqref{eq:gs-alt},
\begin{align}
\gamma>\gamma_s(\mathrm{alt})|_{N=4}=(\mu^{2}-5\beta^{2})/(\mu^{2}+5\beta^{2}).
\end{align}
For $\beta \geq \mu/\sqrt5$ the threshold is non-positive and the ring is uniform for all $\gamma>0$. In the uniform regime, Eq.~\eqref{eq:D_sym} with $N=4$ yields
\begin{align}
  \bar D_{\mathrm{ring}}
  = \frac{8^{1/\gamma}}{\kappa}\big(2\mu^{2}+10\beta^{2}\big).
  \label{eq:cube_D8}
\end{align}
For $\gamma < \gamma_s(\mathrm{alt})|_{N=4}$, the capacities are alternating, and the dissipation is given by Eq.~\eqref{eq:D_alt} with $N=4$, minimized under $4(k_a^\gamma + k_b^\gamma) = \kappa^\gamma$. Alternating capacities do not affect the residual symmetry, since the stabilizer of the ring in $\Gamma$, $C_4^{(i)}$, acts on it as a rotation by two nodes, which preserves the alternating pattern.

\paragraph{Remaining configurations.} The remaining optimal configurations of the cube do not have a simple closed form.  The nine-edge optimum retains only a threefold rotation about a space diagonal, $\Gamma^* \cong C_3$, so all mirror symmetries are broken. For this reason we refer to it as the chiral phase. Above a threshold $\beta_{\rm h}(\gamma)$, the optimal eight-edge support is no longer the alternating ring, but a hexagon with a tail consisting of two edges, which we call a kite~\cite{KimPark2006}. The ten-edge optimum is the cube from which two parallel edges that do not share a face are removed.

\begin{table}[b!]
\caption{\label{tab:cube_configs}
Optimal configurations of the cube, with the number of active edges, the order of the residual symmetry group $\Gamma^*$, the number $|\Gamma|/|\Gamma^*|$ of equivalent optima, the cycle rank $\varsigma$, and the expected dissipation $\bar D$.}
\begin{ruledtabular}
\begin{tabular}{lccccc}
configuration & edges & $|\Gamma^*|$ & optima & $\varsigma$ & $\bar D$ \\
\hline
symmetric          & 12 & 24 & 1  & 5 & Eq.~\eqref{eq:cube_D12} \\
eleven-edge        & 11 & 2  & 12 & 4 & numerical \\
ten-edge           & 10 & 4  & 6  & 3 & numerical \\
chiral             & 9  & 3  & 8  & 2 & numerical \\
ring, uniform      & 8  & 4  & 6  & 1 & Eq.~\eqref{eq:cube_D8} \\
ring, alternating  & 8  & 4  & 6  & 1 & Eq.~\eqref{eq:D_alt}, $N=4$ \\
kite       & 8  & 1  & 24 & 1 & numerical \\
tree               & 7  & 1  & 24 & 0 & Eq.~\eqref{eq:cube_D7} \\
\end{tabular}
\end{ruledtabular}
\end{table}

\newpage
\subsection{Supplementary Methods 10: Group-theoretic stability analysis of complete bipartite networks}
\label{sm:bipartite}

Here we identify the symmetry group $\Gamma$ of the complete bipartite network $K_{N_g,N_c}$, decompose the space of capacity perturbations into blocks, and evaluate the stability threshold of each block. Every source $i\in\mathcal{V}_g$ is connected to every sink $j\in\mathcal{V}_c$ by an edge $e=(i,j)$, oriented from the source to the sink. 

\paragraph{The symmetry group of the optimization problem.}
The automorphisms of $K_{N_g,N_c}$ permute the sources among themselves and the sinks among themselves. For $N_g=N_c$, exchanging the sources and the sinks is an additional automorphism. A symmetry of the optimization problem must also preserve the injection statistics, i.e., $\matr{R}_g\matr{\Sigma}\matr{R}_g^\top=\matr{\Sigma}$, Eq.~(8) of the main text. For $\beta>0$, only the sources have fluctuations, so exchanging the sources and the sinks is excluded. Thus we have
\begin{equation}
    \Gamma\cong S_{N_g}\times S_{N_c}, \qquad |\Gamma| = N_g!\,N_c! .
    \label{eq:SM10_group}
\end{equation}
An element $g = (\sigma, \tau) \in \Gamma$ consists of a permutation $\sigma$ of the sources and a permutation $\tau$ of the sinks. It maps any edge $(i,j)$ to $(\sigma(i), \tau(j))$, i.e., it permutes the rows and the columns of the edge array independently. Any edge can be mapped onto any other edge, so $\Gamma$ acts transitively on $\EE$. By the orbit-stabilizer theorem, the stabilizer of an edge $(i,j)$ has order $|\Gamma|/|\EE| = (N_g-1)!\,(N_c-1)!$. It consists of all the permutations that leave source $i$ and sink $j$ invariant, $S_{N_g-1} \times S_{N_c-1}$.

Since every $g\in\Gamma$ maps sources onto sources, it preserves the source-to-sink orientation of every edge. Just like in the cube scenario, the matrices $\matr{\Pi}_g$ representing $\Gamma$ on edge space (Supplementary Methods~2) are thus unsigned permutation matrices, and their character $\chi(g)=\operatorname{Tr}\matr{\Pi}_g$ is equal to the number of edges left invariant by $g$. An edge $(i,j)$ is invariant if and only if $\sigma(i)=i$ and $\tau(j)=j$. Writing $f(\sigma)$ and $f(\tau)$ for the numbers of sources and sinks left invariant by $\sigma$ and $\tau$, we obtain
\begin{equation}
    \chi(\sigma,\tau) = f(\sigma)\, f(\tau).
    \label{eq:SM10_character}
\end{equation}
The character depends on $(\sigma,\tau)$ only through the numbers of invariant sources and sinks, which are the same for all elements of a conjugacy class. We list representative element types together with their invariant nodes and edges in Supplementary Table~\ref{tab:SM10_elements}.

\begin{table}[b]
\caption{Representative element types of the symmetry group $\Gamma\cong S_{N_g}\times S_{N_c}$ of the complete bipartite network. In the last two columns, we list the number of nodes and edges left invariant. The edge counts are equal to the character $\chi$ of the permutation representation on edge space, Eq.~\eqref{eq:SM10_character}.}
\label{tab:SM10_elements}
\begin{ruledtabular}
\begin{tabular}{lllll}
element $(\sigma,\tau)$ & action & number of elements & invariant nodes & invariant edges \\
\hline
$(\mathrm{id},\mathrm{id})$ & identity & $1$ & $N_g+N_c$ & $N_gN_c$ \\
$((i\,k),\mathrm{id})$ & exchanges two sources & $\binom{N_g}{2}$ & $N_g+N_c-2$ & $(N_g-2)N_c$ \\
$(\mathrm{id},(j\,l))$ & exchanges two sinks & $\binom{N_c}{2}$ & $N_g+N_c-2$ & $N_g(N_c-2)$ \\
$((i\,k),(j\,l))$ & exchanges two sources and two sinks & $\binom{N_g}{2}\binom{N_c}{2}$ & $N_g+N_c-4$ & $(N_g-2)(N_c-2)$ \\
$(\sigma,\tau)$ & general & & $f(\sigma)+f(\tau)$ & $f(\sigma)f(\tau)$
\end{tabular}
\end{ruledtabular}
\end{table}

\paragraph{Decomposition of the space of capacity perturbations.} 
Like in Supplementary Methods~8, we arrange a perturbation $\vec{\delta}\in\mathbb{R}^{N_gN_c}$ as an $N_g\times N_c$ matrix $(\delta_{ij})$, with one row per source and one column per sink,
\begin{equation}
    (\delta_{ij}) =
    \begin{pmatrix}
        \delta_{11} & \cdots & \delta_{1N_c} \\
        \vdots & \ddots & \vdots \\
        \delta_{N_g1} & \cdots & \delta_{N_gN_c} \label{eq:pert-matrix-bipartite}
    \end{pmatrix}.
\end{equation}
An element $(\sigma,\tau) \in \Gamma$ permutes the rows according to $\sigma$ and the columns according to $\tau$, independently of each other. We will first decompose the action of a single symmetric group, and then combine the rows and the columns.

Consider the symmetric group $S_n$, $n\geq 2$, i.e., the group of all permutations of $n$ objects. It acts on $\mathbb{R}^n$ by permuting the entries of a vector $\vec{a}$, $(\matr{\Pi}_\sigma\vec{a})_{\sigma(i)}=a_i$, where $\matr{\Pi}_\sigma$ is the permutation matrix of $\sigma$. The map $\sigma\mapsto\matr{\Pi}_\sigma$ is the permutation representation of $S_n$~\cite{Serre1977, FultonHarris1991}. Its character, $\operatorname{Tr}\matr{\Pi}_\sigma$, is the number $f(\sigma)$ of entries left invariant by $\sigma$. As for the edge space of the cube, we decompose this representation by identifying the subspaces that are invariant under all permutations. For $S_n$, there are exactly two nontrivial proper invariant subspaces, which are orthogonal to each other. The first orthogonal subspace is the line of constant vectors, spanned by $\vec{1}$. Every permutation leaves each constant vector invariant. The restriction of the permutation representation to this line is the unit representation, which we denote by $1$. Its character is equal to $1$ for every $\sigma$. The second subspace is the orthogonal complement of this line, i.e., the vectors whose entries sum to zero. It is invariant because permuting the entries of a vector does not change their sum. We denote the representation of $S_n$ on this $(n-1)$-dimensional subspace by $\xi$, the standard representation. Since the two subspaces are orthogonal complements of each other,
\begin{equation}
    \mathbb{R}^n = 1\oplus\xi ,
    \label{eq:SM10_standard}
\end{equation}
i.e., every vector $\vec{a}$ is the sum of the constant vector $\bar{a}\vec{1}$ with the mean $\bar{a}=\frac{1}{n}\sum_i a_i$, and the vector $\vec{a}-\bar{a}\vec{1}$, whose entries sum to zero. The character of a direct sum is the sum of the characters of its parts, such that $\xi$ has the character $f(\sigma)-1$. The standard representation is irreducible~\cite{FultonHarris1991}, so Eq.~\eqref{eq:SM10_standard} is the decomposition of the permutation representation into irreducible representations.

We now combine the sources and the sinks. Ordering the edges as $e=(i,j)\mapsto(i-1)N_c+j$, the edge space is given by the tensor product $\mathbb{R}^{N_g}\otimes\mathbb{R}^{N_c}$, and $\matr{\Pi}_{(\sigma,\tau)}=\matr{\Pi}_\sigma\otimes\matr{\Pi}_\tau$, since $\sigma$ acts on the rows and $\tau$ on the columns independently. This is consistent with Eq.~\eqref{eq:SM10_character}, since the trace of a tensor product is the product of the traces. Then, using Eq.~\eqref{eq:SM10_standard}, the edge space is decomposed as
\begin{equation}
    \mathbb{R}^{N_gN_c} = (1\otimes 1)\oplus(\xi\otimes 1)\oplus(1\otimes\xi)\oplus(\xi\otimes\xi).
    \label{eq:SM10_decomposition}
\end{equation}
Because tensor products of irreducible representations of $S_{N_g}$ and $S_{N_c}$ are irreducible and pairwise non-isomorphic representations of $\Gamma$~\cite{Serre1977}, each block in Eq.~\eqref{eq:SM10_decomposition} has multiplicity one. It is thus an eigenspace of the constrained Hessian with a single stability threshold (Supplementary Methods~2).

We will now apply the representation-theoretic treatment of the two-way analysis of variance in Ref.~\cite{Diaconis1988} to the capacity perturbation matrix~\eqref{eq:pert-matrix-bipartite}. 
The idea is to represent a matrix as the sum of an overall mean, row effects, column effects and an interaction. 
To identify the perturbations in each block, we construct the orthogonal projectors from Eq.~\eqref{eq:projector}. For a single $S_n$, setting $\chi_\rho=1$ and $d_\rho=1$ demonstrates that the projector onto the unit representation is the average of all permutation matrices, $\frac{1}{n!}\sum_\sigma\matr{\Pi}_\sigma$. Its entry $(a,b)$ is the fraction of permutations that map $b$ onto $a$, i.e., $(n-1)!/n!=1/n$. Denoting the identity matrix by $\matr{I}_n$ and the all-ones matrix by $\matr{J}_n$, the projector onto $1$ is thus $\matr{J}_n/n$. It replaces every entry of a vector by the mean of all entries. The projector onto $\xi$ is its complement, $\matr{I}_n-\matr{J}_n/n$, which subtracts this mean.

For $\Gamma$, the characters and the dimensions are products, $\chi_{\rho\otimes\rho'}(\sigma,\tau)=\chi_\rho(\sigma)\chi_{\rho'}(\tau)$ and $d_{\rho\otimes\rho'}=d_\rho d_{\rho'}$. The same holds for the permutation matrices $\matr{\Pi}_{(\sigma,\tau)}=\matr{\Pi}_\sigma\otimes\matr{\Pi}_\tau$. Inserting these products into Eq.~\eqref{eq:projector}, the sum over $\Gamma$ becomes a product of a sum over $S_{N_g}$ and a sum over $S_{N_c}$, such that $\matr{P}^{\rho\otimes\rho'}=\matr{P}^{\rho}\otimes\matr{P}^{\rho'}$. We obtain the projectors:
\begin{align}
    \matr{P}^{1\otimes 1} &= \frac{\matr{J}_{N_g}}{N_g}\otimes\frac{\matr{J}_{N_c}}{N_c}, &
    \matr{P}^{\xi\otimes 1} &= \Big(\matr{I}_{N_g}-\frac{\matr{J}_{N_g}}{N_g}\Big)\otimes\frac{\matr{J}_{N_c}}{N_c}, \\
    \matr{P}^{1\otimes\xi} &= \frac{\matr{J}_{N_g}}{N_g}\otimes\Big(\matr{I}_{N_c}-\frac{\matr{J}_{N_c}}{N_c}\Big), &
    \matr{P}^{\xi\otimes\xi} &= \Big(\matr{I}_{N_g}-\frac{\matr{J}_{N_g}}{N_g}\Big)\otimes\Big(\matr{I}_{N_c}-\frac{\matr{J}_{N_c}}{N_c}\Big).
\label{eq:SM10_projectors}
\end{align}

To see how these projectors act, we return to the matrix $(\delta_{ij})$, Eq.~\eqref{eq:pert-matrix-bipartite}. For the chosen edge ordering, the tensor product $\matr{A}\otimes\matr{B}$ acts on $\vec{\delta}$ as $(\delta_{ij})\mapsto\matr{A}\,(\delta_{ij})\,\matr{B}^\top$. The first factor of each projector acts on the columns of $(\delta_{ij})$, i.e., on the sources, and the second factor on the rows, i.e., on the sinks. Multiplying from the left by $\matr{J}_{N_g}/N_g$ averages over the sources for a fixed sink. Multiplying from the right by $\matr{J}_{N_c}/N_c$ averages over the sinks for a fixed source. Conversely, multiplying from the left by $\matr{I}_{N_g}-\matr{J}_{N_g}/N_g$ subtracts from every entry the mean of its column, such that every column of the result sums to zero, while multiplying from the right by $\matr{I}_{N_c}-\matr{J}_{N_c}/N_c$ makes every row sum to zero. Applying the four projectors given by Eq.~\eqref{eq:SM10_projectors} to Eq.~\eqref{eq:pert-matrix-bipartite} decomposes every perturbation as
\begin{equation}
    (\delta_{ij}) = \bar{\delta}\,\vec{1}\vec{1}^\top + \vec{x}\,\vec{1}^\top + \vec{1}\,\vec{y}^\top + \matr{Z},
    \qquad
    \vec{1}^\top\vec{x} = 0, \quad \vec{1}^\top\vec{y} = 0, \quad \matr{Z}\vec{1} = \vec{0}, \quad \vec{1}^\top\matr{Z} = \vec{0}^\top,
    \label{eq:SM10_projections}
\end{equation}
where $\vec{x}\in\mathbb{R}^{N_g}$, $\vec{y}\in\mathbb{R}^{N_c}$, and $\vec{1}$ denotes the all-ones vector of the length required by each product. In the language of the analysis of variance, $\vec{x}$ and $\vec{y}$ are the row and column effects, and $\matr{Z}$ is the interaction~\cite{Diaconis1988}. Denoting the overall mean by $\bar{\delta}$, the row means by $\bar{\delta}_{i\cdot}$ and the column means by $\bar{\delta}_{\cdot j}$, the entries are equal to
\begin{equation}
    x_i = \bar{\delta}_{i\cdot}-\bar{\delta} \qquad
    y_j = \bar{\delta}_{\cdot j}-\bar{\delta} \qquad
    Z_{ij} = \delta_{ij}-\bar{\delta}_{i\cdot}-\bar{\delta}_{\cdot j}+\bar{\delta}.
    \label{eq:SM10_entries}
\end{equation}
The uniform perturbation $\bar{\delta}\,\vec{1}\vec{1}^\top$ is excluded by the resource constraint. The remaining three terms define the blocks:
\begin{equation}
\begin{aligned}
    W_{\mathrm{src}} = \xi\otimes 1 &: \ \delta_{ij}=x_i \text{ depends only on the source, with } \textstyle\sum_i x_i = 0; \\
    W_{\mathrm{sink}} = 1\otimes\xi &: \ \delta_{ij}=y_j \text{ depends only on the sink, with } \textstyle\sum_j y_j = 0; \\
    W_{\mathrm{loop}} = \xi\otimes\xi &: \ \delta_{ij}=Z_{ij}, \text{ with } \textstyle\sum_j Z_{ij}=0 \text{ for every } i \text{ and } \sum_i Z_{ij}=0 \text{ for every } j.
\end{aligned}
\label{eq:SM10_blocks}
\end{equation}

The row sum $\sum_j\delta_{ij}$ is the total change of capacity of the edges at source $i$, and the column sum $\sum_i\delta_{ij}$ is the total change at sink $j$. Thus, the perturbations in the loop block redistribute capacity between edges without changing the total capacity at any node. This is similar to the cube scenario, where we had perturbations that depend only on the source, those that depend only on the direction, and those with vanishing row and column sums (Supplementary Methods~8). However, here the columns label sinks instead of directions, and the third type forms a single irreducible block.

The loop block coincides with the cycle space, i.e., the space of loop flows. If we arrange the flows $F_{ij}$ as an $N_g \times N_c$ matrix with one row per source and one column per sink, like the perturbations in Eq.~\eqref{eq:pert-matrix-bipartite}, the row sum $\sum_j F_{ij}$ is the total current leaving source $i$, and the column sum $\sum_i F_{ij}$ is the total current entering sink $j$. Kirchhoff's current law yields $\sum_j F_{ij} = S_i$ and $\sum_i F_{ij} = -S_j$. Loop flows have zero net injection at all nodes, i.e. the row and column sums vanish. This is the defining property of $W_{\mathrm{loop}}$ in Eq.~\eqref{eq:SM10_blocks}. The dimension $\dim W_{\mathrm{loop}}=(N_g-1)(N_c-1)$ is equal to the cycle rank $\varsigma$.

\paragraph{Stability thresholds.}
At unit capacities, $\matr K = \matr I$, Kirchhoff's current law at sink $j$ reads $\sum_i (\theta_i - \theta_j) = \mu$. Taking the difference of these equations for two sinks $j$ and $l$ yields $N_g(\theta_j - \theta_l) = 0$. All sinks are at equal potential, which can be set to zero. Kirchhoff's current law at source $i$, $\sum_j (\theta_i - \theta_j) = N_c \theta_i = S_i$, then yields
\begin{equation}
  \theta_i = \frac{\mu}{N_g} + \frac{X_i}{N_c} \quad (i \in \VV_g),
  \qquad
  \theta_j = 0 \quad (j \in \VV_c).
  \label{eq:SM10_potentials}
\end{equation}
Each source divides its mean injection and fluctuation evenly over its edges. 
The potential difference across edge $e = (i,j)$ is $\Delta\theta_e = \theta_i$. 
Using $\langle X_i \rangle = 0$ and $\langle X_i X_k \rangle = \beta^2 (N_g \delta_{ik} - 1)$, Eq.~(32) of the Methods, the second moments for edges $e = (i,j)$ and $f = (k,l)$ are 
\begin{equation}
    \tilde{\Psi}_{ef} = \frac{\mu^2}{N_g^2}+\frac{\beta^2}{N_c^2}\big(N_g\delta_{ik}-1\big), \qquad
    \tilde{\Psi}_{ee} = \frac{\mu^2}{N_g^2}+\frac{(N_g-1)\beta^2}{N_c^2}.
    \label{eq:SM10_Psi}
\end{equation}
At unit capacities, $\tilde{\matr{G}}$ is the orthogonal projector onto the row space of $\matr{E}$, i.e., onto the orthogonal complement of the cycle space (Supplementary Methods~8). Since the cycle space is the loop block, Eq.~\eqref{eq:SM10_projectors} gives $\tilde{\matr{G}}=\matr{I}-\matr{P}^{\xi\otimes\xi}$, with entries
\begin{equation}
    \tilde{G}_{ef} = \frac{\delta_{ik}}{N_c}+\frac{\delta_{jl}}{N_g}-\frac{1}{N_gN_c}.
    \label{eq:SM10_G}
\end{equation}
To obtain the eigenvalues of $\matr{H} = 2\,\tilde{\matr{G}} \circ \tilde{\matr{\Psi}}$, we express its quadratic form through the current $w_e = \delta_e \Delta\theta_e$ induced by a capacity perturbation $\vec{\delta}$ at fixed potentials. Since $\tilde{\Psi}_{ef} = \langle \Delta\theta_e \Delta\theta_f \rangle$ and $\tilde{\matr{G}}$ is a projector,
\begin{equation}
    \vec{\delta}^\top\matr{H}\vec{\delta} = 2\sum_{e,f}\delta_e\tilde{G}_{ef}\langle\Delta\theta_e\Delta\theta_f\rangle\delta_f = 2\big\langle\lVert\tilde{\matr{G}}\vec{w}\rVert^2\big\rangle.
    \label{eq:SM10_quadform}
\end{equation}
The loop flows in $\vec{w}$ circulate without affecting potentials and are annihilated by $\tilde{\matr{G}}$. The quadratic form is determined by the remaining part of $\vec{w}$, which changes the net current at some nodes, shifting the potentials. Since the blocks are eigenspaces, their eigenvalues are given by $h_\rho=\vec{\delta}^\top\matr{H}\vec{\delta}/\lVert\vec{\delta}\rVert^2$ for any $\vec{\delta}\in W_\rho$. Inserting $\Delta\theta_e=\theta_i$ for $e=(i,j)$ from Eq.~\eqref{eq:SM10_potentials} yields
\begin{equation}
    w_{ij} = \frac{\mu}{N_g}\delta_{ij}+\frac{1}{N_c}\delta_{ij}X_i .
    \label{eq:SM10_current}
\end{equation}
We now apply Eq.~\eqref{eq:SM10_current} to each block.

In the source block, $\delta_{ij}=x_i$, the current $w_{ij}$ depends only on the source. It therefore contains no loop flows, i.e., $\tilde{\matr{G}}\vec{w}=\vec{w}$. Using $\lVert\vec{\delta}\rVert^2=N_c\sum_i x_i^2$, we obtain $h_{\mathrm{src}}=2\tilde{\Psi}_{ee}$.

In the sink block, $\delta_{ij}=y_j$, the mean-load part of $w_{ij}$ depends only on the sink and $\tilde{\matr{G}}$ does not affect it. The row and column sums of the fluctuating part $y_jX_i/N_c$ are zero, since $\sum_j y_j=\sum_i X_i=0$. It is thus a loop flow, annihilated by $\tilde{\matr{G}}$. Using $\lVert\vec{\delta}\rVert^2=N_g\sum_j y_j^2$, we obtain $h_{\mathrm{sink}}=2\mu^2/N_g^2$.

In the loop block, $\delta_{ij}=Z_{ij}$, the roles are reversed. The mean-load part, $\mu Z_{ij}/N_g$, is a loop flow and is annihilated. The fluctuating part, $Z_{ij}X_i/N_c$, has vanishing row sums but nonvanishing column sums $c_j=\sum_i Z_{ij}X_i/N_c$. Since its row means and its overall mean vanish, Eq.~\eqref{eq:SM10_entries} shows that $\tilde{\matr{G}}$ replaces every entry by the mean of its column, $c_j/N_g$. Then,
\begin{equation}
    \big\langle \lVert\tilde{\matr{G}}\vec{w}\rVert^2\big\rangle = \frac{1}{N_g}\sum_j\langle c_j^2\rangle = \frac{\beta^2}{N_c^2}\lVert\matr{Z}\rVert^2,
    \label{eq:SM10_loop}
\end{equation}
where we used $\langle X_iX_k\rangle=\beta^2(N_g\delta_{ik}-1)$ and $\sum_i Z_{ij}=0$. Finally, we obtain $h_{\mathrm{loop}}=2\beta^2/N_c^2$.

We conclude that the sink block is stabilized only by the mean load, and the loop block only by the fluctuations. Inserting the eigenvalues into $\gamma_s(\rho)=1-h_\rho/\tilde{\Psi}_{ee}$, with $\tilde{\Psi}_{ee}$ from Eq.~\eqref{eq:SM10_Psi} we obtain the thresholds
\begin{align}
    \gamma_s(\mathrm{src}) &= -1, \label{eq:SM10_gamma_src}\\
    \gamma_s(\mathrm{sink}) &= \frac{(N_g-1)\beta^2-\beta^{*2}}{(N_g-1)\beta^2+\beta^{*2}}, \label{eq:SM10_gamma_sink}\\
    \gamma_s(\mathrm{loop}) &= \frac{(N_g-3)\beta^2+\beta^{*2}}{(N_g-1)\beta^2+\beta^{*2}},\label{eq:SM10_gamma_loop}
\end{align}
written in terms of the mean source injection $\beta^*=N_c\mu/N_g$. Since $\gamma_s(\mathrm{src})=-1$, the source block cannot destabilize the symmetric configuration for $\gamma>0$.

As $\beta$ increases, $\gamma_s(\mathrm{loop})$ decreases from $1$, whereas $\gamma_s(\mathrm{sink})$ increases from $-1$ to $1$. The two thresholds cross at $\beta=\beta^*$, where both are equal to $1-2/N_g$. The loop block is critical for $\beta<\beta^*$, the sink block for $\beta>\beta^*$, and $1-2/N_g$ is the minimum of $\gamma_s$ over $\beta$. The threshold $\gamma_s$ depends on $N_c$ only through $\beta^*$. Solving Eqs.~\eqref{eq:SM10_gamma_sink} and \eqref{eq:SM10_gamma_loop} for $\beta$ yields the stability boundaries shown in Fig.~5 of the main text,
\begin{align}
    \beta^2 &= \beta^{*2}\,\frac{1+\gamma}{(N_g-1)(1-\gamma)} \quad (\mathrm{sink}), \label{eq:SM10_boundary_sink}\\
    \beta^2 &= \beta^{*2}\,\frac{1-\gamma}{2-(N_g-1)(1-\gamma)} \quad (\mathrm{loop}), \label{eq:SM10_boundary_loop}
\end{align}
where the loop boundary exists for $\gamma>(N_g-3)/(N_g-1)$. For every $\gamma\in(1-2/N_g,\,1)$, the local stability of the symmetric configuration is reentrant in $\beta$: it is unstable for weak fluctuations, stable for intermediate fluctuations, and unstable again for strong fluctuations.

\newpage
\subsection{Supplementary Methods 11: Phases of the complete bipartite network $K_{4,4}$}

Here we provide the expected dissipation of the phases of $K_{4,4}$ shown in Fig.~\textbf{5} of the main text in closed form, where possible. We obtain the phase diagram from coarse and fine parameter scans with the self-consistent iteration of the Methods, starting from a seed bank of initial conditions that contains representatives of every support class found under the $|\Gamma|=4!\,4!=576$ relabelings of sources and sinks. The seed bank also contains configurations obtained by perturbing the symmetric configuration along the sink and loop blocks of Supplementary Methods~10. Configurations related by a symmetry are identified by mapping their supports onto a canonical representative. We set $\kappa=\mu=1$ in the numerics, but keep $\kappa$ and $\mu$ explicit in the formulas.

\paragraph{Symmetric configuration.} 
All capacities are equal to $k_0=(N_gN_c)^{-1/\gamma}\kappa$. Using $\sum_e k_e\Psi_{ee}=\bar D$ (Supplementary Methods~3) with $\tilde{\Psi}_{ee}$ from Eq.~\eqref{eq:SM10_Psi}, we obtain
\begin{equation}
    \bar D_{\mathrm{s}}=\frac{(N_gN_c)^{1/\gamma}}{\kappa}\left[\frac{N_c\mu^2}{N_g}+\frac{N_g(N_g-1)\beta^2}{N_c}\right].
    \label{eq:bip-sym}
\end{equation}

\paragraph{Hub tree.}
The optimal tree (Fig.~5\textbf{h,l} of the main text) consists of a hub sink connected to all four sources, while each of the three remaining sinks is attached as a leaf to a different source. With $N_g=N_c=4$, the flows can be obtained from Eq.~\eqref{eq:tree_flow_constants}. Each leaf edge supplies the constant demand of its sink, $\langle F_e^2\rangle=\mu^2$. A source with a leaf supplies its mean injection to that leaf, so only its fluctuations flow through its edge to the hub, $\langle F_e^2\rangle=3\beta^2$. The remaining source supplies the hub, $\langle F_e^2\rangle=\mu^2+3\beta^2$. Equation~\eqref{eq:D_open} then yields
\begin{equation}
    \bar D_{\mathrm{hub\,tree}}=\frac{1}{\kappa}\left[3\mu^{\frac{2\gamma}{1+\gamma}}+\left(\mu^2+3\beta^2\right)^{\frac{\gamma}{1+\gamma}}+3\left(3\beta^2\right)^{\frac{\gamma}{1+\gamma}}\right]^{\frac{1+\gamma}{\gamma}} .
    \label{eq:hub-tree}
\end{equation}
The mean load is balanced locally through the leaves, and the fluctuations globally through the hub. Since the capacities on a tree increase with $\langle F_e^2\rangle$, Eq.~(31) of the Methods, the leaf edges ($\mu^2$) are stronger than the hub edges of the sources with a leaf ($3\beta^2$) for $\beta<\mu/\sqrt{3}=\beta^*/\sqrt{3}$, and weaker for larger $\beta$ (Fig.~5\textbf{l} and \textbf{h}). Any permutation of the three sources with a leaf, together with their leaves, leaves the hub tree invariant, i.e., $\Gamma^*\cong S_3$, $|\Gamma^*|=6$. There are 96 equivalent optima.

\paragraph{Remaining phases}

At strong fluctuations, the symmetric configuration loses stability in the sink block, and a hub forms continuously at $\gamma_s(\mathrm{sink})$, Eq.~\eqref{eq:SM10_gamma_sink}.  This forms the phase boundary between the symmetric phase and the hub phase, shown with a dash-dotted line in Fig.~5\textbf{a},\textbf{b}. As $\gamma$ decreases further, the hub network becomes sparser. It passes through configurations with 14 (Fig.~5\textbf{e}) and 10 (Fig.~5\textbf{f}) active edges, and through the hub tree with one additional edge, which closes a single cycle (Fig.~5\textbf{g}, $\varsigma=1$), before the hub tree becomes optimal.

At weak fluctuations, the cube (Fig.~5\textbf{i}), the nine-edge configuration (Fig.~5\textbf{j}) and the alternating ring (Fig.~5\textbf{k}) are equivalent to the phases of the cube embedded in $K_{4,4}$ (Supplementary Note~5). Further support classes with 13, 11, 10 and 9 active edges occupy narrow regions, each below $1\,\%$ of the parameter plane, and are not shown. Supplementary Table~\ref{tab:bipartite-phases} summarizes the phases shown in Fig.~5\textbf{c-l}.

\begin{table*}[b]
\caption{Phases of $K_{4,4}$ shown in Fig.~5\textbf{c-l} of the main text, with the number of active edges $\mathcal{N}^*$, the order of the residual symmetry group $\Gamma^*$, the number $|\Gamma|/|\Gamma^*|$ of equivalent optima, and the expression for the expected dissipation.}
\label{tab:bipartite-phases}
\begin{ruledtabular}
\begin{tabular}{llcccl}
panel & phase & $\mathcal{N}^*$ & $|\Gamma^*|$ & optima & $\bar D$ \\
\hline
c & symmetric & 16 & 576 & 1 & Eq.~\eqref{eq:bip-sym} \\
d & hub, weakly broken & 16 & 144 & 4 & numerical \\
e & hub, 14 edges & 14 & 4 & 144 & numerical \\
f & hub, 10 edges & 10 & 2 & 288 & numerical \\
g & hub tree with one cycle & 8 & 4 & 144 & numerical \\
h, l & hub tree & 7 & 6 & 96 & Eq.~\eqref{eq:hub-tree} \\
i & cube & 12 & 24 & 24 & Eq.~\eqref{eq:cube_D12} \\
j & nine-edge & 9 & 6 & 96 & numerical \\
k & alternating ring & 8 & 4 & 144 & Eq.~\eqref{eq:D_alt}, $N=4$, minimized over $k_a$, $k_b$
\end{tabular}
\end{ruledtabular}
\end{table*}

\newpage 


%